\documentclass[
preprint,
superscriptaddress,
showpacs,
preprintnumbers,
nofootinbib,
amsmath,
amssymb,
aps,
onecolumn,
floatfix,
]{revtex4-1}

\usepackage{graphicx}% Include figure files
\usepackage{dcolumn}% Align table columns on decimal point
\usepackage{bm}% bold math
\usepackage{color}
\usepackage{verbatim}
\usepackage{amsfonts}
\usepackage{geometry}
\usepackage{longtable}
\usepackage{natbib}
\usepackage{hyperref}% add hypertext capabilities
\usepackage[mathlines]{lineno}% Enable numbering of text and display math
\graphicspath{{figs/}}% Put all figures in a `figs' folder

\newcommand{\dls}{\Delta_{l,s}}
\newcommand{\tdls}{\widetilde{\Delta}_{l,s}}

\newcommand{\chid}{\chi_\text{disc}}

\newcommand{\chigd}{\chi_\text{5,disc}}
\newcommand{\chit}{\chi_\text{top}}

\newcommand{\MS}{\overline{\mathrm{MS}}}

\newcommand{\df}{\mathrm{d}}

\newcommand{\mtot}{\widetilde{m}}
\newcommand{\mres}{m_\text{res}}
\newcommand{\ua}{U(1)_A}
\newcommand{\Nt}{N_\tau}
\newcommand{\Ns}{N_\sigma}
\newcommand{\pms}{\chi_\pi-\chi_\sigma}
\newcommand{\dps}{\Delta_{\pi,\sigma}}
\newcommand{\emd}{\chi_\eta-\chi_\delta}
\newcommand{\ded}{\Delta_{\eta,\delta}}
\newcommand{\pmd}{\chi_\pi-\chi_\delta}
\newcommand{\dpd}{\Delta_{\pi,\delta}}
\newcommand{\sme}{\chi_\sigma-\chi_\eta}
\newcommand{\dse}{\Delta_{\sigma,\eta}}
\newcommand{\qtop}{Q_\text{top}}
\newcommand{\sua}{SU(2)_L\times SU(2)_R}

\def\slashed#1{\kern+0.1em /\kern-0.65em #1}

\newcommand{\Tspace}{\rule{0pt}{2.6ex}}
\newcommand{\Bspace}{\rule[-1.2ex]{0pt}{0pt}}

\newcounter{run}
\newenvironment{run}{\refstepcounter{run}\therun}{}
\newcommand{\runlabel}[1]{\begin{run}\label{run:#1}\end{run}}
\newcommand{\runref}[1]{run~\#\,\ref{run:#1}}
\newcommand{\runn}[1]{\ref{run:#1}}

\begin{document}
\title{The QCD chiral transition, \boldmath $\ua$ symmetry and the Dirac spectrum using domain wall fermions}

\collaboration{LLNL/RBC Collaboration}

\author{Michael I.~Buchoff}\affiliation{Physics Division, Lawrence Livermore National Laboratory, Livermore CA 94550, USA}\affiliation{Institute for Nuclear Theory, Box 351550, Seattle, WA 98195-1550, USA}
\author{Michael Cheng}\affiliation{Center for Computational Science, Boston University, Boston, MA 02215, USA}
\author{Norman H.~Christ}\affiliation{Physics Department, Columbia University, New York, NY 10027, USA}
\author{H.-T.~Ding}\affiliation{Physics Department, Columbia University, New York, NY 10027, USA}\affiliation{Physics Department, Brookhaven National Laboratory,Upton, NY 11973, USA}
\author{Chulwoo Jung}\affiliation{Physics Department, Brookhaven National Laboratory,Upton, NY 11973, USA}
\author{F.~Karsch}\affiliation{Physics Department, Brookhaven National Laboratory,Upton, NY 11973, USA}\affiliation{Fakult\"at f\"ur Physik, Universit\"at Bielefeld, D-33615 Bielefeld, Germany}
\author{Zhongjie Lin}\affiliation{Physics Department, Columbia University, New York, NY 10027, USA}
\author{R.~D.~Mawhinney}\affiliation{Physics Department, Columbia University, New York, NY 10027, USA}
\author{Swagato Mukherjee}\affiliation{Physics Department, Brookhaven National Laboratory,Upton, NY 11973, USA}
\author{P.~Petreczky}\affiliation{Physics Department, Brookhaven National Laboratory,Upton, NY 11973, USA}
\author{Dwight Renfrew}\affiliation{Physics Department, Columbia University, New York, NY 10027, USA}
\author{Chris Schroeder}\affiliation{Physics Division, Lawrence Livermore National Laboratory, Livermore CA 94550, USA}
\author{P.~M.~Vranas}\affiliation{Physics Division, Lawrence Livermore National Laboratory, Livermore CA 94550, USA}
\author{Hantao Yin}\affiliation{Physics Department, Columbia University, New York, NY 10027, USA}

\date{September 15, 2013}

\begin{abstract}
We report on a study of the finite-temperature QCD transition region for temperatures between 139 and 196 MeV, with a pion mass of 200 MeV and two space-time volumes: $24^3\times8$ and $32^3\times8$, where the larger volume varies in linear size between 5.6 fm (at T=139 MeV) and 4.0 fm (at T=195 MeV).  These results are compared with the results of an earlier calculation using the same action and quark masses but a smaller, $16^3\times8$ volume.   The chiral domain wall fermion formulation with a combined Iwasaki and dislocation suppressing determinant ratio gauge action are used.  This lattice action accurately reproduces the $\sua$ and $\ua$ symmetries of the continuum.   Results are reported for the chiral condensates, connected and disconnected susceptibilities and the Dirac eigenvalue spectrum.   We find a pseudo-critical temperature, $T_c$, of approximately 165 MeV consistent with previous results and strong finite volume dependence below $T_c$.  Clear evidence is seen for $\ua$ symmetry breaking above $T_c$ which is quantitatively explained by the measured density of near-zero modes in accordance with the dilute instanton gas approximation.
\end{abstract}

\preprint{CU-TP-1204, INT-PUB-13-034, LLNL-JRNL-642513}

%\keywords{Suggested keywords}%Use showkeys class option if keyword
                              %display desired
\maketitle

\section{Introduction}
\label{sec:intro}

The QCD phase transition, separating the low-temperature phase in which the (approximate) $\sua$ symmetry of QCD with two light flavors is broken by the vacuum and the high-temperature phase in which this symmetry is restored, has been the subject of active experimental and theoretical study for more than 30 years.  The present expectation is that this is a second-order transition belonging to the $O(4)$ universality class when the up and down quark masses are zero~\cite{Pisarski:1983ms} and a possibly rapid cross-over for non-zero, physical light quark mass.

However, the order of the transition may depend on the degree to which the anomalous $\ua$ symmetry is realized in QCD.  As pointed out in Ref.~\cite{Pisarski:1983ms}, if the $\ua$ breaking is significant near the phase transition, then the resulting four massless degrees of freedom ($\vec \pi$ and $\sigma$) can support $O(4)$ critical behavior at $T_c$, the location of the phase transition.  However, if anomalous breaking of the $\ua$ is small so there are eight light degrees of freedom at $T_c$ ($\vec \pi$, $\sigma$, $\vec \delta$ and $\eta$) then the chiral transition is expected to be first order, although a second order phase transition may still be permitted with a different $\sua/U(2)_V$ universality class as suggested in Refs.~\cite{Butti:2003nu, Pelissetto:2013hqa}. Thus, a thorough study of the behavior of the anomalous $\ua$ symmetry has essential consequences on the nature of the chiral phase transition.  (For a recent investigation of this question using an effective Lagrangian approach see Ref.~\cite{Meggiolaro:2013swa}.)

In this paper we study the temperature region $139\ \mathrm{MeV} \le T \le 195$ MeV using chiral, domain wall fermions (DWF) with a lattice volume having a fixed time extent of 8 in lattice units and a spatial volume of either $24^3$ or $32^3$.  The temperature is varied by varying the inverse gauge coupling $\beta$ between 1.633 and 1.829 using the Iwasaki gauge action combined with a dislocation suppressing determinant ratio (DSDR)~\cite{Vranas:1999rz, Vranas:2006zk, Fukaya:2006vs,Renfrew:2009wu} to reduce the effects of residual chiral symmetry breaking at these relatively strong couplings.  The light quark mass is chosen so that the pion mass is held fixed at a heavier-than-physical 200 MeV value while the strange quark mass is set to its physical value.  This calculation  extends previous work~\cite{Bazavov:2012qja} that used the same action and studied the same quark masses and temperatures but used a smaller $16^3\times 8$ volume.

While the QCD phase transition has been extensively studied using the staggered formulation of lattice fermions, calculations employing chiral fermions are more difficult and less frequent~\cite{Chen:2000zu, Cheng:2009be, Borsanyi:2012xf, Bazavov:2012qja, Cossu:2013uua}. However, in contrast to the staggered formulation in which finite lattice spacing effects explicitly break the anomalous $\ua$ symmetry and all but one of the six $\sua$ symmetry directions, the DWF formulation accurately reproduces these symmetries.  At low temperatures one finds three degenerate light pions and the $\ua$ current obeys an anomalous conservation law identical to that in the continuum up to small, controlled residual chiral symmetry breaking effects.

We will now briefly summarize our results.  The disconnected chiral susceptibility $\chi_{\mathrm{disc}}$ shows a dramatic peak as the temperature increases through the critical region.  This is the quantity of choice for locating the pseudo-critical temperature and showed a quite broad peak when studied earlier on the $16^3\times 8$ volume.  The $24^3$ and $32^3$ results presented here show a significant volume dependence with the large shoulder just below $T_c$ decreasing by between 30 and 50\% as the volume is increased and the peak itself moving to higher temperature and decreasing in height by approximately 15\%.  The $24^3$ and $32^3$ volumes give similar results.  This behavior is predicted by finite size scaling in O(4) models in the presence of an external symmetry breaking field~\cite{Engels:2013xx} and could be anticipated from the first comparison made with QCD data \cite{Engels:2001bq} and the recent work of Braun {\it et al.} \cite{Braun:2010vd}.  

We investigate $\ua$ symmetry breaking above $T_c$ by examining the two $\ua$ symmetry breaking differences $\pmd$ and $\sme$.  These vanish if $\ua$ symmetry is realized and are clearly non-zero at $T=177$ MeV, although they decrease quickly as $T$ is increased above this value.  These two quantities are related by $\sua$ symmetry and are equal within errors for $T \ge 177$ MeV.  We conclude that for temperatures at which $\sua$ symmetry has been restored, $\ua$ symmetry breaking is still present.

The Dirac eigenvalue spectra per unit space-time volume seen on the $16^3\times 8$ and $32^3\times 8$ volumes are very similar.  However, the larger volume results are more accurate in the region of small eigenvalues.   We find that appropriately convergent combinations of spectral integrals agree well with the observed Green's functions to which they are related in continuum field theory.  Of particular importance is the agreement between a spectral integral and $\pmd$.  For $T=177$ MeV we find a small cluster of near-zero Dirac eigenvalues, such as are expected from the dilute instanton gas approximation (DIGA)~\cite{Callan:1976je, Gross:1980br} and it is these eigenvalues which, when included in the spectral formula, reproduce the measured result for $\pmd$.  This relation continues to hold, although within larger errors, at $T=186$ and 195 MeV.  The number of these near-zero modes is found to be proportional to the volume and their chiralities show a mixture of positive and negative values per configuration, as is expected in the DIGA.  We conclude that $\ua$ symmetry is broken in the region immediately above $T_c$ and this breaking is explained by the DIGA.  No additional mechanism is necessary.

In addition to these physics results, we also present two technical improvements to the study of finite temperature phenomena using DWF.  The first is an improved observable representing the chiral condensate, $\langle\overline{\psi}_l\psi_l\rangle$.  This new quantity, the difference of light and strange quark chiral susceptibilites, is equivalent in the continuum to the usual difference of light and strange quark condensates but does not contain the residual chiral symmetry breaking ambiguities present in the usual DWF evaluation of such a difference.   The second development is the recognition that the quantities usually computed when evaluating susceptiblities and computing residual DWF chiral symmetry breaking, and hence fundamental to this and earlier calculations, are related by an exact DWF Ward identity and the demonstration that this relation is satisfied.

This paper is organized as follows. Section~\ref{sec:setup} briefly describes the lattice formulation used, ensembles generated and the input parameters chosen.  In Sec.~\ref{sec:chiral}, we introduce a variety of observables that are associated with the $\sua$ and $\ua$ symmetries and review their properties and the symmetry relations that connect them.  We present and discuss the results for these observables over our $139-195$ MeV temperature range.  Section~\ref{sec:eigen} gives results for the low-lying eigenvalue spectrum of the Dirac operator and examines the relations between this spectrum and various measures of the chiral condensate and $\pmd$.  Finally in Sec.~\ref{sec:concl}, we summarize our results and compare with earlier work.

\section{Ensemble details}
\label{sec:setup}

In this calculation we extend the $16^3\times8$ results reported in Ref.~\cite{Bazavov:2012qja} to larger  $24^3\times8$  and  $32^3\times8$  volumes, keeping all other parameters fixed.  We therefore adopt the same Iwasaki gauge action augmented with dislocation suppression determinant ratio (DSDR)~\cite{Vranas:2006zk,Fukaya:2006vs,Renfrew:2009wu} and the domain wall fermion (DWF) action with $2+1$ flavors. With this choice of action, we are able to simulate a relatively light pion mass and to accurately respect the important continuum chiral and $\ua$ symmetries.

Table~\ref{tab:para} lists the basic parameters for these three sets of ensembles.  The first two sets of ensembles are new and reported here for the first time, with space-time volumes of $32^3\times8$ and $24^3\times8$ respectively. The third set of ensembles, with lattice volume $16^3\times8$, was studied extensively in Ref.~\cite{Bazavov:2012qja} and is listed here (with improved statistics at $T=$195~MeV (\runref{200})) for comparison and later reference. 

The input light quark masses are adjusted so that all the ensembles lie on a line of constant physics with $m_\pi\approx200~\text{MeV}$ and the ratio $\mtot_l/\mtot_s =0.088$ is fixed to ensure a kaon with physical mass.  Here and later in the text, a tilde indicates the total bare quark mass, given by the sum of the input and the residual quark masses $\mtot=m_\text{input}+\mres$, where the residual mass, $\mres$ is the small additive shift to the input quark mass that results from the residual chiral symmetry breaking with DWF with a finite extent $L_s$ in the fifth dimension.  A detailed description of the determination of the line of constant physics can be found in Ref.~\cite{Bazavov:2012qja}.  Here we recalculate the pion masses at each temperature from updated values of the residual mass computed on the $32^3\times8$ and $24^3\times8$ ensembles.  As can be seen in column nine of Tab.~\ref{tab:para}, in all but one case these new values for $m_\pi$ lie within 3\% of the target value of 200 MeV. Determined as it is here from the sum of input and residual light quark masses and the assumed linearity of $m_\pi^2$ on this sum, the pion mass should be independent of the volume and difference of the calculated pion masses between different volumes can be regarded as a measure of systematic errors.

Because of the rapidly increasing residual mass with decreasing temperature, for the two lowest temperature ensembles ($T=139$ and 149 MeV), we use a negative input quark mass.  While much larger negative input quark masses are standard for Wilson fermion calculations, the use of negative $m_\text{input}$ is uncommon in a DWF calculation and, as in the Wilson case, could potentially jeopardize the stability of the evolution because of a singularity in the Dirac operator. Fortunately, we observed no such "exceptional configurations" in any of our evolutions. This use of a negative input quark mass was tested in a study reported in Ref.~\cite{Bazavov:2012qja} where two streams at $T=149$ MeV with a $16^3\times8$ volume were generated: one with $L_s=32$ and a negative input quark mass (\runref{150_1} in Tab.~\ref{tab:para}) and a second with $L_s=48$ and a positive input quark mass (\runref{150} in Tab.~\ref{tab:para}), adjusted to give the same value of $\mtot_l$.  Both ensembles gave consistent results for all the quantities we computed, providing strong support that our interpretation of  $\mtot_l$ and choice of negative input quark mass is solid and correct. 

\begin{table}[hbt]
  \centering
    \begin{tabular}{c|ccccc|ccccc}%\hline
      \#&$T\,(\textrm{MeV})$&$\beta$&$\Ns$&$\Nt$&$L_s$
      &$m_l$&$m_s$&$\mres$&$m_\pi$(MeV)&$N_\text{traj}^\text{equil}$\\\hline
      \runlabel{140_a}&139(6)&1.633&32&8&48&-0.00136&0.0519&0.00657(2) &205(8)&2700\\
      \runlabel{150_a}&149(5)&1.671&32&8&32&-0.00189&0.0464&0.00653(2) &201(5)&2700\\
      \runlabel{160_a}&159(4)&1.707&32&8&32&0.000551&0.0449&0.00366(2) &200(3)&2643\\
      \runlabel{164_a}&164(4)&1.725&32&8&32&0.00138 &0.0436&0.00277(1) &202(3)&2700\\
      \runlabel{170_a}&168(4)&1.740&32&8&32&0.00175 &0.0427&0.00220(2) &200(2)&2708\\
      \runlabel{180_a}&177(4)&1.771&32&8&32&0.00232 &0.0403&0.00135(1) &198(2)&2700\\
      \runlabel{190_a}&186(5)&1.801&32&8&32&0.00258 &0.0379&0.00083(2) &197(3)&2729\\
      \runlabel{200_a}&195(6)&1.829&32&8&32&0.00265 &0.0357&0.00049(1) &195(4)&3112\\
      \hline\hline
      \runlabel{150_b}&149(5)&1.671&24&8&32&-0.00189&0.0464&0.00659(6) &202(5)&4721\\
      \runlabel{160_b}&159(4)&1.707&24&8&32&0.000551&0.0449&0.00370(4) &200(3)&2265\\
      \runlabel{170_b}&168(4)&1.740&24&8&32&0.00175 &0.0427&0.00216(3) &199(2)&2423\\
      \runlabel{180_b}&177(4)&1.771&24&8&32&0.00232 &0.0403&0.00129(3) &197(2)&2892\\
      \runlabel{190_b}&186(5)&1.801&24&8&32&0.00258 &0.0379&0.00084(3) &197(3)&3142\\
      \hline\hline
      \runlabel{140}  &139(6)&1.633&16&8&48&-0.00136&0.0519&0.00588(39)&191(7)&2696\\
      \runlabel{150_1}&149(5)&1.671&16&8&32&-0.00189&0.0464&0.00643(9) &199(5)&5700\\
      \runlabel{150}  &149(5)&1.671&16&8&48&0.00173 &0.0500&0.00295(3) &202(5)&6700\\
      \runlabel{160}  &159(4)&1.707&16&8&32&0.000551&0.0449&0.00377(11)&202(3)&3359\\
      \runlabel{170}  &168(4)&1.740&16&8&32&0.00175 &0.0427&0.00209(9) &197(2)&3043\\
      \runlabel{180}  &177(4)&1.771&16&8&32&0.00232 &0.0403&0.00132(6) &198(2)&3240\\
      \runlabel{190}  &186(5)&1.801&16&8&32&0.00258 &0.0379&0.00076(3) &195(3)&4415\\
      \runlabel{200}  &195(6)&1.829&16&8&32&0.00265 &0.0357&0.00047(1) &194(4)&8830\\
      \hline
    \end{tabular}
  \caption{
  \label{tab:para}
Summary of input parameters ($\beta$, $N_\sigma$, $N_\tau$, $L_s$, $m_l$ and $m_s$) and the measured result for $\mres$ for each ensembles.  Each is assigned a label in the first column for later reference. The final $N_\text{traj}^\text{equil}$ column lists the number of equilibrated trajectories that remain after the imposition of the thermalization and decorrelation cuts described in the text.
  }
\end{table}

The number of effective trajectories for each ensemble that are used in the measurement reported later  is also in the right-most column of Table~\ref{tab:para}. For ensembles with volume $16^3\times8$ and $32^3\times8$, we discard the first 300 trajectories to account for thermalization. However, because we
changed the evolution algorithm during the early stages of the generation of  the $24^3\times8$
ensembles, a larger number of initial trajectories were discarded for those.  For each ensemble a trajectory has a length of one molecular dynamics time unit.

In order to increase the statistics, we have evolved multiple streams for ensembles~\runref{150_b} and~\runref{160_b}.  Ensemble~\runref{150_b} is composed of 8 streams, two of which began from an ordered start, another two from a disordered start and the remaining four were split from the previous four streams after thermalization.  Ensemble~\runref{160_b} is composed of two streams one beginning from an ordered and the other from a disordered configuration. The multiple streams in each ensemble are pooled together after removing an initial 300 trajectories from each stream which began with an ordered or disordered start.  For streams that were split from a previously thermalized stream, the first 100 trajectories of that new stream are discarded to insure that the new stream is not correlated with its parent.

We do not adopt a single set of units in this paper.  When dimensionful quantities are given in physical units, such as MeV, the unit used will be specified.  However, when expressed in lattice units, often no explicit unit will be written.  Occasionally, for clarity or emphasis, explicit powers of the lattice spacing will be shown, with the power given by the length dimension of the quantity being described.

\section{Chiral observables}
\label{sec:chiral}
In this Section we will discuss Green's functions constructed from the eight scalar and pseudoscalar operators: $\overline{\psi}_l\psi_l$, $\overline{\psi}_l\tau^i\psi_l$, $\overline{\psi}_l\gamma^5\psi_l$, $\overline{\psi}_l\tau^i\gamma^5\psi_l$.  Here $\psi_l$ is a doublet of up and down quark fields and $\{\tau_i\}_{1 \le i \le 3}$ the usual Pauli matrices.  These operators are related by the $\sua$ chiral symmetry of QCD and the anomalously broken $\ua$ symmetry.  In Sec.~\ref{sec:prelim} we review the relations among these eight operators and their Green's functions implied by the $\sua$ and $\ua$ symmetries, paying particular attention to the degree to which these relations should hold at finite lattice spacing for the DWF formulation.

In Sec.~\ref{sec:su2} we present our numerical results, focusing on those relations implied by $SU(2) \times SU(2)$ chiral symmetry and examining their dependence on temperature.  In the final subsection, Sec.~\ref{sec:ua1}, we examine the relations implied by $\ua$ symmetry, including evidence for non-zero anomalous, $\ua$ symmetry breaking above the pseudo-critical temperature $T_c$, a non-vanishing asymmetry which disappears rapidly as the temperature increases above $T_c$.

\subsection{Preliminaries}
\label{sec:prelim}

In this section, we present a brief review of a variety of chiral observables and the relations among them implied by the $\sua$ and $\ua$ symmetries.  A more detailed description can be found in Ref.~\cite{Bazavov:2012qja}.

The standard order parameter for the chiral phase transition is the single-flavor, light-quark chiral condensate,
\begin{eqnarray}
  \Sigma_l &\equiv& -\frac{1}{2}\left\langle\overline{\psi}_l\psi_l\right\rangle          \label{eq:pbp1} \\
                &=&           \frac{1}{2}\frac{T}{V}\frac{\partial \ln Z}{\partial m_l}             \label{eq:pbp2} \\
                &=&           \frac{1}{\Ns^3\Nt}\left\langle\textrm{Tr}M_l^{-1}\right\rangle, \label{eq:pbp3}
\end{eqnarray}
where $M_l$ is the single-flavor, light-quark Dirac matrix and the brackets $\langle\ldots\rangle$ in the bottom equation indicate an average over gauge fields.   However, this quantity contains an ultraviolet divergent contribution that is proportional to $m_q/a^2$ for the case of a lattice regularization.  In order to remove this ultraviolet divergence, it is standard to introduce a subtracted chiral condensate constructed from a weighted difference between the chiral condensates of the light and strange quarks~\cite{Cheng:2007jq}:
\begin{equation}
  \dls=\Sigma_l-\frac{\mtot_l}{\mtot_s}\Sigma_s.
  \label{eq:spbp}
\end{equation}
Here $\Sigma_s$ is defined using the strange quark Dirac matrix in a manner analogous to Eq.~\eqref{eq:pbp3}.  For domain wall fermions there is a further difficulty associated with the short distance contributions to $\Sigma_q$ and the subtracted quantity $\dls$.    For a finite fifth dimensional extent,  $L_s < \infty$, the DWF chiral symmetry is only approximate and residual chirally symmetry breaking effects appear.  The largest such effect is a small additive shift in the quark mass: the residual mass $\mres$  mentioned above.   Similar residual chiral breaking will appear in $\Sigma_q$ and will be of order $\mres/a^2$ if we express $\mres$ in physical units.  However, since the detailed mechanism which generates the residual mass is not directly related to that which introduces the additive constant into  $\Sigma_q$, the subtraction coefficient $\alpha$ that would be needed to remove both the $m_q/a^2$ and the $O(\mres/a^2)$ terms in $\Sigma_l - \alpha\Sigma_s$ is not known.  

Thus, the subtracted quantity $\dls$ defined in Eq.~\eqref{eq:spbp} will contain an unphysical, $O(\mres/a^2)$ constant which will decrease the utility of $\dls$ computed in a DWF simulation.  In particular, we cannot compare $\dls$ with the same difference of chiral condensates obtained from other lattice fermion formulations.  While this added unphysical constant does not depend on temperature, it does depend strongly on the gauge coupling $g$ so the usual procedure of varying the temperature by varying $g$ at fixed $N_\tau$ will induce an apparent temperature dependence in this unphysical contribution to $\dls$.   However, the definition of $\dls$ given in Eq.~\eqref{eq:spbp} (which differs from that used in the earlier paper~\cite{Bazavov:2012qja}) does have a useful property.  As is discussed in Sec.~\ref{sec:eigen}, this subtraction using for $\alpha$ the physical quark mass ratio, $\alpha=\mtot_l/\mtot_s$ will lead to a more convergent spectral expression for $\dls$.

Results for the quantities $\Sigma_l$, $\Sigma_s$ and $\dls$ are given in Tab.~\ref{tab:pbp}.  For each configuration used in the calculation, the volume-averaged, chiral condensate is computed from the right hand side of Eq.~\eqref{eq:pbp3}, using 10 Gaussian random volume sources to estimate the trace.  In Sec.~\ref{sec:chiral} we will use the Gell-Mann-Oakes-Renner (GMOR) relation to define an improved, subtracted chiral condensate $\tdls$, which contains a much smaller unknown correction and can be compared with the results from other formulations of lattice fermions.

The chiral condensate $\Sigma_l$ and the various subtracted versions discussed above can be used to explore the vacuum breaking of $\sua$ and $\ua$ symmetry and their restoration (or partial restoration) as the temperature is increased.  However, much more information can be obtained from the susceptibilities defined as integrated correlation functions of the eight local operators,
\begin{eqnarray}
  \sigma &=& \overline{\psi}_l \psi_l \\
  \delta^i &=& \overline{\psi}_l \tau^i \psi_l \\
  \eta      &=& i\overline{\psi}_l \gamma^5 \psi_l \\
  \pi^i      &=& i\overline{\psi}_l \tau^i \gamma^5 \psi_l.
\label{eq:op_def}
\end{eqnarray}
Such susceptibilities are both much more sensitive to the transition from the ordered to the disordered state and also allow independent measures of $\sua$ and $\ua$ symmetry breaking.  The operator quadruplets $(\sigma,\pi^i)$ and $(\eta,\delta^i)$ each transform as an irreducible 4-dimensional representation of $\sua$.  The four pairs, $(\sigma,\eta)$, $(\delta^i,\pi^i)_{1 \le i \le 3}$ each transform the simple, two-dimensional representation of $\ua$.   We then identify the four distinct susceptibilities which are allowed by isospin symmetry:
\begin{eqnarray}
\chi_\sigma &=& \frac{1}{2}\int d^4 x \left\langle \sigma(x) \sigma(0) \right\rangle \label{eq:chi_sigma_def} \\
\chi_\delta   &=& \frac{1}{2}\int d^4 x \left\langle \delta^i(x) \delta^i(0) \right\rangle  \\
\chi_\eta      &=& \frac{1}{2}\int d^4 x \left\langle \eta(x)     \eta(0) \right\rangle       \\
\chi_\pi        &=& \frac{1}{2}\int d^4 x \left\langle \pi^i(x)      \pi^i(0) \right\rangle
\end{eqnarray}
where the factor 1/2 has been introduced so that these correspond to the single flavor quantities that are typically computed using lattice methods and no sum over the repeated index $i$ is intended.  In light of the multiplet structure defined above, the following relations are implied by $\sua$ and $\ua$ symmetry:
\begin{eqnarray}
\left.
\begin{array}{lcr}
\chi_\sigma & = & \chi_\pi \\
\chi_\eta      & = & \chi_\delta
\end{array} \right\}& \quad& \sua, \label{eq:su2_sym} \\
\left.
\begin{array}{lcr}
\chi_\sigma & = & \chi_\eta \\
\chi_\pi        & = & \chi_\delta
\end{array} \right\} &\quad& \ua.                          \label{eq:ua1_sym} 
\end{eqnarray}

These susceptibilities can be written in terms of the Dirac operator $M_l$.  For the correlators of the operators $\pi^i$ and $\delta^i$, which introduce non-zero isospin, only connected combinations appear:
\begin{eqnarray}
\chi_\pi      &=&  \frac{1}{\Ns^3\Nt} \textrm{Tr}\left\langle\gamma^5 M_l^{-1}\gamma^5 M_l^{-1}\right\rangle \\
\chi_\delta &=& -\frac{1}{\Ns^3\Nt}\textrm{Tr}\left\langle M_l^{-1}M_l^{-1}\right\rangle
\label{eq:chide}
\end{eqnarray}
where the notation ``Tr'' indicates a trace over spinor and color indices as well as the space-time volume.  The $\sigma$ and $\eta$ susceptibilities are a combination of the connected parts which appear in $\chi_\delta$ and $\chi_\pi$ respectively and a disconnected part:
\begin{eqnarray}
  \chi_\sigma &=& \chi_\delta + 2 \chid     \label{eq:chid}  \label{eq:chid1}\\
  \chi_\eta &=& \chi_\pi           - 2 \chigd    \label{eq:chi5d} \label{eq:chid2}
\end{eqnarray}
where the disconnected parts $\chid$ and $\chigd$ are given by
\begin{eqnarray} 
\chid   &=& \frac{1}{N_\sigma^3 N_\tau}\left\{  \left\langle\left(\textrm{Tr}M_l^{-1}\right)^2\right\rangle
                                                       - \left(\left\langle\textrm{Tr}M_l^{-1}\right\rangle\right)^2\right\}  \\
\chigd &=& \frac{1}{N_\sigma^3 N_\tau}  \left\langle\left(\textrm{Tr}M_l^{-1}\gamma^5\right)^2\right\rangle.
\label{eq:chisi}
\end{eqnarray}
As is conventional, we have removed the truly disconnected piece $2N_s^3N_\tau\Sigma_l^2$ from the expression for $\chi_\sigma$ given in Eq.~\eqref{eq:chid1}.  This extra term would appear if the right hand side of the definition given by Eq.~\eqref{eq:chi_sigma_def} where completely evaluated. The factor of two that appears in Eqs.~\eqref{eq:chid} and \eqref{eq:chi5d} was mistakenly omitted from the published version of Ref.~\cite{Bazavov:2012qja} and arises when these relations are written in terms of single flavor quantities.  The signs of $\chid$ and $\chigd$ have been chosen so that each is positive.

We can combine Eqs.~\eqref{eq:su2_sym}, \eqref{eq:chid1} and \eqref{eq:chid2} to obtain relations between the $\ua$ symmetry breaking difference $\chi_\pi - \chi_\delta$ and $\chid$ and $\chigd$ if $\sua$ symmetry is assumed:
\begin{eqnarray}
\chi_\pi - \chi_\delta &=& (\chi_\pi - \chi_\sigma) + (\chi_\sigma-\chi_\delta) \\
                                &=& 2\chid \\
                                &=& 2\chigd 
\end{eqnarray}
where the second equation is true if the $\sua$ relation $\chi_\pi  = \chi_\sigma$ of
Eq.~\eqref{eq:su2_sym} is valid while the third is obtained by a similar manipulation and the second $\sua$ relation $\chi_\delta = \chi_\eta$.

The connected Green's functions can be computed from the lattice by integrating the two-point correlators from a point source over the whole volume. This method was used for the calculations on the $24^3\times8$ ensembles as well as our earlier study of the $16^3\times8$ ensembles in~\cite{Bazavov:2012qja}. On the $32^3\times8$ ensembles and for the $16^3\times8$ results presented in this report, we achieved a reduced statistical error by using instead a random $Z_2$ wall source. The disconnected parts are calculated by averaging products of chiral condensates where the stochastic evaluation of the trace appearing in each factor is obtained from  different stochastic sources.

The $\sua$ relations given in Eq.~\eqref{eq:su2_sym} should be valid in the continuum for $T >T_c$ when $\sua$ becomes an accurate symmetry.  They should also be true when $T> T_c$ in a lattice formulation which preserves chiral symmetry.   However, for our DWF formulation we should expect deviations arising from residual chiral symmetry breaking.  For low energy quantities, $\mres$ should provide a good measure of this residual chiral symmetry breaking, with effects that are well described as arising simply from the total bare quark mass $\mtot = m_l+\mres$.  

However, the four susceptibilities being discussed are not simple long-distance quantities since the space-time integrals that appear in their definitions include points where the two local operators collide.  In fact, the connected parts of the susceptibilities contain quadratic divergences while the disconnected parts diverge logarithmically.  The presence of quadratic divergences in the connected susceptibilities, {\it e.g.}  $\chi_\pi$ and $\chi_\delta$, can be easily deduced from the Wilson operator product expansion and dimensional arguments.  The product of two dimension-three fermion bilinears separated by a space-time distance $x$ should contain a constant behaving as $1/x^6$ as $x \to 0$.  When integrated over space-time to form the susceptibility, this $1/x^6$ term will give a quadratic divergence.  For the disconnected parts of the susceptibilities, a similar dimensional argument applies.   However, the disconnected parts are constructed from the product of two independent fermion loops, each evaluated as a separate trace.  For the case of scalar or pseudoscalar susceptibilities, chiral symmetry requires that each trace be proportional to $m_l$ so the product will behave as $m_l^2/x^4$ leading to a logarithmic divergence multiplied by the very small factor $m_l^2$.   Thus, if the continuum regulator respects chiral symmetry, then the $\sua$ and $\ua$ breaking differences $\chi_\pi - \chi_\sigma$, $\chi_\delta - \chi_\eta$, $\chi_\pi - \chi_\delta$ and $\chi_\eta - \chi_\sigma$ will all contain only small, logarithmic singularities proportional to $m_l^2 \ln(m_l/\Lambda)$ if evaluated in order-by-order in QCD perturbation theory, where $\Lambda$ is the continuum cutoff scale.

In our lattice-regulated domain wall theory, the residual chiral symmetry breaking will result in these same differences containing small unphysical pieces of order $\mres^2$.  As in the case of the chiral condensate, $\mres$ does not literally enter these differences but instead we expect that $\mres^2$ will provide a reasonable estimate of their size.   Note, when expressed in physical units $\mres \sim e^{-\alpha L_s}/a$ so that our estimate $\mres^2 \sim e^{-2\alpha L_s}/a^2$ of a chiral symmetry breaking difference remains quadratically divergent but is suppressed by the same factor that makes $\mres^2$ small.   (Here, for simplicity, we assume that the residual chiral symmetry breaking effects fall exponentially with increasing $L_s$, with an exponent $\alpha$, unrelated to the $\alpha$ used earlier in this Section.)  For the purposes of this paper $\mres^2 \sim (10\; \mathrm{MeV})^2$, a quantity that is negligible on the $(\Lambda_\mathrm{QCD})^2 \approx (300\; \mathrm{MeV})^2$ scale of the physical parts of the susceptibilities being subtracted.

Finally we examine two additional identities that hold in the continuum limit.  The first is the relation between $\chigd$ and the topological susceptibility $\chit$.  This relation begins with the identity
\begin{equation}
\qtop =  m_l^c\, \textrm{Tr}\left\{ \gamma^5 \frac{1}{M_l}\right\}
\label{eq:AS}
\end{equation}
which for the continuum theory will hold for each gauge configuration.  Here for clarity we have introduced the quantity $m_l^c$ to represent the light quark mass in the continuum theory.  This is easily understood by using a sum over Dirac operator eigenvectors to evaluate the trace and recognizing that the result is simply the number of right- minus the number of left-handed zero modes~\cite{Brown:1977bj} which is equal to $\qtop$ by the Atiyah-Singer theorem.   Recall that 
\begin{equation}
  \qtop=\frac{g^2}{32\pi^2}\int\df^4x F^a_{\mu\nu}(x)\widetilde{F}^a_{\mu\nu}(x).
  \label{eq:qtop}
\end{equation}
Here $\widetilde{F}_{\mu\nu} = \frac{1}{2}\sum_{\rho\sigma}\epsilon_{\mu\nu\rho\sigma}F_{\rho\sigma}$ where $\epsilon_{\mu\nu\rho\sigma}$ is the usual anti-symmetric Levi-Civita tensor with $\epsilon_{1234}=1$.

The desired identity:
\begin{equation}
\chit = (m_l^c)^2 \chigd
\label{eq:qsuscp}
\end{equation}
is simply the ensemble average of the square of Eq.~\eqref{eq:AS}.  This continuum equation should also relate DWF lattice quantities provided the total bare quark mass $\mtot$ is used in place of the continuum mass $m_l^c$.  As was explored at length in Ref.~\cite{Bazavov:2012qja}, this relation is badly violated for our lattice calculation because at our relatively coarse lattice spacing the quantity $\qtop$ is difficult to compute directly.  The right hand side of Eq.~\eqref{eq:qsuscp} appears to nicely define the topological susceptibility giving the same answer even when the light quark quantity $\mtot_l^2 \chigd$ is replaced with the corresponding strange quark quantity or the product of strange and light quark expressions.  (Note the right hand side of Eq.~\eqref{eq:AS} is expected to give the same result on a given gauge configuration independent of the quark mass.)  For completeness $\chigd/T^2$ and $\chit/(\mtot_l T_c)^2$ are tabulated in the two right-most columns of Tab.~\ref{tab:pbp}, where $\chit$ is computed using the procedure described in Ref.~\cite{Bazavov:2012qja}.  As can be seen in Tab.~\ref{tab:pbp} , their disagreement is substantial.  However, the fractional discrepancy does decrease with increasing temperature (and decreasingly lattice spacing) as should be expected if this is a finite lattice spacing artifact.  We will not make further use of $\chit$.

The second identity is the usual Ward identity connecting $\chi_\pi$ and the chiral condensate.  This can be derived in the continuum for non-zero quark mass by evaluating the following integrated divergence:
\begin{eqnarray}
0 &=& \int d^4 x \partial_\mu \left\langle 0| T\left(A^{a\mu}(x) \pi^b(0) \right)|0\right\rangle \\
   &=&  \int d^4 x \left\langle 0| T\left(-2 m_l^c i \pi^a(x) \pi^b(0) \right)|0\right\rangle 
             -2i\left\langle 0| \sigma(0)|0 \right\rangle\delta^{ab}
\end{eqnarray}
where $a$ and $b$ are isospin indexes.  Here the left term in the second line comes from the divergence of the axial current, $\partial_\mu A(x)^{a\mu}$, while the right term results from the equal-time commutator that arises when the partial derivative with respect to the time is brought inside the time-ordered product.  The result is the Gell-Mann-Oakes-Renner relation~\cite{Gell-Mann:1968rz}:
\begin{equation}
m_l^c \chi_\pi = \Sigma_l.
\label{eq:GMOR}
\end{equation}
While this relation should be true in a continuum theory which has been regulated in a chirally symmetric way, both the right- and left-hand sides of Eq.~\eqref{eq:GMOR} contain quadratic divergences as discussed earlier.  Thus, we should not expect this equation to be obeyed in our DWF theory unless we take the limit of infinite $L_s$ at finite $a$ so that our theory has an exact chiral symmetry.

\begin{figure}[htb]
  \begin{center}
    \begin{minipage}[t]{0.5\linewidth}
      \centering
      \resizebox{\linewidth}{!}{\includegraphics{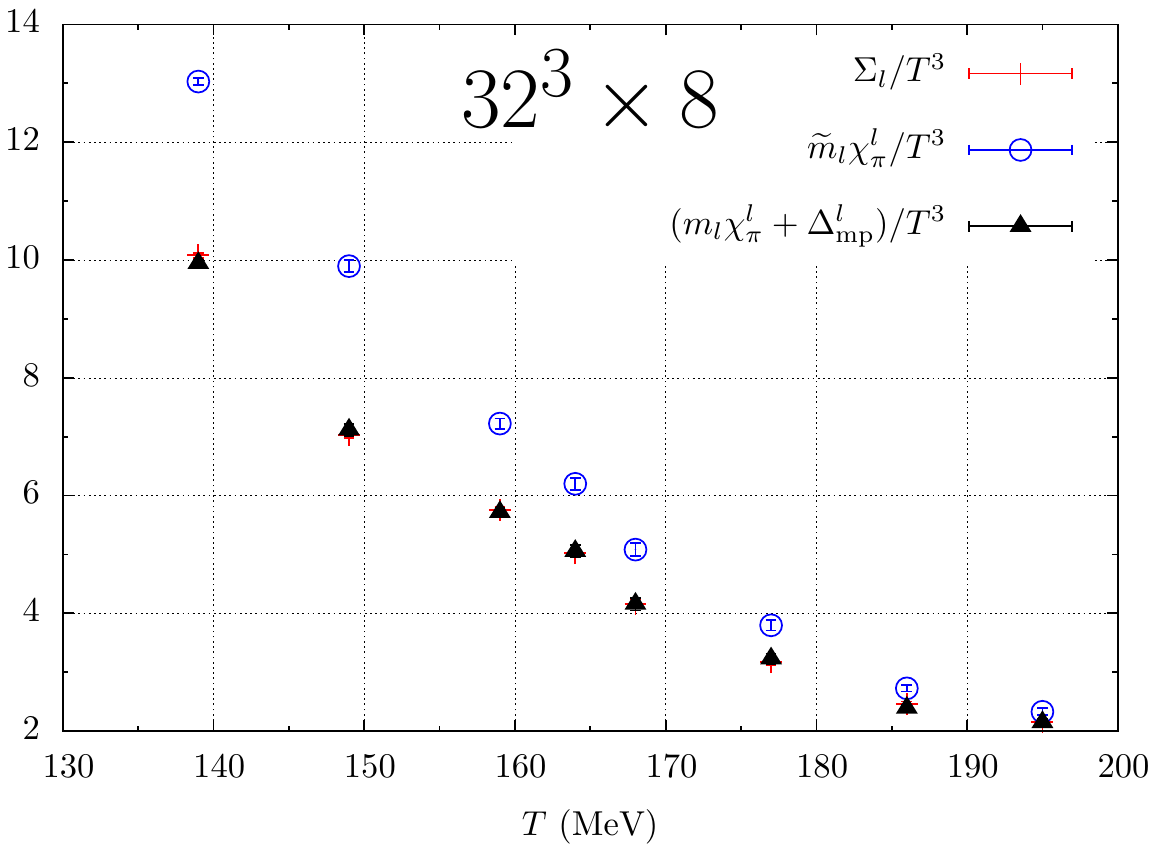}}
    \end{minipage}%
    \begin{minipage}[t]{0.5\linewidth}
      \centering
      \resizebox{\linewidth}{!}{\includegraphics{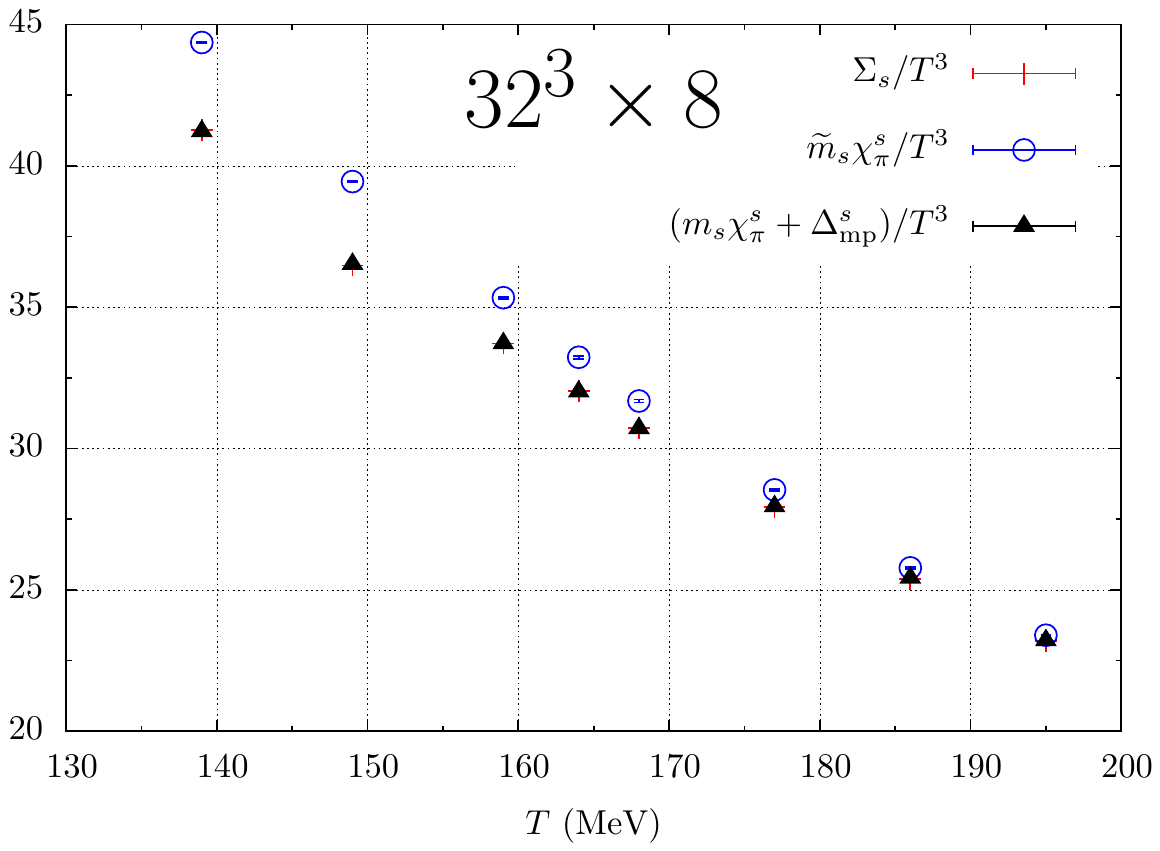}}
    \end{minipage}
  \end{center} 
  \caption{The left panel shows the light-quark chiral condensate, $\Sigma_l$, and the sum of $m_l\chi_\pi$ and the mixed $\pi-J_{5q}/2$ susceptibility to which it should be equal according to the Ward identity in Eq.~\eqref{eq:GMOR_DWF}.  Also shown is $(m_l+\mres)\chi_\pi\equiv\mtot_l\chi_\pi$ which would equal $\Sigma_l$ if $\mres$ were the only effect of residual chiral symmetry breaking.  The right panel shows the same quantities computed using the strange instead of the light quark.  Similar agreement between the right and left hand sides of Eq.~\eqref{eq:GMOR_DWF} is found for the $24^3$ and $16^3$ volumes, as can be seen from Tab.~\ref{tab:WI}}
  \label{fig:WI}
\end{figure}

However, this equation has two important uses.  First, we can repeat its derivation in our lattice theory using the partially conserved, 5-dimensional axial current ${\cal A}^{a\mu}$ constructed by Furman and Shamir~\cite{Furman:1995ky} and the divergence equation obeyed by ${\cal A}^{a\mu}$:
\begin{equation}
\partial_\mu {\cal A}^{a\mu} = -2i m_l \pi^a + 2J_{5q}^a
\label{eq:axial_div}
\end{equation}
where the definition of the ``mid-point term'' $J_{5q}^a$ can be found in Ref.~\cite{Blum:2000kn}.  When used in the above derivation this relation yields the lattice identity:
\begin{equation}
2 m_l \chi_\pi + \int d^4 x \left\langle 0| T\left(iJ_{5 q}(x)^a \pi^a(0) \right)\right\rangle = 2 \Sigma_l
\label{eq:GMOR_DWF}
\end{equation}
for $a=1$, 2 and 3.  In the usual application of Eq.~\eqref{eq:axial_div}, $iJ_{5q}^a$ is replaced in Eq.~\eqref{eq:GMOR_DWF} by $\mres \pi^a$ which would provide a DWF derivation of Eq.~\eqref{eq:GMOR} in which the continuum light quark mass $m_l^c$ is replaced by $\mtot=m_l+\mres$.  However, the low-energy relation $iJ_{5q}^a \approx \mres \pi^a$ cannot be used here because short-distances are involved.  Never-the-less, we can simply evaluate both sides of Eq.~\eqref{eq:GMOR_DWF} in our lattice calculation as a check of this discussion and find agreement within errors.  Our numerical results for the three quantities which appear in Eq.~\eqref{eq:GMOR_DWF}  are tabulated in Tab.~\ref{tab:WI} for each of the seven temperatures studied as well as the right- and left-hand sides of  Eq.~\eqref{eq:GMOR_DWF} after a common factor of 2 has been removed.  We also plot in Fig.~\ref{fig:WI} both the left- and right-hand sides of Eq.~\eqref{eq:GMOR_DWF} as well $(m_l+\mres)\chi_\pi$, as the result of the naive use of the low-energy relation $iJ_{5q}^a \approx \mres \pi^a$.  The left panel of Fig.~\ref{fig:WI} shows these quantities for the light-quark case discussed here while the right panel shows  the same quantities computed using the strange quark. In both Tab.~\ref{tab:WI} and Fig.~\ref{fig:WI}, the mixed susceptibility appearing in Eq.~\eqref{eq:GMOR_DWF} is represented by $\Delta_{\rm mp}^f$ where
\begin{equation}
\Delta_{\rm mp}^f = \int d^4 x \left\langle 0| T\left(iJ_{5 q}^{(f)}(x) \pi^{(f)}(0) \right)\right\rangle.
\end{equation}
where in this equation we construct the quark bi-linears $J_{5q}^{(f)}$ and $\pi^{(f)}$ from a single flavor of quark specified by $f=l$ or $s$ and include only connected graphs, in which the quark fields are contracted between $J_{5q}$ and $\pi$.  In these tables and figures and those which follow, when a combination of quantities that were computed separately are combined, such as $m_l\chi_\pi^l + \Delta_{\rm mp}^l$, we will use the jackknife method with data that has been averaged over bins of 50 configurations to compute the error on the combined quantity so that the effects of statistical correlations between the quantities being combined are included.  However, for simplicity, if a computed renormalization factor, factor of $a$ expressed in physical units or factor of $\mres$ appears, these factors usually have smaller errors than the quantities they multiply and their fluctuations will be ignored.

\begin{table}[H]
  \centering
  \vskip -0.75in
  \begin{tabular}{c|cc|cc|cc|cc|c}\hline
    \#&
    $T\,(\textrm{MeV})$
    &$\beta$
    &$\chi_{\pi}^l/T^2$
    &$\chi_{\pi}^s/T^2$
    &$\Delta_{\rm mp}^l/T^3$
    &$\Delta_{\rm mp}^s/T^3$
    &$\frac{m_l\chi_\pi^l + \Delta_{\rm mp}^l}{T^3}$
    &$\frac{m_s\chi_\pi^s + \Delta_{\rm mp}^s}{T^3}$
    &$\Sigma_l/T^3$
    \\ \hline
    \runn{140_a}&139&1.633&313(2) &94.83(7) &13.34(8) &1.833(11)&9.94(6) &41.21(2)&10.07(4) \\
    \runn{150_a}&149&1.671&267(3) &93.15(7) &11.14(14)&1.939(10)&7.11(10)&36.52(3)& 7.03(6) \\
    \runn{160_a}&159&1.707&214(3) &90.96(10)& 4.77(7) &1.038(6) &5.71(9) &33.71(5)& 5.80(6) \\
    \runn{164_a}&164&1.725&187(3) &89.57(12)& 2.99(7) &0.757(5) &5.05(10)&32.00(5)& 5.02(7) \\
    \runn{170_a}&168&1.740&161(3) &88.20(14)& 1.91(6) &0.576(5) &4.16(11)&30.70(7)& 4.16(8) \\
    \runn{180_a}&177&1.771&129(3) &85.64(11)& 0.83(3) &0.329(2) &3.23(9) &27.94(3)& 3.17(5) \\
    \runn{190_a}&186&1.801&100(2) &83.20(11)& 0.33(1) &0.193(2) &2.39(6) &25.42(4)& 2.46(4) \\
    \runn{200_a}&195&1.829& 93(2) &80.81(9) & 0.18(1) &0.118(1) &2.15(6) &23.20(2)& 2.15(3) \\
    \hline\hline                                                
    \runn{150_b}&149&1.671&270(13)&93.2(7)  &11.6(7)  &2.02(7)  &7.5(5)  &36.6(3) & 7.10(6) \\
    \runn{160_b}&159&1.707&198(11)&90.6(6)  & 4.3(3)  &1.05(4)  &5.2(4)  &33.6(3) & 5.58(10)\\
    \runn{170_b}&168&1.740&164(8) &89.6(6)  & 1.96(15)&0.61(3)  &4.3(3)  &31.2(2) & 4.40(10)\\
    \runn{180_b}&177&1.771&124(10)&85.7(5)  & 0.79(12)&0.33(2)  &3.1(3)  &28.0(2) & 3.03(7) \\
    \runn{190_b}&186&1.801& 99(3) &82.6(4)  & 0.31(2) &0.184(7) &2.35(8) &25.2(1) & 2.58(6) \\
    \hline\hline
    \runn{140}  &139&1.633&302(5) &95.0(2)  &12.6(2)  &1.825(21)&9.30(18)&41.26(8)& 9.26(13)\\
    \runn{150_1}&149&1.671&247(5) &93.0(1)  &10.1(2)  &1.922(13)&6.34(14)&36.43(6)& 6.26(12)\\
    \runn{150}  &149&1.671&257(3) &93.6(1)  & 4.84(8) &0.815(7) &8.40(12)&38.24(6)& 8.39(10)\\
    \runn{160}  &159&1.707&189(5) &90.8(2)  & 4.09(16)&1.034(10)&4.92(19)&33.64(7)& 5.25(17)\\
    \runn{170}  &168&1.740&155(6) &88.3(2)  & 1.83(11)&0.573(7) &4.00(19)&30.73(8)& 4.03(18)\\ 
    \runn{180}  &177&1.771&127(7) &85.5(2)  & 0.80(7) &0.326(4) &3.15(19)&27.89(7)& 3.16(15)\\ 
    \runn{190}  &186&1.801&102(4) &83.5(2)  & 0.35(3) &0.196(3) &2.46(11)&25.50(6)& 2.44(9) \\
    \runn{200}  &195&1.829& 91(2) &80.9(1)  & 0.17(1) &0.118(1) &2.10(5) &23.22(4)& 2.10(5) \\
    \hline
  \end{tabular}
  \caption{
  \label{tab:WI}
The unrenormalized iso-vector pseudoscalar and mixed pseudoscalar/mid-point susceptibilities for the light and strange quarks as well as the combinations $(m_q\chi_\pi^q + \Delta_{\rm mp}^q)/T^3$ for $q=l,s$, which appear in the Ward identity, Eq.~\eqref{eq:GMOR_DWF}. The Ward identity requires the right and third-from-right columns to agree as well as agreement between the column second from the right above and the fifth column from the left in Tab.~\ref{tab:pbp}.  Moving from top to bottom, the three sections in this table correspond to the volumes $32^3\times8$, $24^3\times8$ and $16^3\times8$.}
\end{table}

%\begin{figure}[htb]
%  \begin{center}
%    \begin{minipage}[t]{0.5\linewidth}
%      \centering
%      \includegraphics[width=\linewidth]{./figs/chipi_pbp_WI_Ns32.eps}
%    \end{minipage}%
%    \begin{minipage}[t]{0.5\linewidth}
%      \centering
%      \includegraphics[width=\linewidth]{./figs/chipis_pbps_WI_Ns32.eps}
%    \end{minipage}%
%  \end{center} 
%  \caption{The left panel shows the light-quark chiral condensate, $\Sigma_l$, and the sum of $m_l\chi_\pi$ and the mixed $\pi-J_{5q}/4$ susceptibility to which it should be equal according to the Ward identity in Eq.~\eqref{eq:GMOR_DWF}.  Also shown is $(m_l+\mres)\chi_\pi$ which would equal $\Sigma_l$ if $\mres$ were the only effect of residual chiral symmetry breaking.  The right panel shows the same quantities computed using the strange instead of the light quark.  Similar agreement between the right and left hand sides of Eq.~\eqref{eq:GMOR_DWF} is found for the $24^3$ and $16^3$ volumes, as can be seen from Tab.~\ref{tab:WI}}
%  \label{fig:WI}
%\end{figure}

A second use of Eq.~\eqref{eq:GMOR} is to provide a method to compute a more physical result for $\dls$ in a DWF calculation.  Since no chiral limit has been taken in the continuum derivation of Eq.~\eqref{eq:GMOR}, it will hold equally well if applied to either strange or light quarks.  If we use the resulting equations for  $\Sigma_l$ and $\Sigma_s$ to determine the weighted difference $\dls$ we obtain:
\begin{equation}
\dls = m_l^c\left(\chi_{\pi_l} -\chi_{\pi_s}\right),
\label{eq:GMOR_sub}
\end{equation}
where we use the symbol $\chi_{\pi_s}$ to represent the ``pion'' susceptibility that results if the light quark mass is replaced by that of the strange quark and add the subscript  $l$ to the usual pion susceptibility for clarity.  From the perspective of the continuum theory both sides of Eq.~\eqref{eq:GMOR_sub} provide an equally good value for the subtracted chiral condensate. Neither quantity contains a quadratic divergence and the much smaller logarithmic divergences present on both sides are equal.  For a DWF theory with residual chiral symmetry breaking this equation does not hold and the left hand side $\dls$ contains an unphysical additive constant $O(\mres/a^2)$.  However, the right-hand side is much better defined with no $1/a^2$ term.  Thus, we can use the right-hand side of Eq.~\eqref{eq:GMOR_sub} to provide a more physical result for $\dls$ which will contain only a small, unphysical piece of order $m_l m_s^2 \ln(m_s a)$.  Thus, we can define an improved value for $\dls$:
\begin{equation}
\tdls = \mtot_l\left(\chi_{\pi_l} -\chi_{\pi_s}\right)
\end{equation}
which we will use to compare with spectral formulae and with the results for $\dls$ from other lattice fermion formulations.

\subsection{Chiral Symmetry Restoration}
\label{sec:su2}

\begin{figure}[htb]
  \begin{center}
    \begin{minipage}[t]{0.5\linewidth}
      \centering
      \resizebox{\linewidth}{!}{\includegraphics{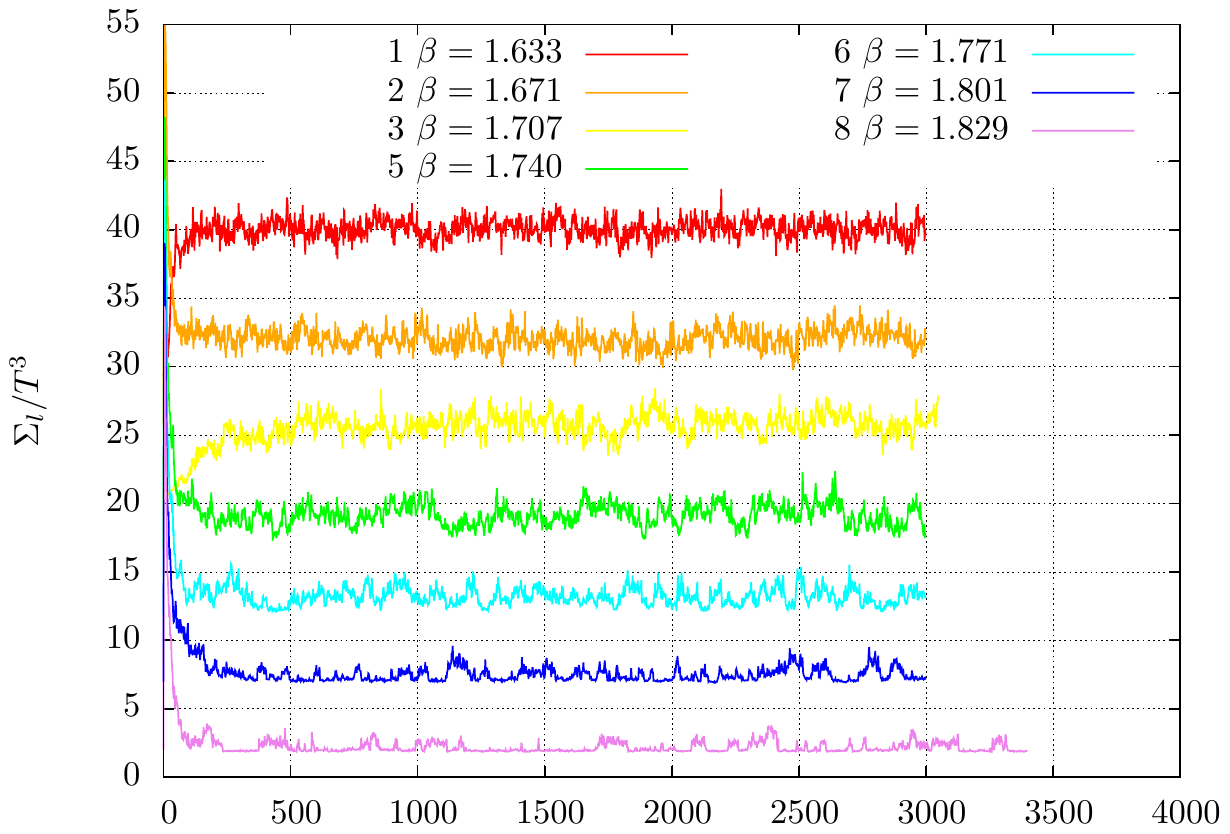}}
    \end{minipage}%
    \begin{minipage}[t]{0.5\linewidth}
      \centering
      \resizebox{\linewidth}{!}{\includegraphics{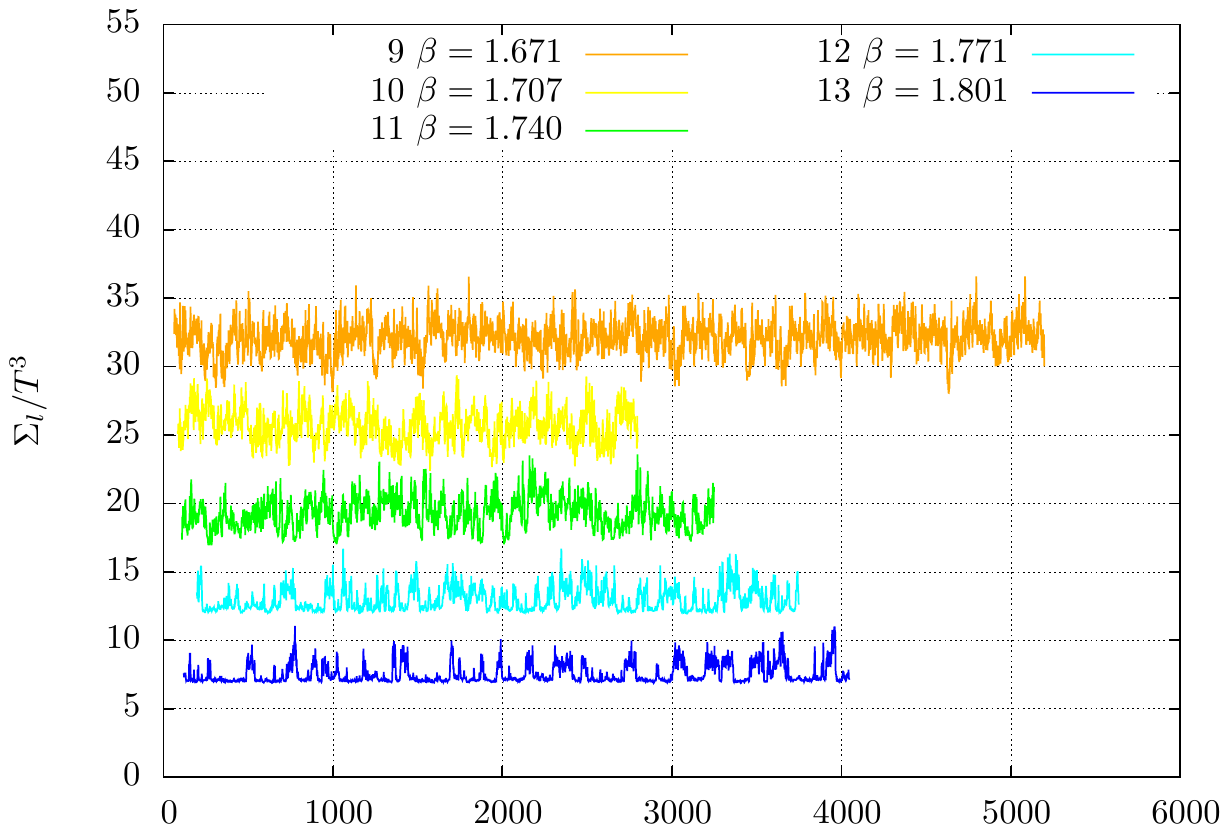}}
    \end{minipage}%
  \end{center} 
  \caption{
  \label{fig:light}
Monte Carlo time histories of the light-quark chiral condensate $\Sigma_l/T^3$ on the $32^3\times8$ (left) and $24^3\times8$ (right) ensembles. (Only the longest streams from~\runref{150_b} and~\#\runn{160_b} are displayed.) There is a vertical offset of 5 units between successive data sets with the $\beta=1.829$ results  unshifted.  Note that the time evolution corresponding to $\beta=1.725,\,32^3\times8$ (\runref{164_a}) behaves in a similar manner to those of its neighboring ensembles, but is omitted from the graph to preserve a uniform separation between each ensemble.}
\end{figure}

In this section we present and discuss our numerical results for the chiral condensate and for the disconnected chiral susceptibility as a function of temperature.  Figure~\ref{fig:light} shows the Monte Carlo time histories of the light-quark chiral condensate for seven of the temperatures studied. The time evolutions for the $32^3\times8$ ensembles are displayed in the left panel and those from $24^3\times8$ in the right. The evolutions of the light-quark condensates from both sets of ensembles appear to follow the same trend. For the lower temperature region ($T\le168$ MeV), the light-quark condensate fluctuates around its average value. However, as temperature grows higher, the fluctuations can better be described as upward spikes added to an otherwise flat base line.  

This behavior is typically seen in finite temperature DWF calculations and arises because above $T_c$ the main contribution to the chiral condensate comes from isolated, near-zero modes~\cite{Christ:2002pg}.  These modes become increasingly infrequent as the temperature is increased but, when present, produce a noisy, non-zero chiral condensate.  The noise results from the relatively small space-time extent of each zero mode which is therefore sampled in our stochastic determination with relatively few random numbers.

Such behavior becomes most pronounced for $T\ge186$ MeV in the $32^3\times 8$ calculations.  At $T=177$ MeV, the $24^3\times8$ Monte Carlo time evolution shows this characteristic plateau-spike structure more distinctly than does the comparable $32^3\times8$ time history.  This suggests a lower pseudo-critical transition temperature for the smaller volume or that the larger $32^3$ volume supports a larger number of such zero modes, reducing the size of the intervals when none are present and the chiral condensate is nearly zero.

\begin{figure}[htb]
  \begin{center}
    \begin{minipage}[t]{0.5\linewidth}
      \centering
      \resizebox{\linewidth}{!}{\includegraphics{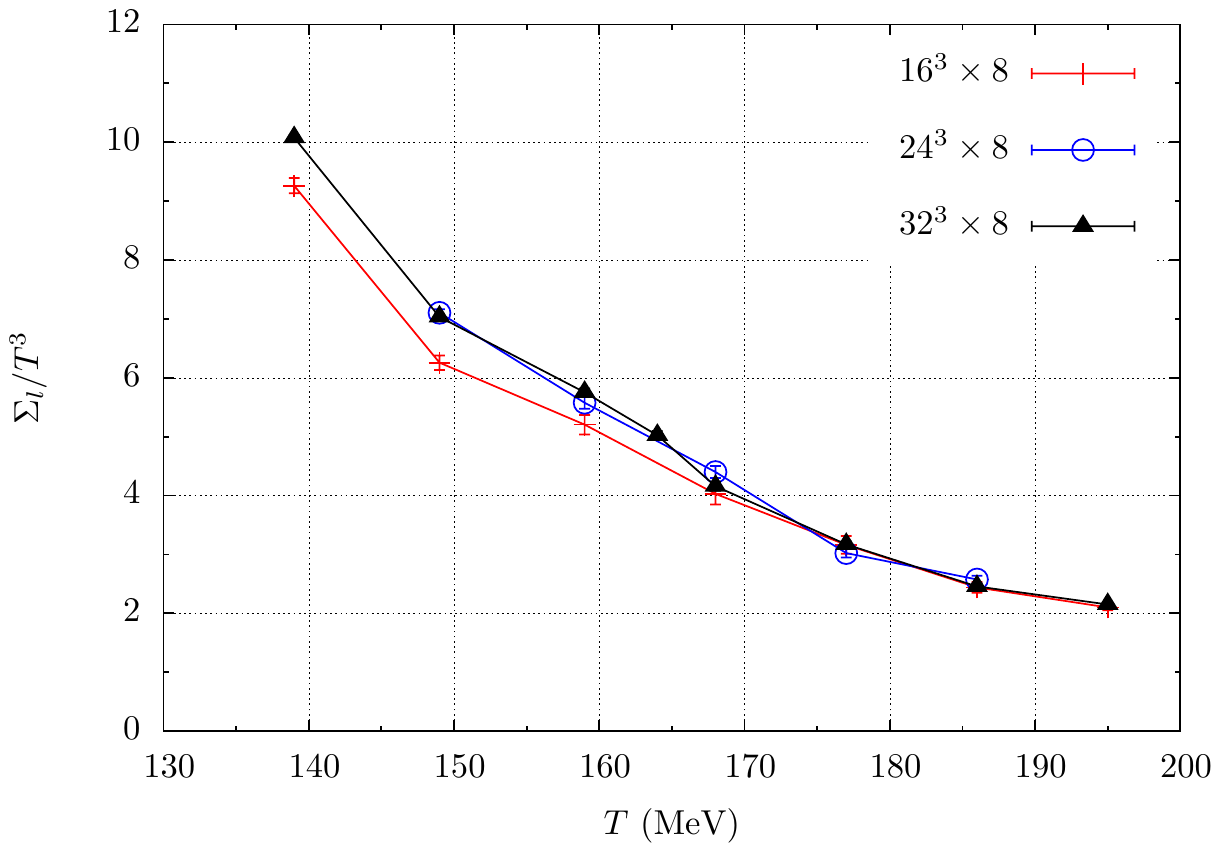}}
    \end{minipage}
    \begin{minipage}[t]{0.5\linewidth}
      \centering
      \resizebox{\linewidth}{!}{\includegraphics{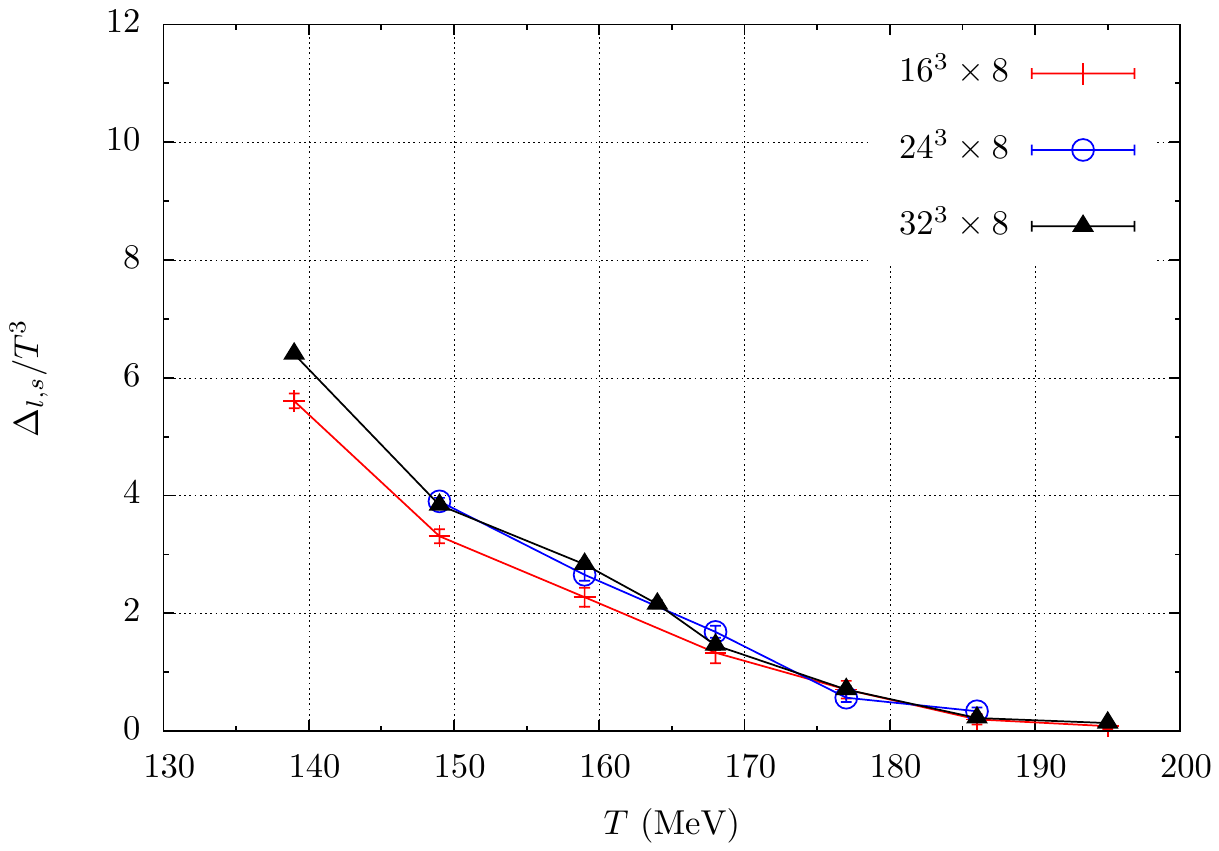}}
    \end{minipage}%
    \begin{minipage}[t]{0.5\linewidth}
      \centering
      \resizebox{\linewidth}{!}{\includegraphics{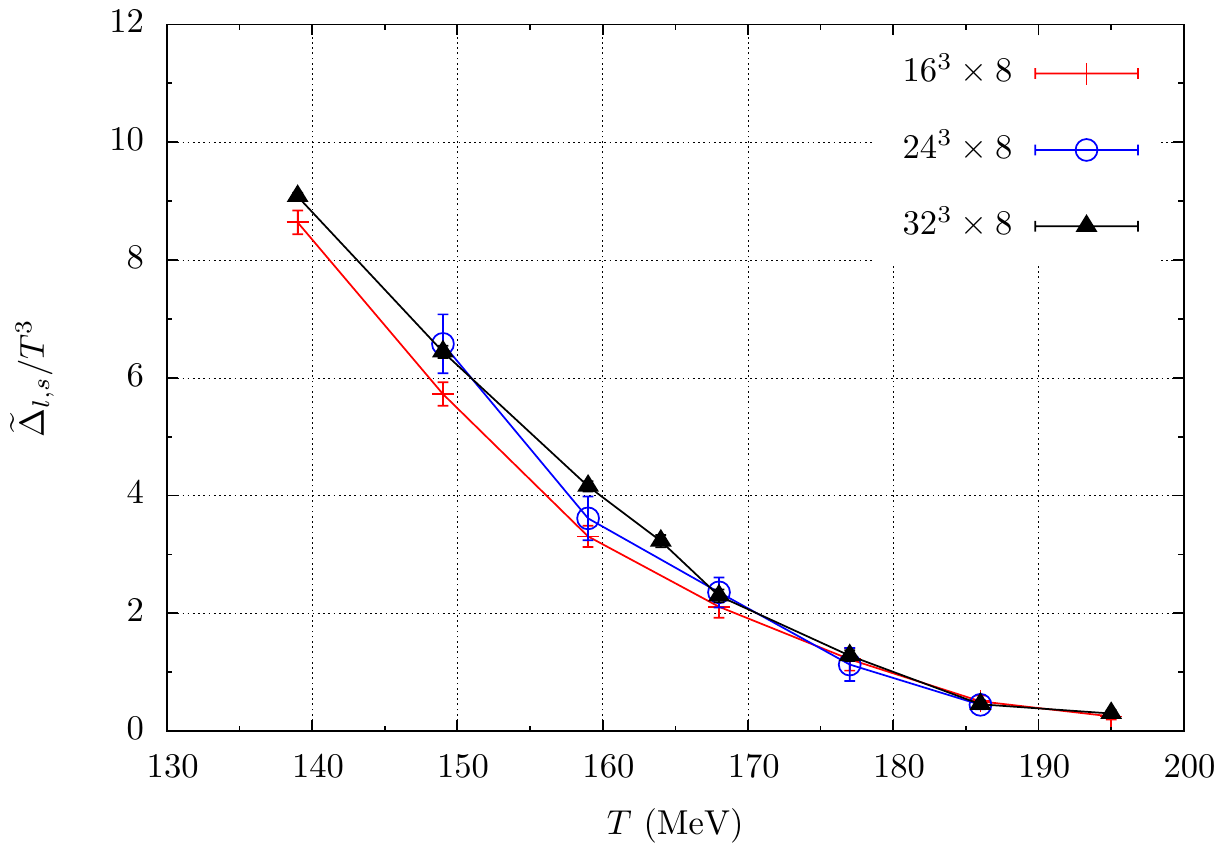}}
    \end{minipage}%
  \end{center} 
  \caption{
  \label{fig:pbpsum}
Comparison of light-quark (upper), subtracted (lower left) and improved subtracted (lower right) chiral condensates computed on different volumes.  The $32^3$ and $24^3$ volumes agree reasonably well for all temperatures but are 5-10\% larger than the corresponding values from the $16^3$ volume for $T < 168$ MeV.  The results appear to be volume independent for $T \ge 168$ MeV.}
\end{figure}

The ensemble averages of the light, subtracted and strange chiral condensates are summarized in Tab.~\ref{tab:pbp}.  The temperature dependence of the light and the subtracted condensates is also illustrated in Fig.~\ref{fig:pbpsum}.  As that figure shows, results from $32^3\times8$ and $24^3\times8$ ensembles agree well throughout the transition region, whereas those from the $16^3\times8$ ensembles show an appreciable discrepancy for $T < 168$ MeV, indicating a small but well-resolved finite volume effect.

\begin{table}[H]
  \centering
  \vskip -0.75 in
  \begin{tabular}{c|cc|cccccc}\hline
    \#&
    $T\,(\textrm{MeV})$
    &$\beta$
    &$\Sigma_l/T^3$
    &$\Sigma_s/T^3$
    &$\dls/T^3$
    &$\chid/T^2$
    &$\chigd/T^2$
    &$\chit/(\mtot_l T)^2$
    \\ \hline
    \runn{140_a}&139&1.633&10.07(4)&41.27(2)&6.40(4) &20(2)&118(7)&261(11)\\
    \runn{150_a}&149&1.671& 7.03(6)&36.48(2)&3.84(5) &28(3)& 94(8)&177(11)\\
    \runn{160_a}&159&1.707& 5.80(6)&33.73(2)&2.83(6) &33(3)& 70(8)&118(10)\\
    \runn{164_a}&164&1.725& 5.02(7)&32.04(3)&2.16(7) &38(3)& 49(4)& 78(4) \\
    \runn{170_a}&168&1.740& 4.16(8)&30.72(3)&1.46(7) &37(3)& 38(5)& 54(4) \\
    \runn{180_a}&177&1.771& 3.17(5)&27.94(2)&0.71(5) &22(2)& 24(3)& 37(3) \\
    \runn{190_a}&186&1.801& 2.46(4)&25.38(2)&0.22(4) &12(2)& 10(2)& 15(2) \\
    \runn{200_a}&195&1.829& 2.15(3)&23.20(1)&0.14(3) & 7(1)& 10(1)& 15(2) \\
    \hline\hline                                                  
    \runn{150_b}&148&1.671&7.10(6) &36.53(2)&3.90(6) &31(2)& 89(5)&165(7) \\
    \runn{160_b}&159&1.707&5.58(10)&33.68(3)&2.66(10)&36(3)& 64(6)&110(6) \\
    \runn{170_b}&168&1.740&4.40(10)&30.84(4)&1.69(10)&32(3)& 47(6)& 67(6) \\
    \runn{180_b}&177&1.771&3.03(7) &27.90(3)&0.57(7) &19(2)& 21(3)& 32(3) \\
    \runn{190_b}&186&1.801&2.58(6) &25.41(2)&0.34(6) &13(2)& 14(2)& 18(2) \\
    \hline\hline
    \runn{140}  &139&1.633&9.26(13)&41.02(4)&5.61(12)&36(3)&113(7)&252(11)\\
    \runn{150_1}&149&1.671&6.26(12)&36.42(5)&3.07(12)&44(3)& 89(6)&159(6) \\
    \runn{150}  &149&1.671&8.39(10)&38.30(3)&5.00(10)&41(2)& 90(6)&168(7) \\
    \runn{160}  &159&1.707&5.25(17)&33.81(6)&2.27(16)&43(4)& 55(6)& 97(7) \\
    \runn{170}  &168&1.740&4.03(18)&30.66(7)&1.33(18)&35(5)& 37(5)& 60(7) \\
    \runn{180}  &177&1.771&3.16(15)&27.88(6)&0.71(15)&25(4)& 24(4)& 36(4) \\
    \runn{190}  &186&1.801&2.44(9) &25.43(4)&0.20(9) &11(4)&  9(3)& 21(6) \\
    \runn{200}  &195&1.829&2.10(5) &23.22(3)&0.09(5) &6(2) &  6(2)& 11(2) \\
    \hline
  \end{tabular}
  \caption{
  \label{tab:pbp}
The unrenormalized chiral condensates and disconnected chiral susceptibilities.  The two right-most columns should agree according to Eq.~\eqref{eq:qsuscp}.  As discussed, we attribute their large difference to inaccuracy in the strong-coupling measurement of $\chit$.  Moving from top to bottom, the three sections correspond to the volumes $32^3\times8$, $24^3\times8$ and $16^3\times8$.}
\end{table}

\begin{figure}[htb]
  \begin{center}
    \begin{minipage}[t]{0.5\linewidth}
      \centering
      \resizebox{\linewidth}{!}{\includegraphics{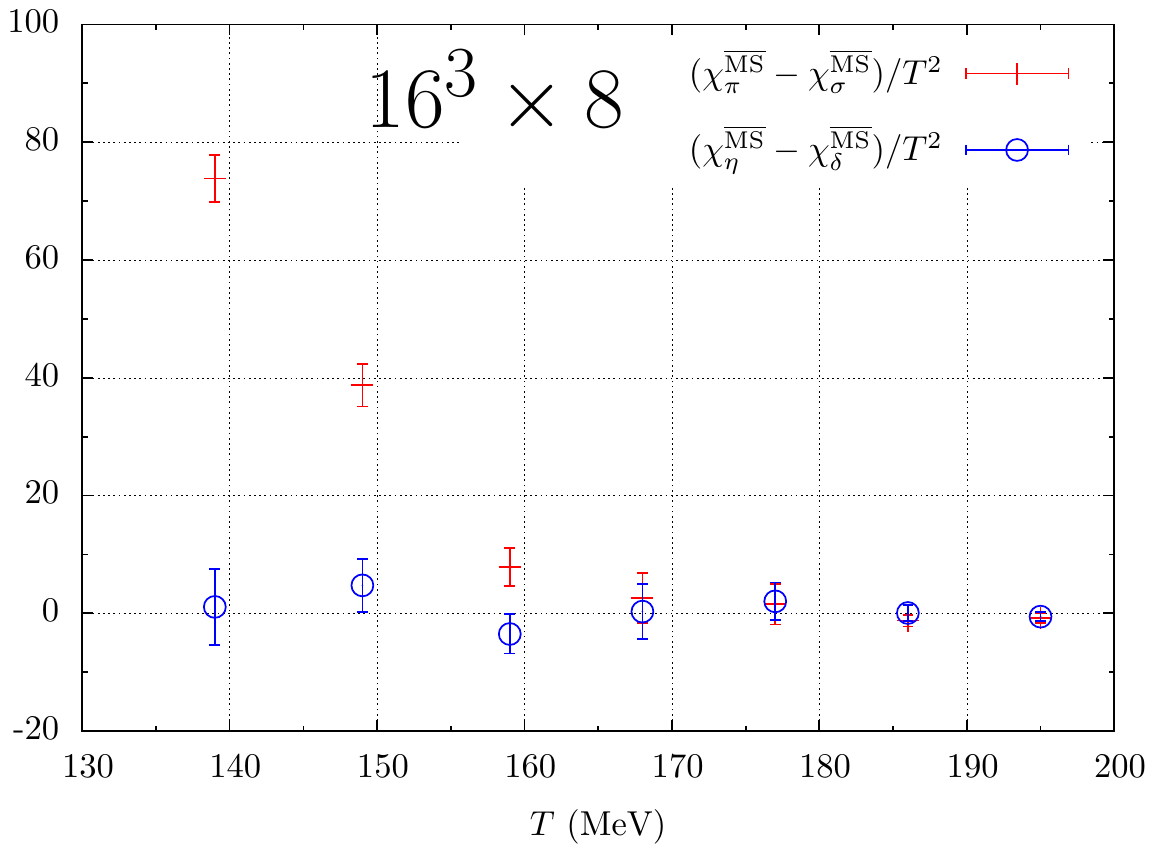}}
    \end{minipage}%
    \begin{minipage}[t]{0.5\linewidth}
      \centering
      \resizebox{\linewidth}{!}{\includegraphics{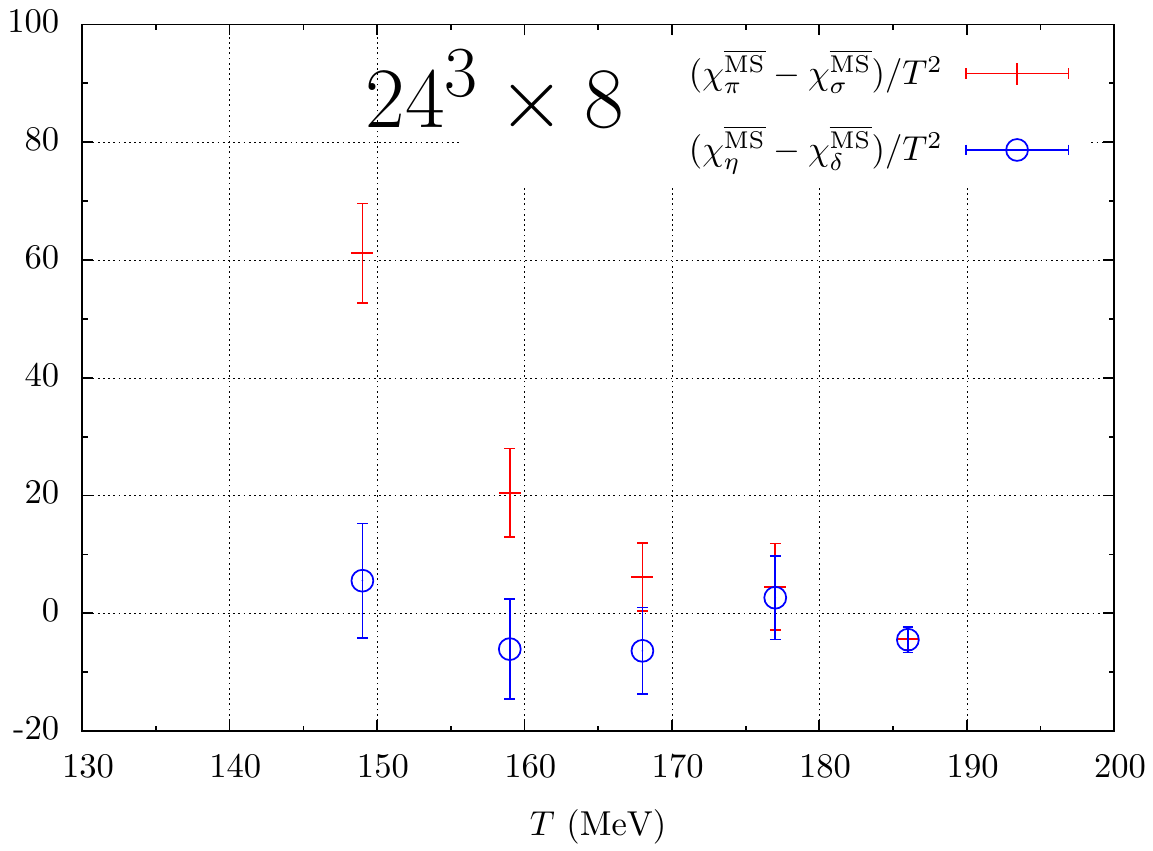}}
    \end{minipage}
    \begin{minipage}[t]{0.5\linewidth}
      \centering
      \resizebox{\linewidth}{!}{\includegraphics{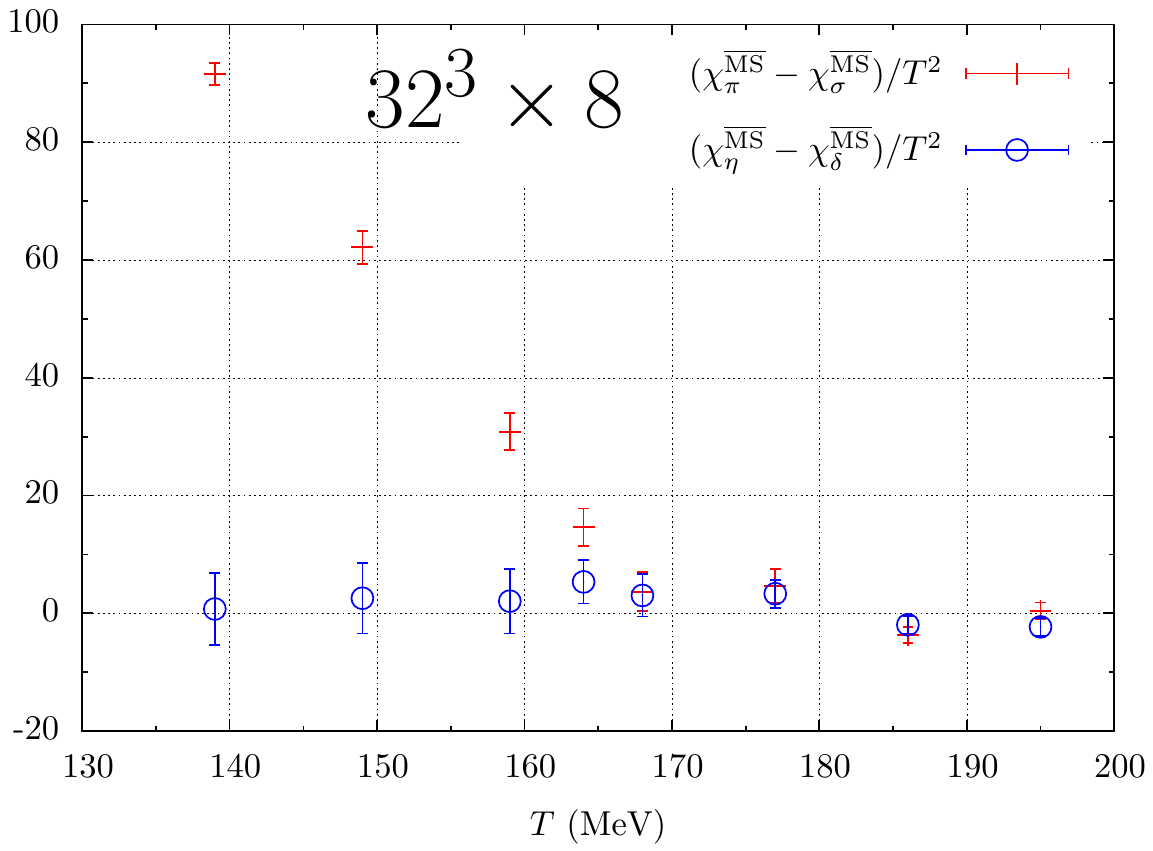}}
    \end{minipage}%
  \end{center} 
  \caption{
  \label{fig:sua32}
The two $\sua$-breaking susceptibility differences $\chi_\pi^{\overline{\textrm{MS}}} - \chi_\sigma^{\overline{\textrm{MS}}}$ and $\chi_\delta^{\overline{\textrm{MS}}} - \chi_\eta^{\overline{\textrm{MS}}}$ plotted as a function of temperature for our three spatial volumes: $16^3$, $24^3$ and $32^3$.  For temperatures of 170 MeV and above these differences are consistent with zero and the expected restoration of chiral symmetry above $T_c$.  The quantity $\pms$ becomes very large below $T_c$ reflecting the small mass of the pseudo-Goldstone $\pi$ meson below $T_c$.  In contrast, the second difference $\emd$ remains relatively small as the temperature decreases below $T_c$, reflecting the relatively large masses of the $\delta$ and $\eta$ mesons.
}
\end{figure}

A second measure of the restoration of $\sua$ symmetry is the two differences $\pms$ and $\emd$, following Eq.~\eqref{eq:su2_sym}.  These two $\sua$-breaking differences are plotted in Fig.~\ref{fig:sua32}.  The quantity $\pms$ shows the behavior that might be expected from the temperature dependence of the chiral condensate shown Fig.~\ref{fig:pbpsum}.  A large $\sua$-breaking difference is seen for $T \le 159$ MeV which becomes zero for $T \ge 168$ MeV.  The second difference $\emd$ is more surprising, being essentially zero throughout our temperature range.  While we do not have a crisp explanation for this unexpected $\sua$ symmetry below $T_c$ we do expect this difference to vanish for $T>T_c$ and to be small relative to $\pms$ for $T<T_c$ since the large value of $\chi_\pi$ reflects the small pion mass while the $\delta$, $\sigma$ and $\eta$ are all expected to be relatively massive below $T_c$.

While the chiral condensate is the order parameter for the chiral transition, its strong apparent temperature dependence results from a combination of the finite temperature physics of interest and its dependence on the lattice scale as a dimension 3 operator.  (This can be recognized by noting that we often discuss the dimensionless quantity $\Sigma_l/T^3$ which will change significantly with temperature simply because of the $1/T^3$ factor.)  The location of the pseudo-critical temperature is much more easily seen by examining the disconnected chiral susceptibility $\chid$.  This has dimension 2 and so varies a little less strongly with the lattice scale (which we are changing to vary $T$ on our $N_\tau=8$ lattice) and shows a dramatic peak near the transition which can be used to define the location of the pseudo-critical temperature $T_c$.  Numerical results for $\chid$ before renormalization are presented in Tab.~\ref{tab:pbp}. In order to allow a comparison with results from the staggered formalism, the susceptibilities should be normalized in the $\overline{\rm MS}$ scheme at 2 GeV. They can be obtained from the directly-computed lattice quantities using the relation:
\begin{equation}
  \chi^{\overline{\rm MS}}=\left(\frac{1}{
  Z_{m_f\to\overline{\rm MS}}}\right)^2\chi^\mathrm{bare}.
  \label{eq:chirn}
\end{equation}
The renormalization factors $Z_{m_f\to\overline{\rm MS}}$ for each temperature are listed in Tab.~\ref{tab:suscl}.  These values for $Z_{m_f\to\overline{\rm MS}}$ were obtained in Ref.~\cite{Bazavov:2012qja} from the dependence of the pion mass, expressed in physical units, on the input quark mass and the known value of $\mtot_l$ which corresponds to the physical value of $m_\pi$~\cite{Aoki:2010dy}.

\begin{figure}[htb]
  \begin{center}
    \begin{minipage}[t]{0.5\linewidth}
      \centering
      \resizebox{\linewidth}{!}{\includegraphics{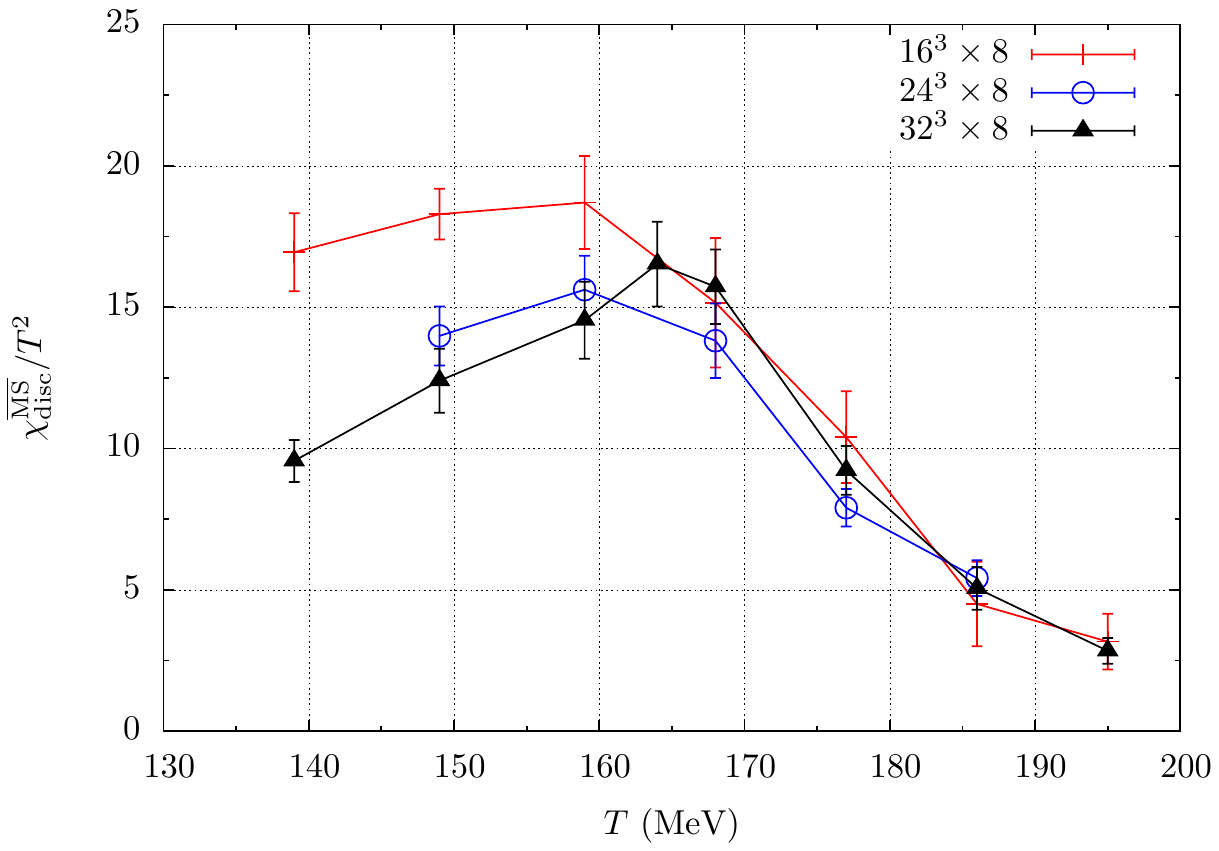}}
    \end{minipage}%
    \begin{minipage}[t]{0.5\linewidth}
      \centering
      \resizebox{\linewidth}{!}{\includegraphics{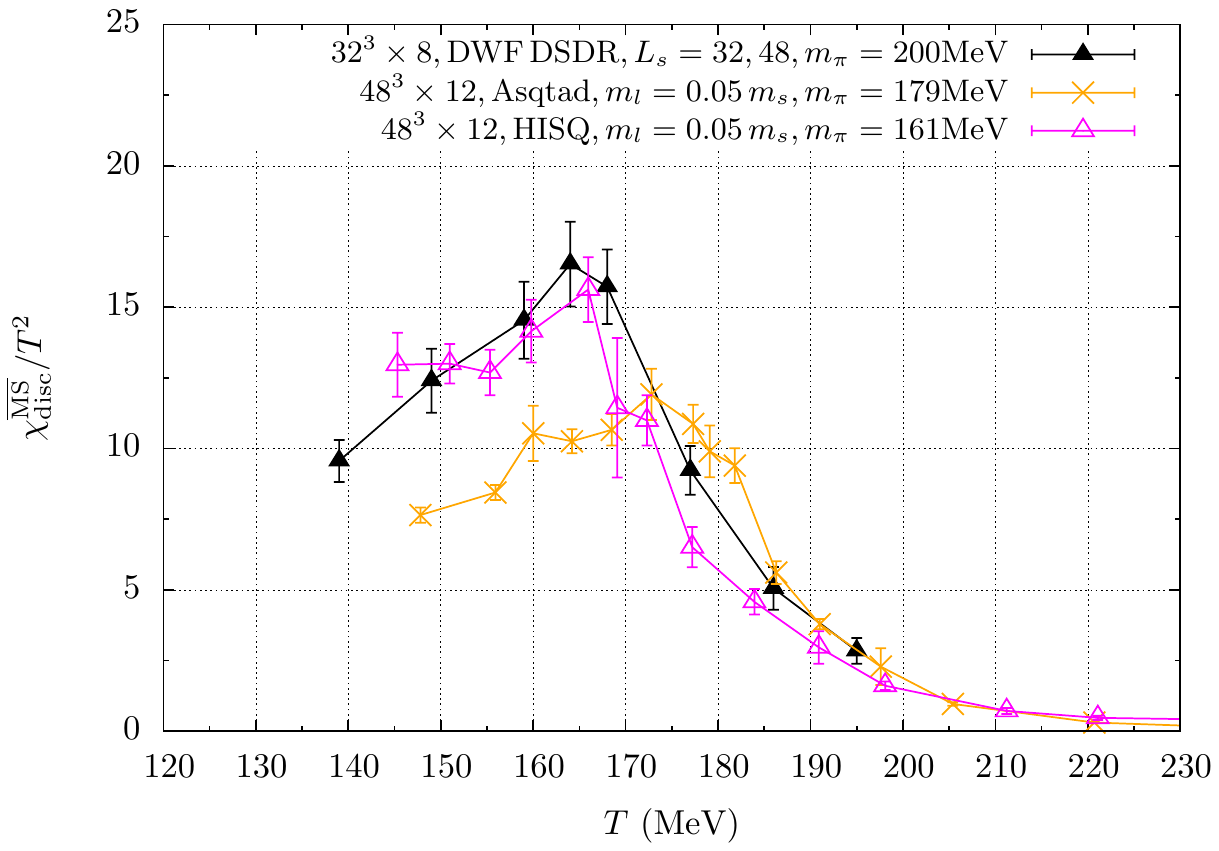}}
    \end{minipage}%
  \end{center} 
  \caption{
  \label{fig:chi}
The left panel compares $\chid$ computed using DWF on $32^3$, $24^3$ and $16^3$ volumes.  Significant volume dependence can be seen between $32^3$ and $16^3$, while the  $24^3$  results agree with those from $32^3$ within errors.  The right panel compares the $32^3$, $N_\tau=8$ DWF results for $\chid$ with those from staggered fermions on a $48^3\times12$ volume using both the ASQTAD and HISQ actions~\cite{Bazavov:2011nk}.  In each case $\chid$ is renormalized in the $\overline{\rm MS}(\mu=2~\textrm{GeV})$ scheme.}
\end{figure}

The dependence of the renormalized $\chid$ on volume is shown in the left panel of Fig.~\ref{fig:chi}.  At  $T=$~168 MeV and above the disconnected chiral susceptibilities from all volumes agree within errors.  However, at lower temperatures there is a large discrepancy between the $16^3\times8$ and the $32^3\times8$ ensembles which becomes larger as temperature decreases.  Results from $24^3\times8$, fall in between, although they tend to lie closer to the $32^3\times8$ points.

Since we are studying only a single value of $N_\tau$ and a pion mass that is larger than physical by a factor of 1.5, it is premature to draw a definite quantitative conclusion about the pseudo-critical transition temperature. However, a qualitative examination of the left panel in Fig.~\ref{fig:chi} suggests that a peak in $\chid$ occurs for the $16^3$ and $24^3$ volumes at approximately 160 MeV and that this peak position increases to slightly above 165 MeV as the volume is increased to $32^3$.

The right panel of Fig.~\ref{fig:chi} compares the $m_\pi=200$ MeV, $32^3\times 8$ DWF results for $\chid$ with those obtained from staggered fermions using an $48^3\times12$ volume  and the HISQ and ASQTAD staggered actions with $m_\pi=161$ and 177 MeV respectively~\cite{Bazavov:2011nk}.   Again, the disconnected chiral condensates are consistent among these three methods for $T\ge175$ MeV.  However, the ASQTAD 
results lie substantially below the DWF and HISQ results for temperatures at and below the transition region.  The HISQ results are in good agreement with the $32^3\times8$ DWF results.  However, this agreement appears to be coincidental, since the HISQ results are obtained for a quoted pion mass of 161 MeV, significantly smaller than the 200 MeV pion mass of the DWF ensembles.  The expected strong dependence of $\chid$ near $T_c$ on the pion mass suggests that $m_\pi=160$ MeV DWF results would lie above those found with HISQ.  The discrepancy between the DWF and ASQTAD results and the expected discrepancy with comparable HISQ results are likely explained by lattice discretization errors associated with staggered taste symmetry breaking.

\subsection{$\ua$ symmetry}
\label{sec:ua1}
We will now discuss the degree to which the anomalous $\ua$ symmetry is restored above $T_c$ by examining the two implications of this symmetry for the four susceptibilities given in Eq.~\eqref{eq:ua1_sym}: $\chi_\pi=\chi_\delta$ and $\chi_\sigma=\chi_\eta$.   The numerical results  for each of these four susceptibilities are summarized in Tab.~\ref{tab:suscl} as well as their $\ua$-breaking differences $\pmd$ and $\sme$ which we will often abbreviate as $\dpd=\pmd$ and $\dse=\sme$.  The integrated susceptibilities $\chi_\pi$ and $\chi_\delta$ are calculated from the corresponding two point correlation functions by summing the position of the sink over the entire space-time volume. For the $24^3\times8$ ensembles, we use a single point source located at $(0,0,0,0)$, while for the $16^3\times8$ and $32^3\times8$ ensembles, we use a random $Z_2$ wall source located on a fixed, 3-dimensional spatial slice, $x_z=0$. 

\begin{figure}[htb]
  \begin{center}
    \begin{minipage}[t]{0.5\linewidth}
      \centering
      \resizebox{\linewidth}{!}{\includegraphics{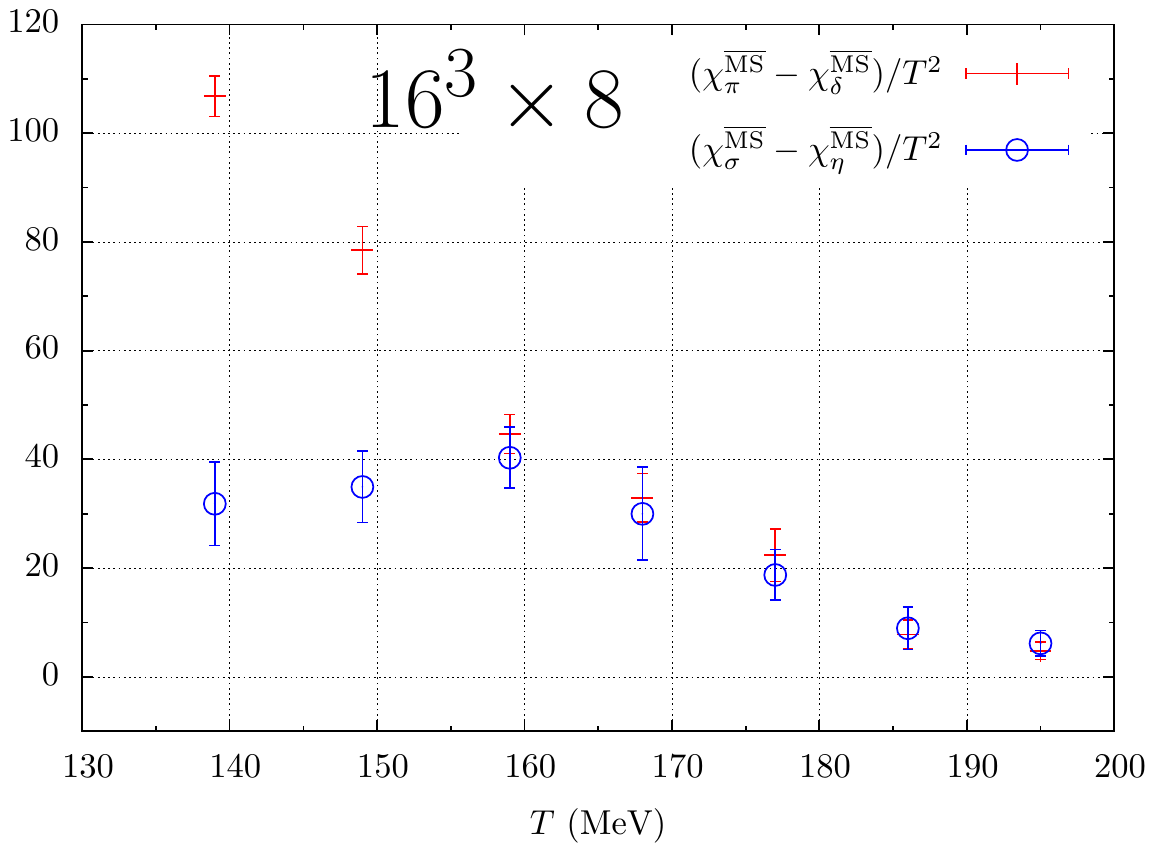}}
    \end{minipage}%
    \begin{minipage}[t]{0.5\linewidth}
      \centering
      \resizebox{\linewidth}{!}{\includegraphics{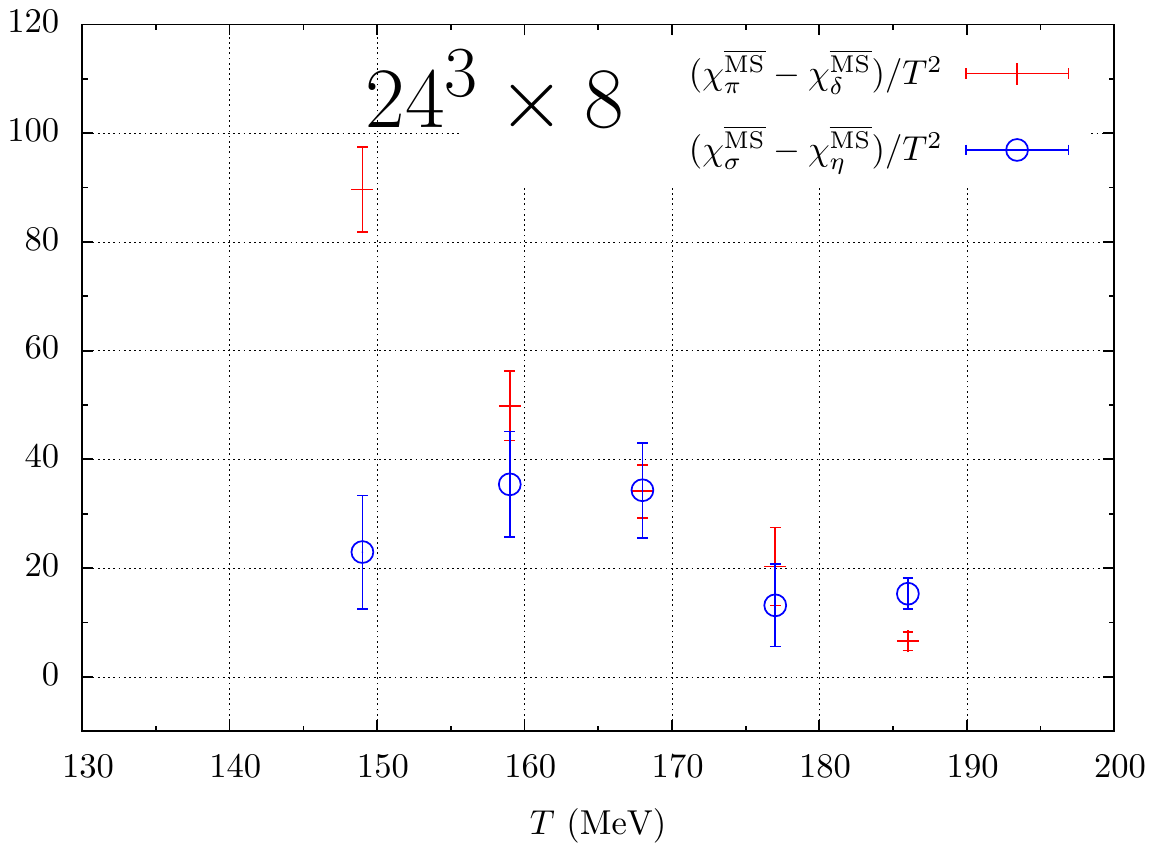}}
    \end{minipage}
    \begin{minipage}[t]{0.5\linewidth}
      \centering
      \resizebox{\linewidth}{!}{\includegraphics{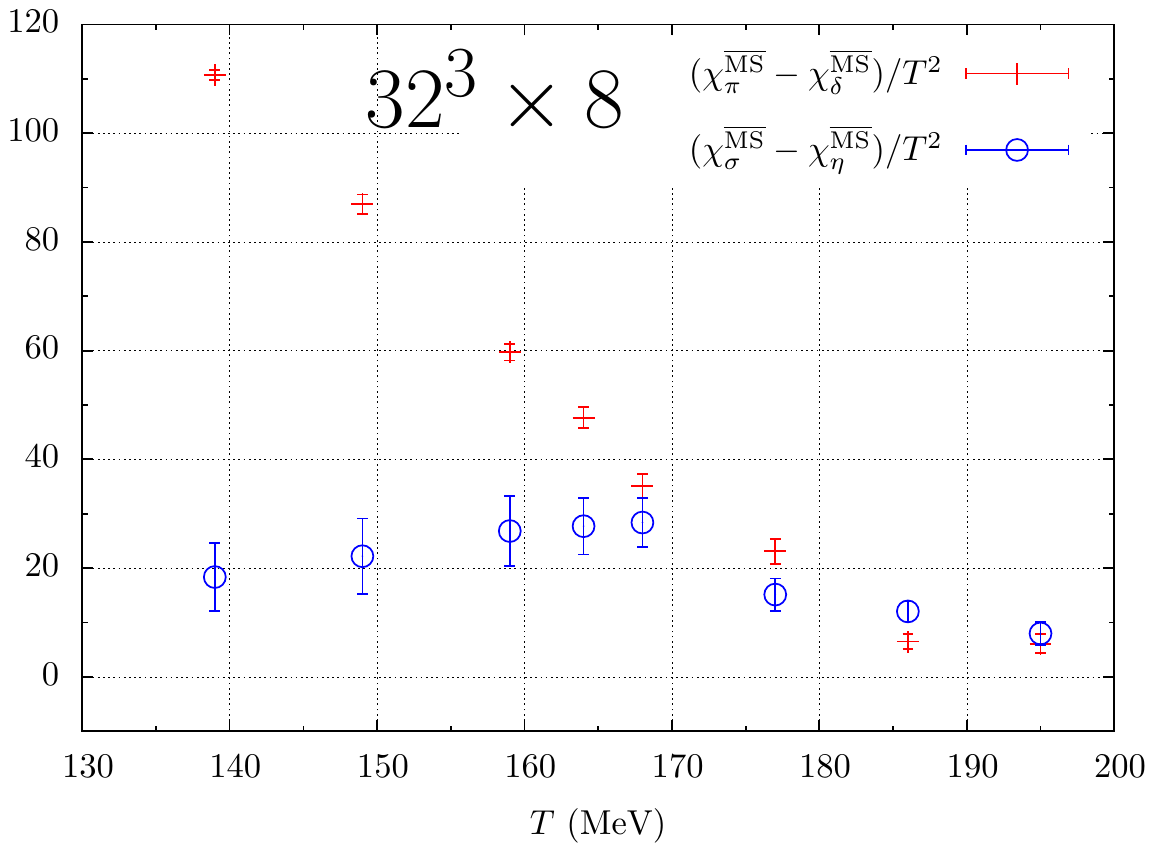}}
    \end{minipage}%
  \end{center} 
  \caption{
  \label{fig:ua32}
The two $\ua$-violating susceptibility differences, $\chi_\pi^{\MS}-\chi_\delta^{\MS}$ and $\chi_\sigma^{\MS}-\chi_\eta^{\MS}$ plotted as a function of temperature for our three spatial volumes.  As expected these quantities are very different below $T_c$.  However, even for temperatures of 160 MeV and above these quantities differ from zero by many standard deviations, providing clear evidence for anomalous symmetry breaking above $T_c$.  The near equality of these two differences above $T_c$, which are related by $\sua$ symmetry suggests that the effects of explicit chiral symmetry breaking are much smaller (as expected) than this anomalous symmetry breaking.}
\end{figure}

 \begin{table}[hbt]
 \centering
 \vskip -0.5 in
 \makebox[\textwidth][c]{%
 \begin{tabular}{c|cc|ccccccccc}\hline
    \#
    &$T\,(\mathrm{MeV})$
    &$Z_{m_f\to\MS}$
    &$\Tspace \Bspace \chi_\pi^{\overline{\rm MS}}/T^2$
    &$\chi_\delta^{\overline{\rm MS}}/T^2$
    &$\chi_\sigma^{\overline{\rm MS}}/T^2$
    &$\chi_\eta^{\overline{\rm MS}}/T^2$
    &$\dps^{\overline{\rm MS}}/T^2$
    &$\ded^{\overline{\rm MS}}/T^2$
    &$\dpd^{\overline{\rm MS}}/T^2$
    &$\dse^{\overline{\rm MS}}/T^2$\\ \hline
    \runn{140_a}&139&1.47&144.7(7)  &34.0(3)  &53(2)&35(6)& 92(2)  & 1(6)    &111(1)&18(6)\\
    \runn{150_a}&149&1.49&120.1(1.3)&33.1(6)  &58(2)&36(6)& 62(3)  & 3(6)    & 87(2)&22(7)\\
    \runn{160_a}&159&1.51& 94.0(1.1)&34.3(5)  &63(3)&36(6)& 31(3)  & 2(5)    & 60(2)&27(6)\\
    \runn{164_a}&164&1.52& 80.8(1.3)&33.2(8)  &66(3)&39(4)& 15(3)  & 5(4)    & 48(2)&28(5)\\
    \runn{170_a}&168&1.53& 68.7(1.4)&33.6(9)  &65(3)&37(4)&  4(3)  & 3(4)    & 35(2)&28(4)\\
    \runn{180_a}&177&1.55& 53.8(1.3)&30.8(1.1)&49(2)&34(2)&  5(3)  & 3(2)    & 23(2)&15(3)\\
    \runn{190_a}&186&1.57& 40.6(8)  &34.1(6)  &44(1)&32(1)& -4(1)  &-2(1)    &  6(1)&12(2)\\
    \runn{200_a}&195&1.58& 37.2(9)  &31.1(8)  &37(1)&29(1)&0.4(1.4)&-2(2)    &  6(2)& 8(2)\\
    \hline\hline
    \runn{150_b}&149&1.49&122(6)    &32(2)    &61(4)&38(9)& 61(8)  & 6(10)   & 90(8)&23(10)\\
    \runn{160_b}&159&1.51& 87(5)    &37(2)    &66(4)&31(7)& 20(8)  &-6(8)    & 50(6)&35(10)\\
    \runn{170_b}&168&1.53& 70(3)    &36(2)    &64(3)&30(7)&  6(6)  &-6(7)    & 34(5)&34(9) \\
    \runn{180_b}&177&1.55& 52(4)    &31(3)    &47(4)&34(4)&  4(7)  & 3(7)    & 20(7)&13(8) \\
    \runn{190_b}&186&1.57& 40(1)    &34(1)    &44(1)&29(2)& -4(2)  &-4(2)    &  7(2)&15(3) \\
    \hline\hline
    \runn{140}  &139&1.47&140(2)    &33(2)    &66(3)&34(7)& 74(4)  & 1(6)    &107(4)&32(8)\\
    \runn{150_1}&149&1.49&111(2)    &33(2)    &73(2)&38(6)& 39(4)  & 5(5)    & 78(4)&35(7)\\
    \runn{160}  &159&1.51& 83(2)    &38(2)    &75(3)&35(4)&  8(3)  &-3(3)    & 45(4)&40(6)\\
    \runn{170}  &168&1.53& 66(3)    &33(2)    &64(4)&34(5)&  3(4)  & 0.3(4.7)& 33(4)&30(9)\\
    \runn{180}  &177&1.55& 53(3)    &31(2)    &51(2)&33(3)&  2(3)  & 2(3)    & 22(5)&19(5)\\
    \runn{190}  &186&1.57& 41(1)    &34(1)    &43(2)&34(2)& -1(1)  & 0.1(1.3)&  8(3)& 9(4)\\
    \runn{200}  &195&1.58& 36(1)    &32(1)    &37(1)&31(1)& -1(1)  &-0.5(8)  &  5(2)& 6(2)\\
    \hline
  \end{tabular}}
  \caption{
  \label{tab:suscl}
Results for the four independent susceptibilities $\chi_\pi$, $\chi_\delta$, $\chi_\sigma$ and $\chi_\eta$ as well as the two pairs of differences, $\dps=\pms$, $\ded=\emd$ and $\dpd=\pmd$, $\dse=\sme$ which measure the degree of $\sua$ and $\ua$ symmetry, respectively.  All of these susceptibilities are renormalized in the $\overline{\rm MS}(\mu=2\textrm{~GeV})$ scheme using the renormalization factor listed in the $Z_{m_f\to\MS}$ column.  Moving from top to bottom, the three sections correspond to the volumes $32^3\times8$, $24^3\times8$ and $16^3\times8$.}
\end{table}

 \begin{table}[hbt]
 \centering
 \vskip -0.5 in
 \makebox[\textwidth][c]{%
 \begin{tabular}{c|c|ccccccccc}\hline
    \#
    &$T\,(\mathrm{MeV})$
    &$\Tspace \Bspace \chi_\pi^{s,\overline{\rm MS}}/T^2$
    &$\chi_\delta^{s,\overline{\rm MS}}/T^2$
    &$\chi_\sigma^{s,\overline{\rm MS}}/T^2$
    &$\chi_\eta^{s,\overline{\rm MS}}/T^2$
    &$\dps^{s,\overline{\rm MS}}/T^2$
    &$\ded^{s,\overline{\rm MS}}/T^2$
    &$\dpd^{s,\overline{\rm MS}}/T^2$
    &$\dse^{s,\overline{\rm MS}}/T^2$\\ \hline
    \runn{140_a}&139&43.89(3) &31.50(2) &33.7(2)&42.9(4)&10.1(2)&11.4(4)&12.39(5) &-9.2(4)\\
    \runn{150_a}&149&41.96(3) &31.70(3) &33.8(2)&41.6(3)& 8.2(2)& 9.9(3)&10.26(5) &-7.9(4)\\
    \runn{160_a}&159&39.89(4) &31.71(3) &34.8(4)&39.0(3)& 5.1(4)& 7.3(3)& 8.18(7) &-4.2(4)\\
    \runn{164_a}&164&38.77(5) &31.74(4) &35.6(4)&38.1(4)& 3.2(4)& 6.4(4)& 7.02(8) &-2.6(5)\\
    \runn{170_a}&168&37.68(6) &31.67(3) &35.3(4)&37.1(3)& 2.4(4)& 5.4(3)& 6.00(9) &-1.8(5)\\
    \runn{180_a}&177&35.65(5) &31.39(2) &33.4(3)&35.1(3)& 2.2(3)& 3.7(4)& 4.26(6) &-1.7(5)\\
    \runn{190_a}&186&33.75(5) &30.83(3) &32.7(3)&33.4(3)& 1.1(3)& 2.5(3)& 2.93(6) &-0.7(3)\\
    \runn{200_a}&195&32.37(4) &30.46(2) &31.7(1)&32.2(2)& 0.7(1)& 1.7(2)& 1.91(4) &-0.5(3)\\
    \hline\hline
    \runn{150_b}&149&42.0(3)  &31.57(16)&34.0(5)&41.5(5)& 7.9(6)&10.0(5)&10.4(4)  &-7.5(7)\\
    \runn{160_b}&159&39.7(3)  &31.82(12)&34.4(3)&39.0(5)& 5.3(4)& 7.2(6)& 7.9(4)  &-4.6(6)\\
    \runn{170_b}&168&38.3(3)  &31.73(11)&33.9(4)&37.7(4)& 4.3(6)& 5.9(4)& 6.5(3)  &-3.7(6)\\
    \runn{180_b}&177&35.7(2)  &31.45(9) &33.5(2)&35.5(4)& 2.2(3)& 4.1(4)& 4.2(3)  &-2.0(5)\\
    \runn{190_b}&186&33.5(1)  &30.84(7) &32.3(2)&32.9(3)& 1.2(2)& 2.0(3)& 2.7(2)  &-0.6(4)\\
    \hline\hline
    \runn{140}  &139&43.95(7) &31.52(5) &33.8(2)&43.6(3)&10.2(2)&11.5(3)&12.44(11)&-9.3(4)\\
    \runn{150_1}&149&41.87(6) &31.79(5) &34.8(3)&41.1(4)& 7.1(3)& 9.4(4)&10.08(9) &-6.4(5)\\
    \runn{160}  &159&39.81(8) &31.72(6) &34.6(3)&39.7(3)& 5.2(3)& 8.0(3)& 8.09(13)&-5.1(4)\\
    \runn{170}  &168&37.72(10)&31.68(6) &34.7(4)&38.0(4)& 3.0(4)& 6.4(4)& 6.04(14)&-3.3(5)\\
    \runn{180}  &177&35.58(9) &31.41(6) &33.9(2)&35.6(3)& 1.6(2)& 4.2(3)& 4.18(13)&-1.7(3)\\
    \runn{190}  &186&33.86(8) &30.87(4) &32.7(1)&34.0(2)& 1.2(2)& 3.1(2)& 2.99(10)&-1.3(3)\\
    \runn{200}  &195&32.41(6) &30.48(3) &31.8(1)&32.2(3)& 0.6(1)& 1.7(3)& 1.92(5) &-0.3(3)\\
    \hline
  \end{tabular}}
  \caption{
  \label{tab:suscs}
The same quantities as tabulated in Tab.~\ref{tab:suscl} but with the light quark replaced by the strange quark.} 
\end{table}

These two $\ua$-breaking differences are plotted in Fig.~\ref{fig:ua32}.  As can be seen, these diminish rapidly with temperature but are many standard deviations from zero even at the temperatures of 177 and 186 MeV, well above $T_c$.   We expect that the effect of explicit chiral symmetry breaking, either from the non-zero input quark mass or finite $L_s$, residual chiral symmetry breaking, on these differences will be much smaller.   Specifically, for $T>T_c$ we might estimate the contribution of explicit $\ua$ breaking to be of order $\mtot_l^2/T^2 \sim (0.004*8)^2 = 0.001$ compared to results between 3 and 7 shown in Tab.~\ref{tab:suscl}.~\footnote{This assumed quadratic dependence on $\mtot_l$ does not allow for a possible combined effect of explicit chiral symmetry breaking and the sort of non-analytic behavior above $T_c$ that we are trying to study.   We do not have sufficient numerical results to study such effects which we view as ``second order'' since they require both non-perturbative chiral breaking above $T_c$ and $\mtot_l \ne 0$.}   Numerical evidence for the absence of explicit chiral symmetry breaking is provided by the near equality of the two differences $\pmd$ and $\sme$ which are related by $\sua$ symmetry, a symmetry also explicitly broken by $m_l$ and $\mres$.  

Strong evidence for the small size of possible explicit chiral symmetry breaking also comes from the results for $\chi_\pi-\chi_\delta$ computed for the strange quark.  It is the explicit breaking of chiral symmetry by the valence propagators which can create a non-anomalous signal for $\chi_\pi-\chi_\delta$.  As can be seen from Tab.~\ref{tab:suscs} the results for $\chi_\pi-\chi_\delta$ are smaller for the strange than for the light quark.  If the strange quark results are interpreted as coming entirely from explicit chiral symmetry breaking, the corresponding effects for the light quarks should be reduced by a factor of $(\mtot_l/\mtot_s)^2 \approx 0.008$.  At $T=179$ MeV, this approach gives explicit chiral symmetry breaking for the light quark quantity  $\chi_\pi-\chi_\delta$ of order $4.26 \cdot 0.008=0.034$.  This is larger than the 0.001 estimate above but only a fraction of a percent of the signal.  Thus, we interpret the results for $\pmd$ and $\sme$ shown in Tab.~\ref{tab:suscl} and Fig.~\ref{fig:ua32} as clear evidence for the anomalous breaking of $\ua$ symmetry for $T>T_c$.  

\section{Low-lying Eigenvalue Spectrum}
\label{sec:eigen}

In Section~\ref{sec:chiral} we studied the QCD transition region by examining the temperature dependence of vacuum expectation values and correlation functions whose behavior is closely related to the $\sua$ and $\ua$ symmetries that are restored, or partially restored, as the temperature is increased through the transition region.  In this section we will examine a different quantity, the spectrum of the light-quark Dirac operator, which is also directly related to the violation of these symmetries.  In the first subsection, Sec.~\ref{sec:eigen_prelim} we review the basic formulae relating the Dirac eigenvalue spectrum to other measures of $\sua$ and $\ua$ symmetry breaking in continuum field theory.  In Sec.~\ref{sec:eigen_results} we present the distribution found for the 100 lowest Dirac eigenvalues for each of the six temperatures studied between 150 - 200 MeV on our largest, $32^3\times 8$ volume.  Finally in Secs.~\ref{sec:eigen_pbp} and \ref{sec:eigen_ua} we make a quantitative connection between this measured eigenvalue spectrum and the subtracted chiral condensate $\Delta_{l,s}$ and the $\ua$-breaking susceptibility difference $\dpd=\pmd$, respectively.  As is discussed in Sec.~\ref{sec:eigen_ua}, at temperatures just above $T_c$ the Dirac spectrum agrees well with the predictions of the dilute instanton gas approximation and this approximation provides a good quantitative description of the anomalous $\ua$ symmetry breaking difference $\pmd$ seen in this region.

\subsection{Preliminaries}
\label{sec:eigen_prelim}

The most familiar relation between the Dirac spectrum and an important QCD observable is the spectral expression for the chiral condensate,
\begin{equation}
  \Sigma_q =
  -\left\langle\overline{\psi}\psi\right\rangle_q=
  \int_0^\infty\textrm{d}\lambda\,
  \rho(\mtot_l,\mtot_s,\lambda)\frac{2\mtot_q}{\mtot_q^2+\lambda^2},\qquad q=l,s.
  \label{eq:pbpms}
\end{equation}
Here we have used the symmetry $\rho(\lambda) = \rho(-\lambda)$, limiting the integral to non-negative values of $\lambda$ and introducing the compensating factor of 2 in the numerator.  In the infinite volume and chiral limits and applied to the light quark condensate, this equation becomes the well-known Banks-Casher relation~\cite{Banks:1979yr}:
\begin{equation}
  -\lim_{\mtot_l\to0}\lim_{V\to\infty} \left\langle\overline{\psi}\psi\right\rangle_l
  = \lim_{\lambda\to0}\lim_{\mtot_l\to0}\lim_{V\to\infty} \pi  \rho(\mtot_l,\mtot_s, \lambda).
  \label{eq:bcr}
\end{equation}
Therefore, if the eigenvalue density $\rho(\mtot,\lambda)$ is non-vanishing in infinite volume at the origin, chiral symmetry will be broken by a non-vanishing quark condensate.  

While we have used the lattice variable $\mtot_q$ to represent the quark mass in this equation, it should be emphasized that this is an equation derived in continuum field theory.  The equivalent expression, derived for DWF in a lattice theory will be quite different.  For example, a spectral expression for $\Sigma_q$ derived from an eigenmode expansion of the DWF lattice propagator will involve wave functions for the five-dimensional modes evaluated on and integrated over the two $s=0$ and $s=L_s-1$, four-dimensional faces, yielding an expression significantly more complex than that given in Eq.~\eqref{eq:pbpms}~\cite{Blum:2000kn}.  However, when appropriately renormalized, the eigenvalue density $\rho(\mtot,\lambda)$ is a physical quantity that can be computed using lattice methods~\cite{Giusti:2008vb}.  Thus, as in Ref.~\cite{Bazavov:2012qja}, we compute the low-lying spectrum $\rho^{\mathrm{latt}}(\lambda)$ of the hermitian DWF Dirac operator, $D_H=\gamma^5 R_5 D_{\mathrm{DWF}}$, where $R_5$ is the reflection operator in the fifth dimension: $s \to L_s-1-s$ for the fifth-dimension coordinate $0 \le s \le L_s-1$.  We then use the $\beta$-dependent renormalization factor $Z_{\mathrm{tw}\to\MS}$ to transform $\rho^{\mathrm{latt}}(\lambda)$ into $\MS$ conventions:
\begin{equation}
\rho(\lambda) = \frac{1}{Z_{\mathrm{tw}\to\MS}}\rho^{\mathrm{latt}}(Z_{\mathrm{tw}\to\MS}\lambda).
\end{equation}
As is discussed in Ref.~\cite{Bazavov:2012qja} the renormalization factor $Z_{\mathrm{tw}\to\MS}$ is given by a product of the factor $Z_{\mathrm{tw}\to m_f}$ given in Tab.~IV of that reference and the factor $Z_{m_f \to \MS}$ listed in Tab.~\ref{tab:suscl} of the present paper.

Since in a lattice calculation the Banks-Casher limit of infinite volume and vanishing quark mass cannot be easily evaluated, we would like to use Eq.~\eqref{eq:bcr} for the case of finite volume and non-zero quark mass.  However, in that case the integral over $\lambda$ diverges quadratically.  As a result, this equation is dominated by the region of large $\lambda$ where the DWF lattice and continuum formalisms should not agree and is well outside the limited range of the 100 lowest eigenvalues which we have computed.  However, much can be learned from Eq.~\eqref{eq:bcr} if we use it to evaluate the difference $\dls$, subtracting the light and strange quark equations.  This difference will be studied in Sec.~\ref{sec:eigen_pbp}, comparing the subtracted spectral integral with both the simple difference of condensates, $\dls$ and the improved quantity $\tdls$.

In a similar manner, the difference between the connected pseudoscalar and scalar light-quark susceptibilities, $\pmd$, which serves as a good indicator of $\ua$ symmetry breaking, can be expressed as a spectral integral~\cite{Chandrasekharan:1995gt}:
\begin{equation}
  \dpd\equiv\chi_\pi-\chi_\delta=\int_0^\infty\textrm{d}\lambda\,
  \rho(\mtot_l,\lambda)\frac{4\mtot_l^2}{(\mtot_l^2+\lambda^2)^2},
  \label{eq:cpmd}
\end{equation}
where again this is a continuum equation which requires that all of the quantities which appear are renormalized in a consistent scheme.   In contrast to Eq.~\eqref{eq:bcr}, this expression is only logarithmically divergent and for our values of the lattice spacing and quark masses, is dominated by the region where $\lambda$ is small -- the region in which we have measured the spectrum and in which the lattice and continuum spectral functions should agree, except for the usual $O(a^2)$ errors inherent in a calculation at non-zero lattice spacing.

In order to distinguish and to better understand the effects of different possible behaviors of $\rho(\mtot_l,\lambda)$ we will also make use of the small $\lambda$ and small $\mtot$ parametrization for $\rho(\mtot_l,\lambda)$,
\begin{equation}
  \rho(\mtot,\lambda) = c_0\mtot^2\delta(\lambda)+c_1|\lambda|+c_2 \mtot+\cdots ,
  \label{eq:expn}
\end{equation}
appropriate for $T \ge T_c$ and introduced in Ref.~\cite{Bazavov:2012qja}.  Each term provides an ansatz for a possible behavior of $\rho(\mtot_l,\lambda)$ and results in a different contribution to the susceptibility difference. In particular, $\dpd$ will receive three corresponding contributions:
\begin{equation}
  \dpd\approx 
  2c_0+2c_1+\pi c_2
  \equiv \dpd^0 +\dpd^1 +\dpd^2.
  \label{eq:contr}
\end{equation}
Once the eigenvalue density has been computed and fit to the form assumed in Eq.~\eqref{eq:expn}, the resulting coefficients can be used to calculate $\dpd$ and discover which of these three behaviors gives the dominant contribution to the spectral integral. 

In addition to allowing a quantitative measure of the relative importance of these three possible behaviors, the use of the analytic expression in Eq.~\eqref{eq:expn} also allows us to potentially correct finite-lattice spacing errors which may be important for small $\lambda$ in our DWF formulation with finite $L_s$.  Although much more accurate, the hermitian DWF spectrum, like the Wilson spectrum, does not have the continuum form $\Lambda = \pm \sqrt{\lambda^2 + \mtot^2}$ where $\mtot= m_l+\mres$, at least for finite volume, finite $L_s$ and non-zero lattice spacing.  For eigenvalues $\Lambda$ of $D_H$  on the order of $\mres$, {\it i.e.} $\Lambda \lessapprox 10$ MeV, we expect deviations from the continuum $\pm \sqrt{\lambda^2 + \mtot^2}$ form because of residual chiral symmetry breaking.   These effects do not occur if we use $\rho(\lambda)$ given by Eq.~\eqref{eq:expn}.  In fact, comparing results obtained by direct summation over the measured spectrum with those obtained using Eq.~\eqref{eq:expn} provides an estimate of the importance of these finite lattice spacing errors. 

Each of the three terms in Eq.~\eqref{eq:expn} corresponds to potentially interesting behavior.  The $\lambda$-independent $c_2 \mtot$ term is expected to dominate the behavior below $T_c$ and should describe the Banks-Casher contribution to the chiral condensate $\Sigma_l$.  For $T<T_c$ the factor of $\mtot$ should not appear but has been introduced here because above $T_c$ the condensate should vanish in the limit $\mtot\to0$.   As can be seen in Eq.~\eqref{eq:contr}, this $c_2\mtot$ term will result in $\dls \ne 0$ and anomalous symmetry breaking.  Likewise, the linear $c_1$ term provides a possible mechanism for $\ua$ symmetry breaking above $T_c$.  Both the $c_1$ and $c_2$ terms are sufficiently regular as $\lambda$ and $\mtot$ approach 0 that they do not result in an explicit $\sua$ symmetry breaking chiral condensate but have sufficient infra-red singularity that the presence of either does result in a non-zero value for $\pmd$.  Thus, either term in $\rho(\lambda)$ could describe the behavior we see for $T>T_c$ where $\Sigma_l$ should vanish as $\mtot_l\to0$ but $\pmd$ is non-zero.  As we will see, neither term appears to be present with a sufficient magnitude to describe $\pmd$ for $T>T_c$.

As is discussed below, the $c_0$ term has the greatest relevance.  This term represents the Dirac spectrum that results from the dilute instanton gas approximation (DIGA)~\cite{Gross:1980br}.  Asymptotic freedom implies that at sufficiently high temperature, the QCD partition function will be governed by weak-coupling phenomena.  These should include a ``dilute gas'' of instantons and anti-instantons of radius $\approx 1/T$ and density $\propto \mtot^2_l\exp\{-8\pi^2/g^2(T)\}$ decreasing with increasing temperature, where $g(T)$ is the running QCD coupling constant evaluated at the energy scale $T$.  The number of such instantons and anti-instantons is proportional to the volume and each will induce a near-zero mode in the Dirac eigenvalue spectrum.  (These eigenvalues will not be exactly zero because of the overlap of the `zero'-mode wave functions associated with neighboring instantons.)  The factor of $\mtot^2$ in the instanton density arises from the fermion determinant for two light flavors of quarks.  The contribution of such a dilute gas of instantons and anti-instantons to the Dirac spectrum will be accurately described by the $c_0$ term in Eq.~\eqref{eq:expn}, at least for sufficiently high temperatures.   As can be seen from Eq.~\eqref{eq:contr}, such a term will result in a non-zero value for the difference $\pmd$ even in the chiral limit, $\mtot_l\to0$. The expected presence of such effects leads to the phrase ``effective restoration of $\ua$ symmetry'', since these effects, which should appear as $T$ becomes very large, will lead to a possibly very small but non-vanishing result of $\pmd$.  

As we will demonstrate in Sec.~\eqref{sec:eigen_ua} we find a significant cluster of near-zero modes in the Dirac spectrum whose number is proportional to the volume with the characteristics expected from the DIGA.  We conclude that the non-zero value of $\pmd$ in the region just above $T_c$ is explained by the DIGA and that this is the dominant mechanism for our observed, non-zero breaking of $\ua$ just above $T_c$.

\subsection{Eigenvalue distributions}
\label{sec:eigen_results}

To compute the Dirac eigenvalue spectrum, we follow closely the method described in detail in Ref.~\cite{Bazavov:2012qja}. The lowest 100 eigenvalues $\{\Lambda_n\}_{1 \le n \le 100}$ of the Hermitian DWF Dirac operator $D_H$ are calculated for each of $\approx 100$ configurations for each of six ensembles ranging in temperature between 149 and 195 MeV using the Kalkreuter-Simma method~\cite{Kalkreuter:1995mm}. The same fermion mass is used in the Dirac operator as was used when the ensemble was generated.

In the continuum, the eigenvalues $\Lambda_n$ of the hermitian Dirac operator have the form $\pm\sqrt{\lambda_n^2+\mtot_l^2}$ and the eigenvalue density is conventionally expressed in terms of the mass-independent eigenvalue $\lambda$.  Here we will attempt to follow the same practice.  However, for the DWF Dirac operator, the quark mass is not a simple additive constant but is embedded within $D_{\mathrm{DWF}}$ in a complex fashion.  The continuum form $\pm\sqrt{\lambda_n^2+\mtot_l^2}$ is therefore not guaranteed by the structure of $D_{\mathrm{DWF}}$ but is expected to emerge in the limit of infinite volume, infinite $L_s$ or vanishing lattice spacing $a$.  Thus, in our circumstances, we will find some eigenvalues $\Lambda_n$ which are smaller than $\mtot_l$ and for which $\lambda_n = \sqrt{\Lambda_n^2 - \mtot_l^2}$ will be imaginary.  As in Ref.~\cite{Bazavov:2012qja}, when we present a histogram showing $\rho(\lambda)$ we include these imaginary values in a separate histogram plotted at negative $\lambda$ with an imaginary value of $\lambda$ added to a bin at $-|\lambda|$.  Plotted in this way, these ``unphysical'' values of $\Lambda$ are made visible and their relative importance can be judged.  We exploit the symmetry between positive and negative values of $\lambda$ and associate each $\Lambda_n$ with magnitude greater than $\mtot_l$ with the positive value $\lambda_n = +\sqrt{\Lambda_n^2 - \mtot_l^2}$.

\begin{figure}[hbt]
  \begin{center}
	\begin{minipage}[t]{0.45\linewidth}
		\centering
        \resizebox{\linewidth}{!}{\includegraphics{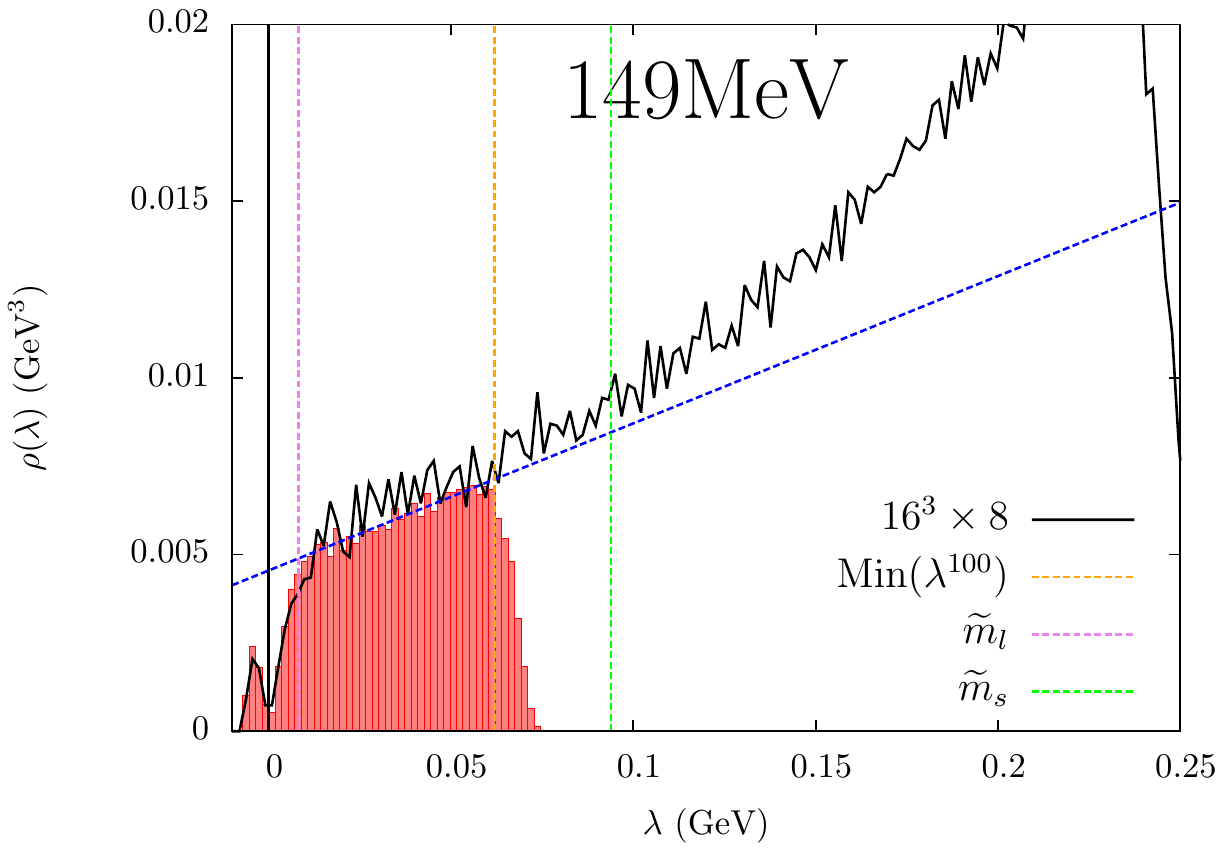}}
	\end{minipage}%
	\begin{minipage}[t]{0.45\linewidth}
		\centering
        \resizebox{\linewidth}{!}{\includegraphics{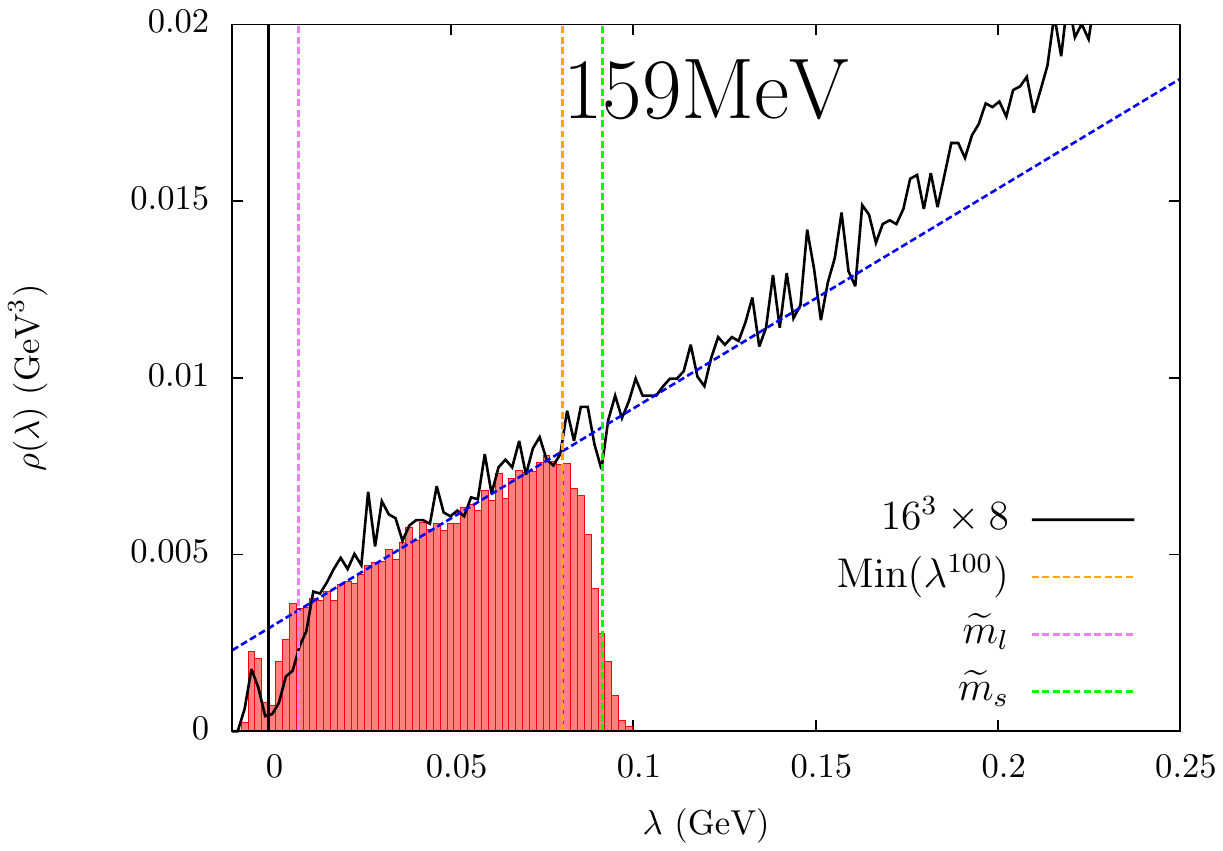}}
	\end{minipage}
	\begin{minipage}[t]{0.45\linewidth}
		\centering
        \resizebox{\linewidth}{!}{\includegraphics{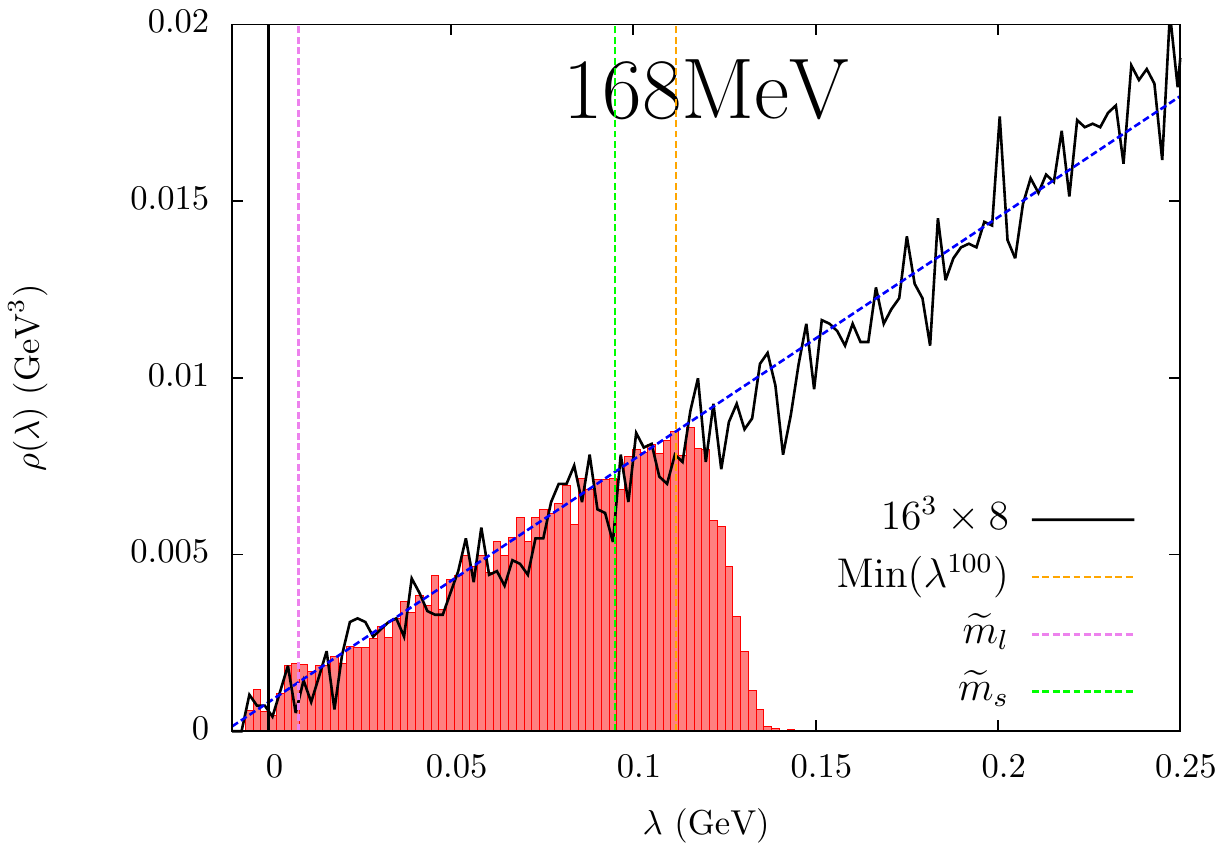}}
	\end{minipage}%
	\begin{minipage}[t]{0.45\linewidth}
		\centering
        \resizebox{\linewidth}{!}{\includegraphics{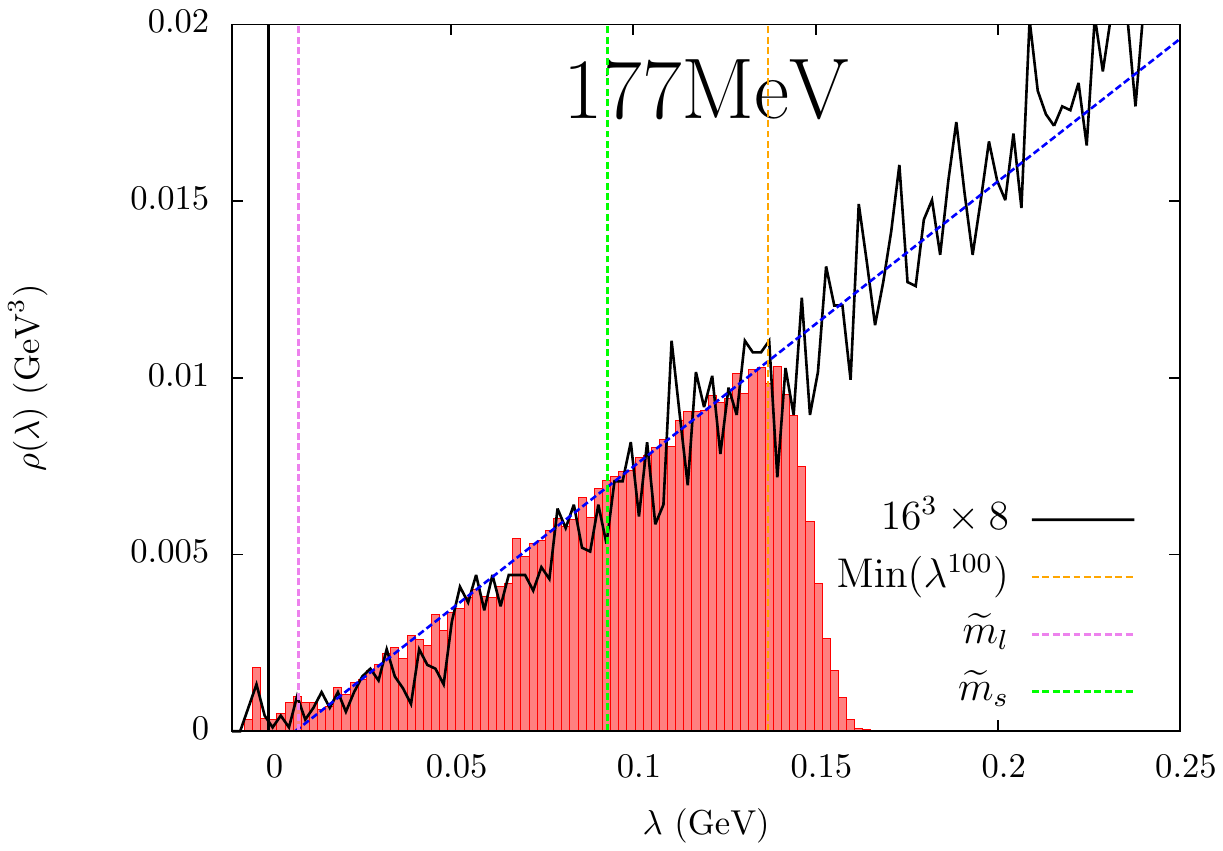}}
	\end{minipage}
	\begin{minipage}[t]{0.45\linewidth}
		\centering
        \resizebox{\linewidth}{!}{\includegraphics{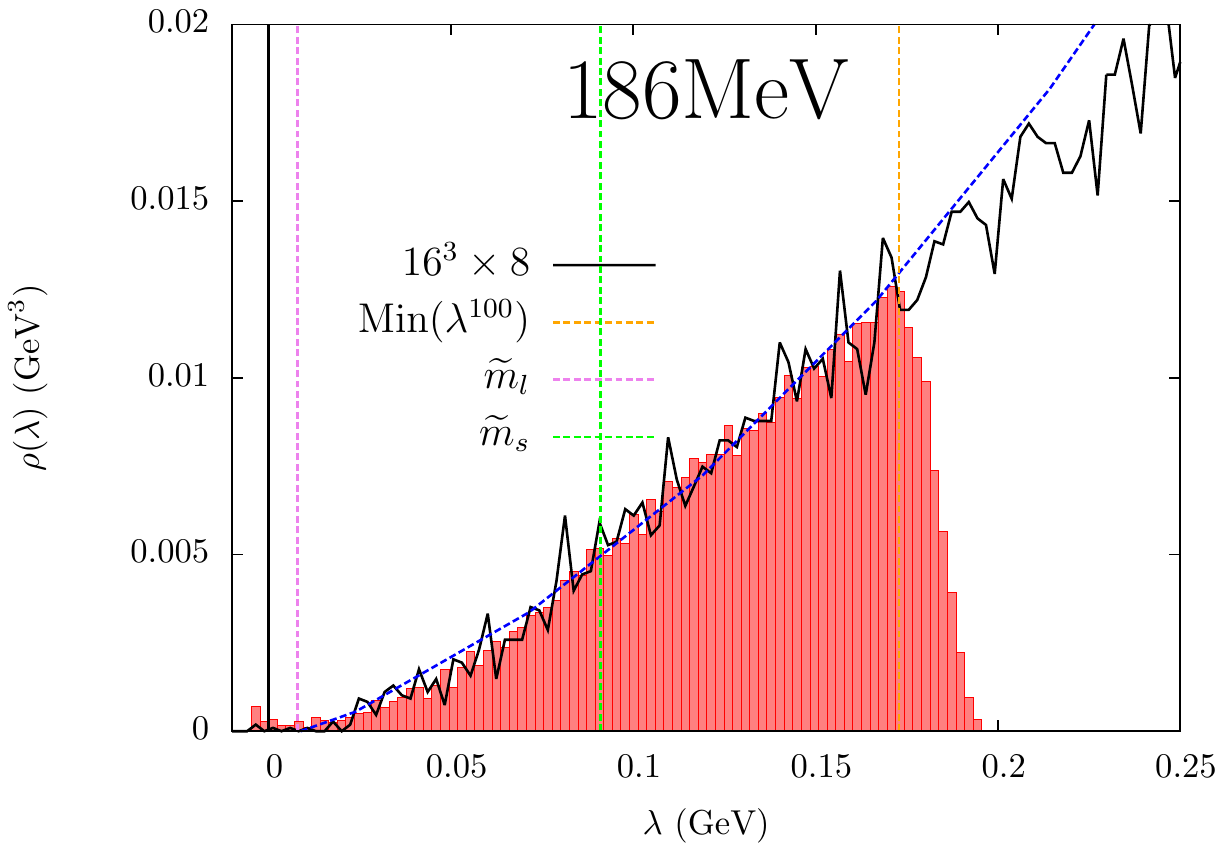}}
	\end{minipage}%
	\begin{minipage}[t]{0.45\linewidth}
		\centering
        \resizebox{\linewidth}{!}{\includegraphics{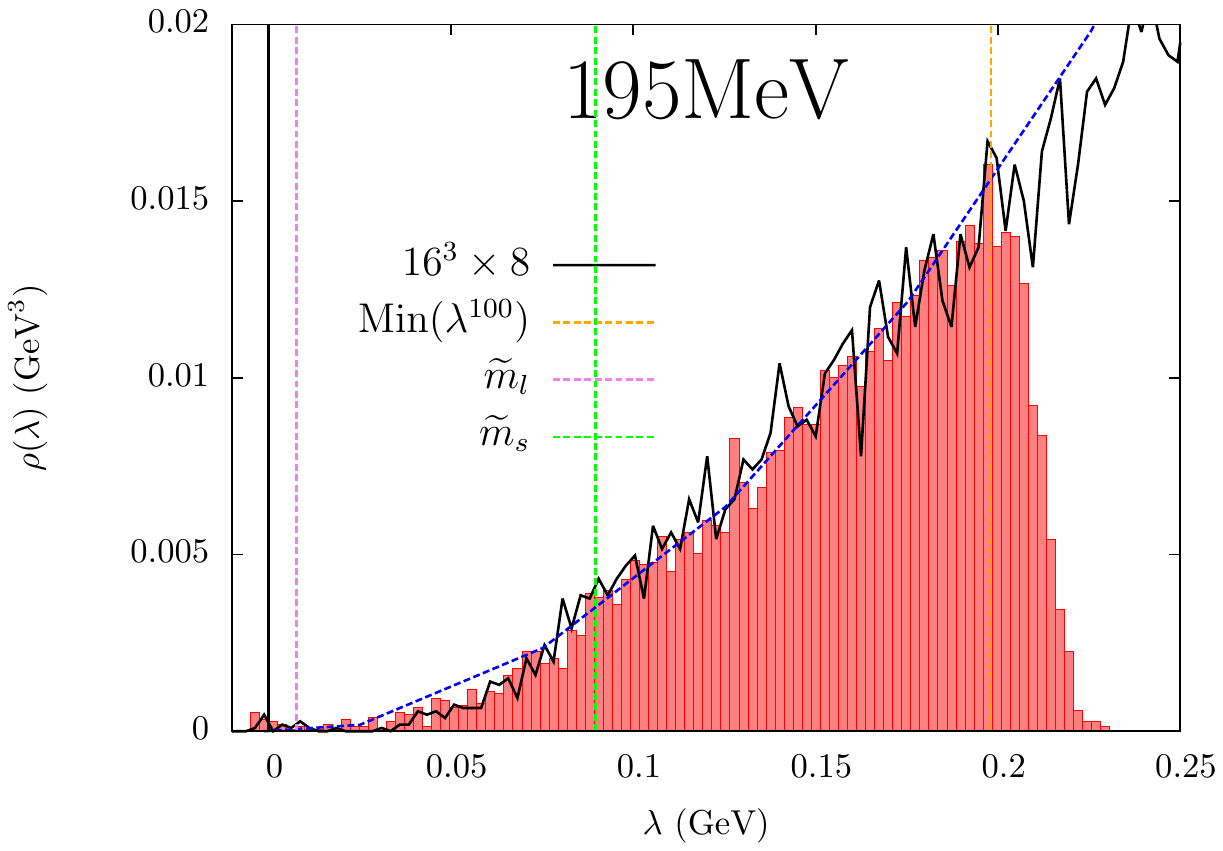}}
	\end{minipage}
  \end{center} 
  \caption{The eigenvalue spectrum  for $T=149-195$ MeV, expressed in the $\MS$ scheme at the scale $\mu=2$ GeV. The imaginary, ``unphysical'' eigenvalues are plotted as $-\sqrt{|\Lambda^2-\mtot^2_l |}$. The spectra from the $32^3\times8$ ensembles are plotted as histograms and fit with a linear ($T=149-178$ MeV) or a quadratic ($T=186-195$ MeV) function (blue dashed line). The spectrum from each of the $16^3\times8$ ensembles~\cite{Bazavov:2012qja} is plotted as a black solid line.}
\label{fig:eig}
\end{figure}

Figure~\ref{fig:eig} shows the distributions, renormalized in the $\MS$ scheme at the scale $\mu=2$ GeV, determined from the lowest 100 eigenvalues ($\lambda$) for six ensembles at temperatures from 149 MeV to 195 MeV.  The eigenvalue densities for the $32^3\times8$ space-time volumes are plotted as solid histograms, while the $16^3\times8$ results are plotted as black, solid lines. The aforementioned imaginary, ``unphysical'' modes are plotted as $-\sqrt{|\Lambda^2-\mtot^2_l |}$ on the negative axis. The values for the total mass of light and strange quarks, $\mtot_l^{\MS}$ and $\mtot_s^{\MS}$, are indicated by vertical dashed lines, which give a physical scale for the eigenvalue distribution.  Since we have determined only a fixed number of eigenvalues, the spectral distributions will be distorted at their upper ends.  The third vertical dashed line in these plots, which appears with various $x$-coordinates, locates the smallest value for $\lambda_{100}$ found for each ensemble.  The spectrum shown to the left of this line will then be undistorted by our failure to include larger eigenvalues in the figure.

\begin{figure}[htb]
  \begin{center}
	\begin{minipage}[t]{0.33\linewidth}
		\centering
        \resizebox{\linewidth}{!}{\includegraphics{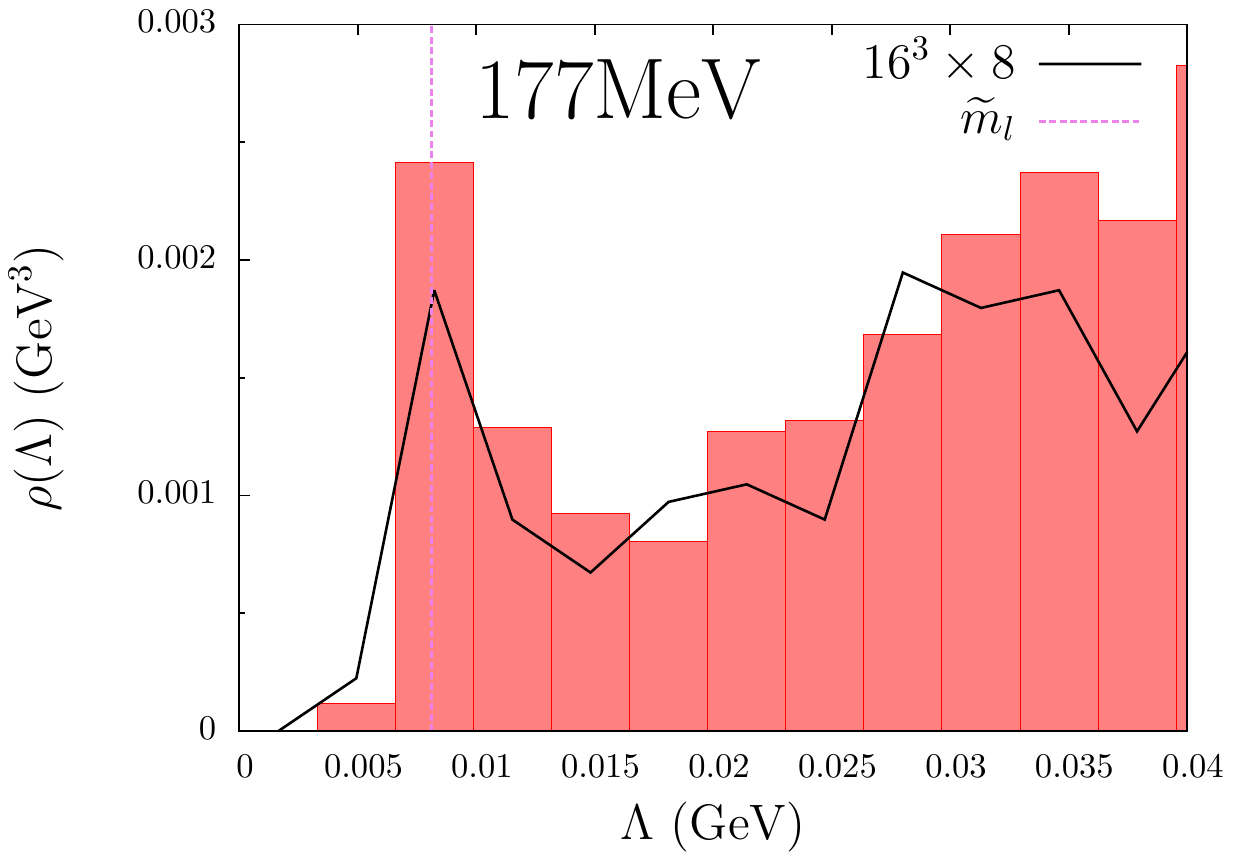}}
	\end{minipage}%
	\begin{minipage}[t]{0.33\linewidth}
		\centering
        \resizebox{\linewidth}{!}{\includegraphics{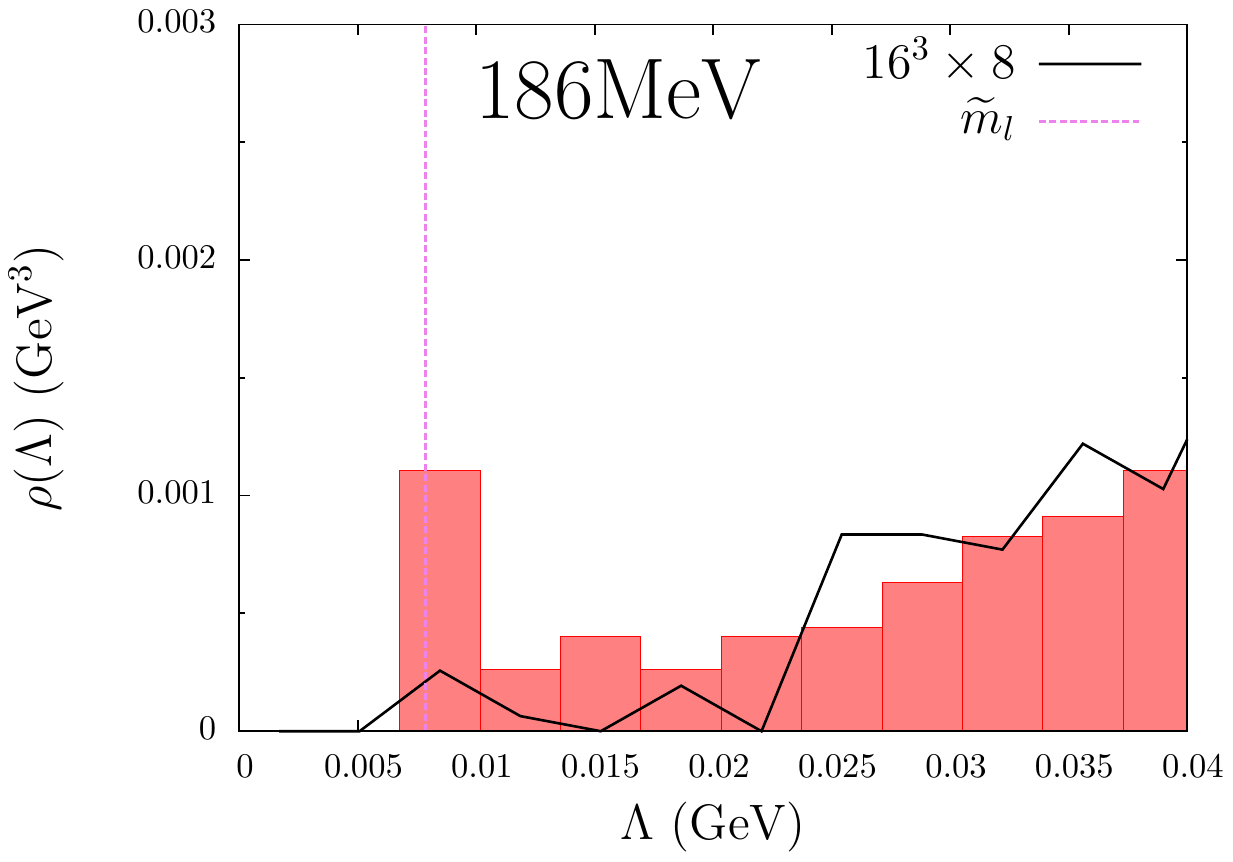}}
	\end{minipage}%
	\begin{minipage}[t]{0.33\linewidth}
		\centering
        \resizebox{\linewidth}{!}{\includegraphics{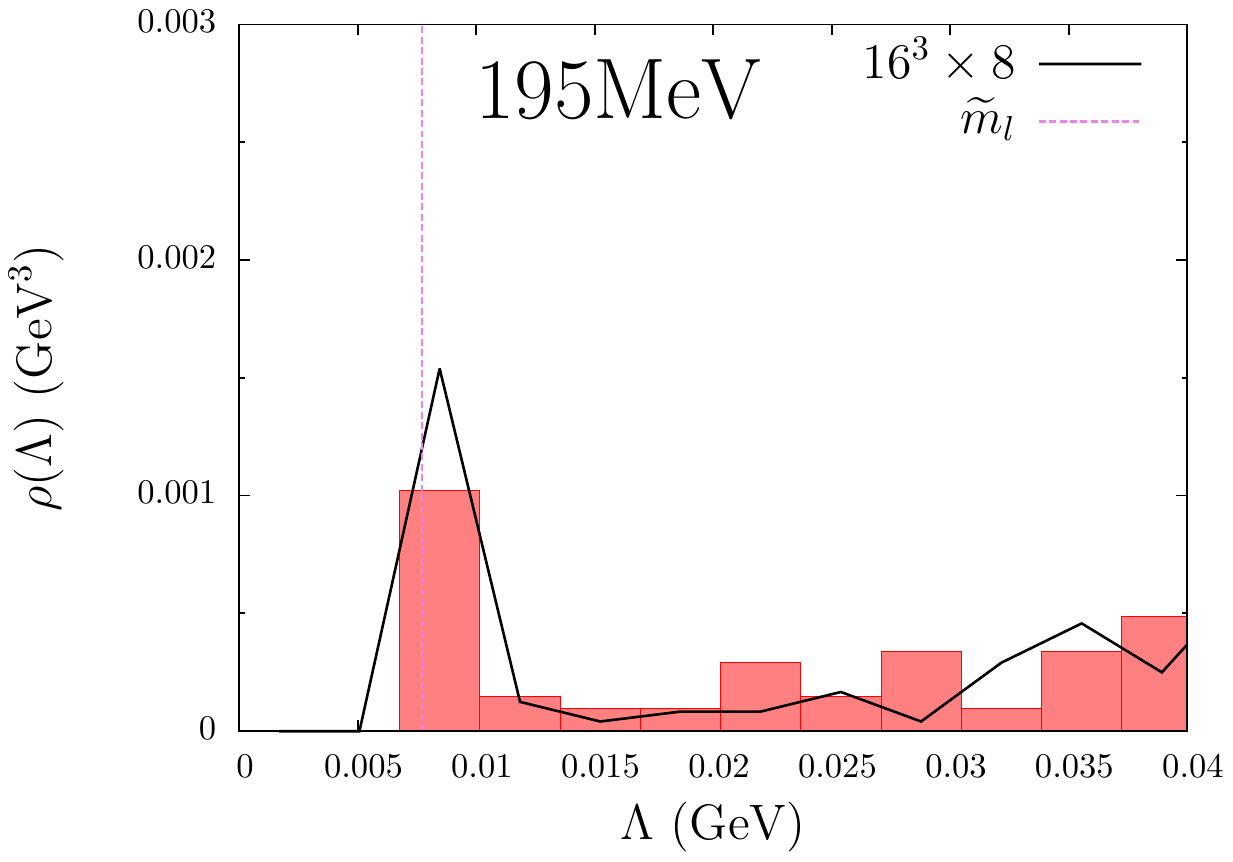}}
	\end{minipage}%
  \end{center} 
  \caption{(Left to right) The renormalized eigenvalue spectrum for $T=177-195$ MeV without the removal of the bare quark mass. The statistics are likely insufficient for $186$ MeV on  the $16^3\times8$ ensemble; only 5 instances of ''near-zero modes'' are collected.}
  \label{fig:low}
\end{figure}

Since the number of eigenmodes is proportional to the space-time volume, a fixed number of the lowest modes will become more concentrated at the lower-end of the spectrum as the volume increases. This phenomena can be easily seen in Fig.~\ref{fig:eig} where the range of eigenvalues studied decreases dramatically as the space-time volume is increased from $16^3\times8$ to $32^3\times8$.  However, while the range of eigenvalues covered by the larger $32^3\times8$ volume is reduced, this larger volume provides a better sampling and more convincing view of the spectrum near zero, the region of greatest interest.

For $T=149$ and 159 MeV, the eigenvalue distributions can be characterized as a linear function with a non-vanishing intercept for eigenvalues of order 10 MeV or larger.  Below 10 MeV the spectrum is distorted by a combination of finite volume and residual chiral symmetry breaking effects. The non-vanishing intercept, interpreted through the Banks-Casher relation, is consistent with the non-vanishing chiral condensate and vacuum chiral symmetry breaking observed at these temperatures which lie below the pseudo-critical temperature.

For $T=168$ MeV, the linear behavior continues to be visible, but the intercept has essentially vanished,
suggesting that 168 MeV is close to the pseudo-critical temperature, consistent with the temperature dependence of the $\sua$-breaking susceptibility difference $\pms$  shown in Fig.~\ref{fig:sua32}.

For $T=177$ MeV, a small peak in $\rho(\lambda)$ near the origin emerges as a cluster of near-zero 
modes. Such a cluster of near-zero modes might result from the Atiyah-Singer theorem and non-vanishing topological charge or from the dilute instanton gas approximation (DIGA).  As is discussed below, the volume dependence of this peak and the distribution of the chirality of these modes is consistent with the DIGA and inconsistent with their arising from non-zero global topology.  This small eigenvalue region can be best seen in the expanded view given in Fig.~\ref{fig:low}.

For $T=186$ and 195 MeV, this small peak survives although it diminishes in size with increasing temperature.  In addition, the peak becomes increasingly separated from the rest of the spectrum by a gap containing few eigenvalues.  As a result the remainder of the spectrum, excluding this peak, can no longer be fit using a linear function.  A quadratic fit is possible at $T=186$ but an even higher power may be needed to describe the 195 MeV spectrum.

\subsection{Subtracted Chiral Condensate}
\label{sec:eigen_pbp}

It is not difficult to see very approximate agreement between the intercept of the spectral density at $\lambda=0$ (ignoring obvious distortions to the spectrum near $\lambda=0$) and the measured value of $\Sigma_l$ implied by the Banks-Casher relation.  However, a careful, quantitative test of Eq.~\eqref{eq:pbpms} must overcome two obstacles: both the finite volume suppression of $\rho(\lambda)$ as $\lambda\to0$ and the quadratic divergence present in $\Sigma_q$ for non-zero quark mass.  For a DWF calculation such a test is further complicated by the contributions of residual chiral symmetry breaking to $\Sigma_q$ and $\rho(\lambda)$ for small $\lambda$.   As suggested above, all of these difficulties can be overcome.  The first step is to consider the subtracted chiral condensate, $\dls$ defined in Eq.~\eqref{eq:spbp}.  If Eq.~\eqref{eq:pbpms} is used to express $\dls$ in terms of the spectral density, we obtain the more convergent result:
\begin{equation}
\dls = \int_0^\infty d\lambda \rho(\lambda) \frac{2 \mtot_l(\mtot_s^2-\mtot_l^2)}
                                                 {(\lambda^2+\mtot_l^2)(\lambda^2+\mtot_s^2)}.
\label{eq:eigen_pbp_sub}
\end{equation}
While this expression still receives a contribution from large eigenvalues, well above the group of low modes studied here, this high-mode contribution is expected to be of order $m_l m_s^2 \ln(m_s a)$ which is possibly 1\% of the $(250 \mathrm{MeV})^3$ value of the zero temperature chiral condensate.  Thus, we expect that for our present quark masses and lattice spacing, we can evaluate the right hand side of Eq.~\eqref{eq:eigen_pbp_sub} using our 100 low modes to at least a few percent accuracy, at least for $T \le T_c$.  

We can evaluate the integral in Eq.~\eqref{eq:eigen_pbp_sub} using our measured eigenvalues in two ways.  First for each measured configuration we can replace the integral over $\lambda$ on the right hand side of Eq.~\eqref{eq:eigen_pbp_sub} by a sum over the measured eigenvalues.  In addition we can express the integrand in Eq.~\eqref{eq:eigen_pbp_sub} in terms of the directly measured eigenvalues $\Lambda_n$ so that the uncertainties associated with those values of $\Lambda_n$ lying below $\mtot_l$ are avoided.  The resulting expression for $\dls$ becomes
\begin{equation}
\dls^{\mathrm{ms}} =\frac{1}{N_\sigma^3N_\tau}\left\langle
                          \sum_{n=1}^{100} \frac{\mtot_l(\mtot_s^2-\mtot_l^2)}
                                    {\Lambda_n^2(\Lambda_n^2+\mtot_s^2-\mtot_l^2)} \right\rangle,
\label{eq:eigen_pbp_sub_ms}
\end{equation}
where $\langle\ldots\rangle$  indicates an average over configurations and we use the notation ``${\mathrm{ms}}$'' (mode sum) to identify the result obtained from this summation over modes. 

In the second approach to Eq.~\eqref{eq:eigen_pbp_sub} we replace the spectral density $\rho(\lambda)$ by the fitted expression given in Eq.~\eqref{eq:expn} and then perform the integration over $\lambda$ analytically with the result:
\begin{equation}
  \dls^{\mathrm{eig}} \equiv c_0\mtot_l +c_1\mtot_l\ln\left(\frac{\mtot_s^2}{\mtot_l^2}\right)
  +c_2\pi\mtot_l ,
  \label{eq:subeig}
\end{equation} 
where terms of order $\mtot_l/\mtot_s$ have been neglected and the label ``eig'' has been introduced to distinguish this expression from those resulting from the three other approaches to the calculation of this quantity.

In Tab.~\ref{tab:dls} we compare these two spectral methods for computing $\dls$ with the results from both the direct subtraction of the measured condensates (which we continue to label as $\dls$) and the improved quantity $\tdls$ which is less contaminated by residual DWF chiral symmetry breaking effects.   As can be seen from the table, for the temperatures at which the fit form given in Eq.~\eqref{eq:expn} provides a good description of the eigenvalue distribution, $139 \mathrm{MeV} \le T \le 168 \mathrm{MeV}$, analytic integration of this three-parameter function and the direct sum over the lowest 100 modes agree reasonably well.  This supports the use of the three-parameter function to provide an interpretation of our results.  This agreement also suggests that the region $|\Lambda| \lessapprox 10$ MeV, which is distorted in our computed Dirac eigenvalue spectrum by finite volume and residual chiral symmetry breaking effects but treated in a fashion consistent with infinite volume, continuum expectations by the fitting function, does not play a large role in these results.  The difference between $\dls^\text{eig}$ and $\dls^\text{ms}$ can serve as an estimate for the systematic error in the fit coefficients, a difference which at its largest is about $15\%$.

A second observation that can be drawn from the data in Tab.~\ref{tab:dls} is that the quantity $\tdls$ agrees reasonably well with the result obtained directly from the Dirac spectrum over the full temperature range.  This suggests that a good representation for the chiral condensate can be obtained by performing the subtraction of light and strange quark Green's functions and that in the case of DWF it is best to use the GMOR relation and subtract connected pseudoscalar susceptibilities rather than the condensates themselves which contain relatively large, uncontrolled residual chiral symmetry breaking effects.  We would like to emphasize that our use of the continuum spectral Eq.~\eqref{eq:expn} combined with the renormalized DWF spectrum makes strong assumptions about the validity of continuum methods in our lattice calculation at reasonably strong coupling.   It is impressive that on the larger $32^3$ volume, where the statistical errors are likely most reliable, Tab.~\ref{tab:dls} shows agreement between $\dls^{\mathrm{ms}}$ and $\tdls$ consistently at the 1 sigma level, which in some cases represent an accuracy of 4\%  or less.

\begin{table}[hbt]
  \centering
  \begin{tabular}{c|ccccc|cccc}\hline
    \#&$T$ (MeV)&$\Ns$&$L_s$&$\mtot_l$&$\mtot_s$&\Tspace\Bspace
    $\dls^\text{eig}/T^3$&$\dls^\text{ms}/T^3$&$\dls/T^3$
    &$\tdls/T^3$\\\hline
    \runn{150_1}&149&16&32&0.00464&0.05293&6.72&6.00&3.07(12)&5.7(2) \\
    \runn{150}  &149&16&48&0.00468&0.05295&6.85&5.65&5.00(10)&6.3(1) \\
    \runn{150}  &149&16&64&0.00459&0.05289&   -&   -&5.57(10)&6.2(1) \\
    \runn{150_a}&149&32&32&0.00464&0.05293&6.45&6.39&3.84(5) &6.4(1) \\
    \runn{160_a}&159&32&32&0.00421&0.04856&3.86&4.28&2.83(6) &4.2(1) \\
    \runn{170_a}&168&32&32&0.00395&0.04490&1.64&2.19&1.46(7) &2.3(1) \\
    \runn{180_a}&177&32&32&0.00367&0.04165&   -&1.21&0.71(5) &1.3(1) \\
    \runn{190_a}&186&32&32&0.00341&0.03873&   -&0.42&0.22(4) &0.46(5)\\
    \runn{200_a}&195&32&32&0.00314&0.03619&   -&0.25&0.14(3) &0.30(6)\\
    \hline
  \end{tabular}
  \caption{
  \label{tab:dls}
Comparison of the unrenormalized results for $\dls$ computed using four different methods at various temperatures and values of $L_s$.  The data in the $16^3\times8,\;L_s=64$ row results from a valence calculation performed on the $L_s=48$, $\beta = 1.671$ (\runref{150}) ensemble.  (While these quantities are all expressed in the scheme defined by the bare lattice mass, $m_q$, this is not the scheme in which the eigenvalues of the 5-dimensional DWF Dirac operator are defined and renormalization using the factor $Z_{\mathrm{tw}\to m_f}$ defined in Ref.~\cite{Bazavov:2012qja} has been carried out.)
}
\end{table}

Finally we examine  the results at $T=149$ MeV where multiple ensembles with different values of $L_s$ are available, shown in the first four lines of Tab.~\ref{tab:dls}.   Here results are shown for three values of $L_s$: 32, 48 and 64.   As expected, the simple difference $\dls$  shows a very strong dependence on $L_s$.   While there should be substantial cancellation between the large, continuum-like modes in this difference, at the very highest energies this cancellation will be distorted by residual chiral symmetry breaking effects.  The use of the factor $(m_l+\mres)/(m_s+\mres)$ in the subtracted strange condensate will not, in general, cause these effects to cancel.  However, this argument suggests that as $L_s$ increases and these residual chiral symmetry breaking effects are suppressed, $\dls$ should approach $\tdls$, behavior that can be seen in Tab.~\ref{tab:dls}.  Less consistent is the apparent increase in the value of $\tdls/T^3$ with increasing $L_s$ seen on the $16^3$ volume, where an increase by more than two standard deviation from 5.7(2) to 6.2(1) is seen as $L_s$ grows from 32 to 64.  Since $\tdls$ is supposed to already be close to its $L_s=\infty$ value such $L_s$ dependence is not expected and we attribute this discrepancy to the under estimation of statistical errors for this small, $16^3$ volume.

\subsection{Near-Zero Modes and $\ua$ Symmetry}
\label{sec:eigen_ua}

We now turn to one of the central questions addressed in this paper, the origin of the observed $\ua$ symmetry breaking above $T_c$.  We will focus on the quantity $\dpd = \pmd$ since this difference of susceptibilites can be expressed in terms of the spectral density using Eq.~\eqref{eq:cpmd}.  Table~\ref{tab:cpmd} shows this difference at six temperatures as determined from the integrated connected Green's functions.  This difference contains only a very small logarithmic singularity after multiplicative renormalization by $1/Z_{m_f\to\MS}^2$ in the continuum, $\sim (m_l+\mres)^2 \ln{m_l a}$, where the sum $m_l+\mres$ represents schematically the effects of both the input quark mass and DWF residual chiral symmetry breaking.  This controlled high-energy behavior is realized by the convergence of the integral in Eq.~\eqref{eq:cpmd}, even when $\rho(\lambda)$ increases linearly or quadratically with $\lambda$.  

Therefore, in Tab.~\ref{tab:cpmd} we also show the contributions to the spectral integral in Eq.~\eqref{eq:cpmd} of each of the three separate ans\"atze in Eq.~\eqref{eq:expn}, given in Eq.~\eqref{eq:contr}.  Some cells are left blank because the corresponding behavior cannot be seen in the spectral data.  For example, at $T\le168$ MeV, there is no visible accumulation of near-zero modes that might be described by a $\delta(\lambda)$ term in $\rho(\lambda)$.   However, at $T\ge177$ MeV and above we can count a number of near-zero modes that form a small but visible peak in $\rho(\lambda)$ near $\lambda=0$.   Assuming a Poisson distribution, we take the square root of the total number of these near-zero modes as a rough estimate of errors for the corresponding contribution.  Similarly the constant contribution or intercept has vanished for $T \ge 177$ MeV and above $T=177$ MeV the linear term is also difficult to determine and the eigenvalue density  is dominated by what appears to be quadratic behavior.

We can also determine the susceptibility difference $\dpd$ by using a direct sum over modes as was done for $\dls$ in Eq.~\eqref{eq:eigen_pbp_sub_ms} and tabulated as $\dls^{\mathrm{ms}}$ in Tab.~\ref{tab:dls}.  Examining the continuum spectral Eq.~\eqref{eq:cpmd}, we can write an expression for $\dpd$ analogous to that in Eq.~\eqref{eq:eigen_pbp_sub_ms} for $\dls$:
\begin{equation}
\dpd^{\mathrm{ms}} =\frac{1}{N_\sigma^3N_\tau}\left\langle
                          \sum_{n=1}^{100} \frac{2\mtot_l^2}
                                    {\Lambda_n^4} \right\rangle.
\label{eq:eigen_chi_sub_ms}
\end{equation}
The results from this mode sum are shown in the second column from the right in Tab.~\ref{tab:cpmd} where very good agreement is seen with the explicit difference of correlation functions.  This substitution of our renormalized DWF eigenvalue spectrum directly into the continuum equation for $\dpd$ is a stringent test of that spectrum.  The infra-red singular factor $1/\Lambda_n^4$ appearing in Eq.~\eqref{eq:eigen_chi_sub_ms} might have shown large, unphysical fluctuations associated with configuration-by-configuration fluctuations in residual chiral symmetry breaking.  In fact, it is possible that the larger values shown in Tab.~\ref{tab:cpmd}  for $\dpd^{\mathrm{ms}}$ relative to the actual correlator difference $\dpd$ at the two lowest temperatures are a result of this effect.  However, overall the agreement between $\dpd^{\mathrm{ms}}$ and $\dpd$ is remarkably good.

The separate contributions to $\dpd$ presented in Table~\ref{tab:cpmd} give a clear, quantitative description of how the contribution of each piece evolves as the temperature increases.   For $T\le T_c$, the constant, or Banks-Casher term, gives the major contribution to $\dpd$.  In contrast, in the region above the pseudo-critical temperature, the delta function term dominates and its contribution alone agrees well with the result from the difference of integrated correlators.   We conclude that the non-zero $\ua$ symmetry breaking that we observe above $T_c$ in the correlator difference $\pmd$ results from this small cluster of near-zero modes which can be seen in the spectral distributions shown in Fig.~\ref{fig:eig} for $T=177$, 186 and 195 MeV and more easily in the expanded plots in Fig.~\ref{fig:low}.

It is possible that these near-zero modes become exact zero modes in the continuum limit and are a result of non-zero global topology and the Atiyah-Singer theorem.   If this is the case, the number of these zero modes should increase in proportion to $\sqrt{V}$ with increasing space-time volume.  Thus, for zero modes resulting from non-zero global topology we expect the corresponding density per space-time volume to be proportional to $1/\sqrt{V}$.  Were such exact zero modes the only contribution to $\ua$ symmetry breaking then we would conclude that $\ua$ symmetry will be restored in the limit of infinite volume.

However if we compare the results for $32^3$ (solid red histograms) and $16^3$ (black lines) in the expanded view of these peaks shown in Fig.~\ref{fig:low}\;\footnote{Here we use the distributions of $\Lambda$ instead of $\lambda$ near the origin, since it allows us to ignore the large relative fluctuations in these small eigenvalues below $\mtot_l$.} for $T=177$, 186 and 195 MeV, we easily see that the density is volume independent, instead of shrinking by a factor of $\sqrt{8}$ as the volume is increased from $16^3$ to $32^3$.  Thus, the volume dependence of these near-zero modes corresponds to what is expected if they result from a relatively dilute gas of instantons and anti-instantons whose number, and whose corresponding near-zero modes, will grow proportional to the volume.  

We have also examined the chirality of these near-zero modes.  In particular, if these modes are the result of non-zero global topology, then, for a single configuration, all these modes should be of the same chirality, that of the global topological charge $\nu$.  If $\nu$ is positive then each of the zero modes should be right-handed and in our DWF case have support primarily on the right-hand, $s=L_s-1$  boundary.  If $\nu$ is negative then all modes should be left-handed and their wave functions should be largest on the left-hand, $s=0$ boundary.  In contrast, if these modes arise from a dilute instanton gas, they are produced by a mixture of instantons and anti-instantons and the chirality of each mode should have an equal probability to be either positive or negative within a single configuration. 

\begin{figure}[htb]
  \begin{center}
	\begin{minipage}[t]{0.33\linewidth}
		\centering
        \resizebox{\linewidth}{!}{\includegraphics{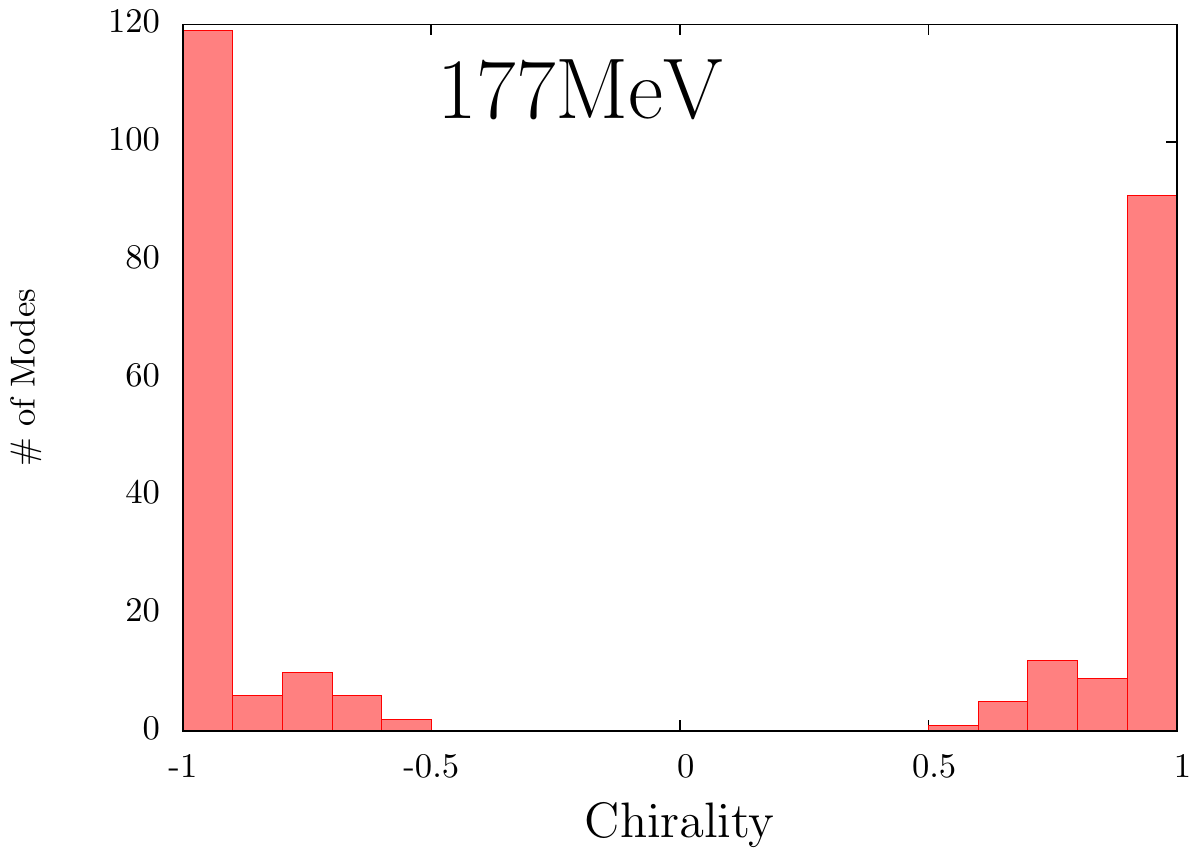}}
	\end{minipage}%
	\begin{minipage}[t]{0.33\linewidth}
		\centering
        \resizebox{\linewidth}{!}{\includegraphics{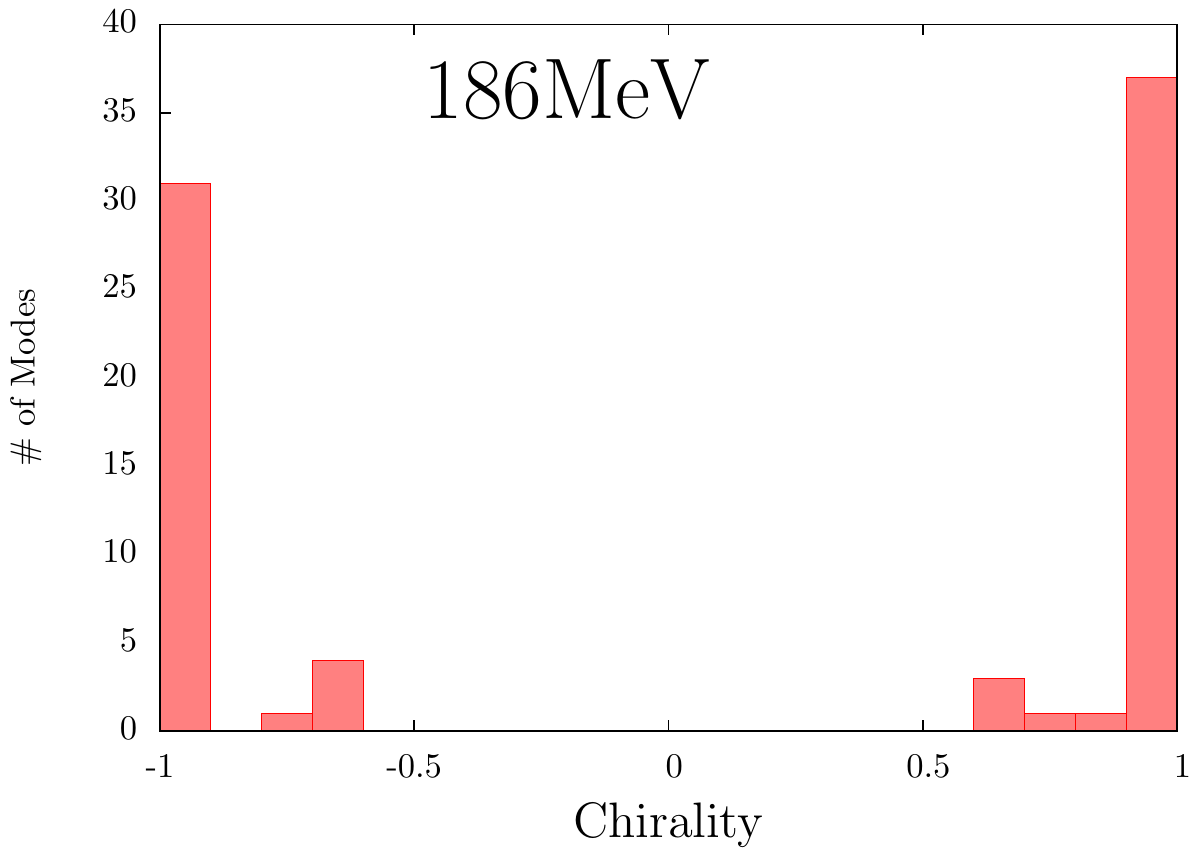}}
	\end{minipage}%
	\begin{minipage}[t]{0.33\linewidth}
		\centering
        \resizebox{\linewidth}{!}{\includegraphics{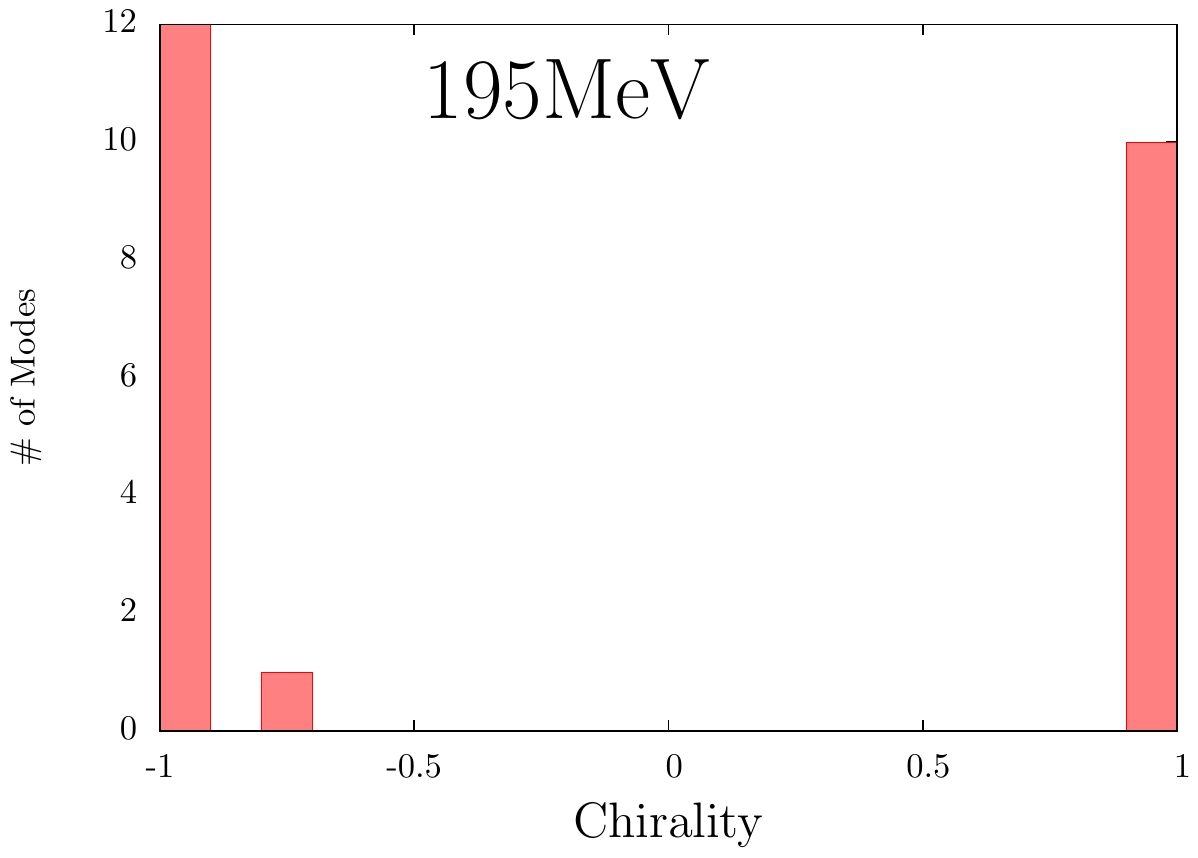}}
	\end{minipage}%
  \end{center} 
  \caption{(Left to right) The distribution of chiralities for the near-zero modes at the three temperatures $T=177$, 186 and 195 MeV and the $32^3\times8$ volume. Here we only use modes lying in the first four histogram bins in Fig.~\ref{fig:low} which corresponds to $\Lambda\lesssim12.5$~MeV. }
  \label{fig:eigchi}
\end{figure}

We choose the $T=177$ MeV ensemble to study the chirality of the near-zero modes since it has the most near-zero modes among the three highest temperature ensembles, where these modes are seen.  We did not save the full five-dimensional eigenfunctions when computing the lowest 100 modes and have available only values for the squared modulus of the five-dimensional wave function, integrated over the left- and right- hand wall for each mode.  Therefore we define the chirality of the $n^{th}$ mode as
\begin{equation}
  \chi_n = \frac{\int d^4 x \overline\Psi_n(x,0)(1+\gamma^5)\Psi_n(x,0)
                         - \int d^4 x \overline\Psi_n(x,L_s-1)(1-\gamma^5)\Psi(x,L_s-1) }
                       {\int d^4 x \overline\Psi_n(x,0)(1+\gamma^5)\Psi(x,0)
                         + \int d^4 x \overline\Psi_n(x,L_s-1)(1-\gamma^5)\Psi(x,L_s-1) }
\end{equation}
which compensates for the fact that even for a chirality eigenstate, the five-dimensional wave function will not be localized solely on one of the four-dimensional walls but will spread into the fifth dimension.  If we examine the zero modes, we find that some of them have chiralilty near zero.  This might be expected for a not-too-dilute instanton gas where the two modes of a nearby instanton-anti-instanton pair will mix so that neither have a definite chirality,  However, such behavior could also be the result of our strong coupling and gauge configurations with changing topology producing zero modes of uncertain chirality.  As a result we choose to examine only those near-zero modes whose chirality is greater than 0.7 in magnitude.  The effects of this choice choice can be seen in Fig.~\ref{fig:eigchi} where we plot the histogram of the near-zero modes for $T=177$, 186 and 195 MeV. It appears that at these temperatures, almost all of the near-zero modes are localized on one of the two four-dimensional walls and thus have a chirality very close to +1 or -1.  Our restriction that the magnitude of the chirality is greater than 0.7 captures approximately 95\% of the near-zero modes.  Figure~\ref{fig:eigchi} suggests that this concentration of chirality at $\pm 1$ increases with increasing temperature.  Determining whether this apparent trend is the result of i) limited statistics at the higher temperatures, ii) increasing spatial localization of the zero modes and therefore less mixing as $T$ increases or iii) better defined gauge field topology at weaker coupling requires further study.

Table~\ref{tab:zero} lists the number of configurations which have $N_0$ near-zero modes, $N_+$ of which have positive chirality.  Those modes included in the counts presented in Tab.~\ref{tab:zero}  must lie in the peak region (first four bins) shown in Fig.~\ref{fig:low}, with $\Lambda$ at or below approximately 12.5 MeV  and with a chirality of magnitude 0.7 or greater.  A binomial distribution consistent with the DIGA describes the data in a more convincing way than the bimodal distribution that would be seen for the exact zero modes resulting from non-zero global topology. 

We conclude that the agreement between the value of $\dpd$ measured from the difference of correlators and the delta-function contribution $\dpd^0$ shown in Tab.~\ref{tab:cpmd} implies that the anomalous breaking of chiral symmetry for $T>T_c$ results from these near-zero modes.   Further, the volume dependence and chirality distribution of the modes making up this delta-function contribution gives strong evidence that the non-zero anomalous symmetry breaking found above $T_c$ is the result of a dilute gas of instantons and anti-instantons and that no new mechanism of anomalous symmetry breaking is needed.

\begin{table}[hbt]
  \centering
  \begin{tabular}{c|cccc|ccccc}\hline
    \#
    &$T\,(\textrm{MeV})$
    &$\beta$
    &$\mtot_l$
    &$N_\text{cfg}$
    &$\dpd^0/T^2$
    &$\dpd^1/T^2$
    &$\dpd^2/T^2$
    &$\dpd^\text{ms}/T^2$
    &$\dpd/T^2$\\\hline
    \runn{150_a}&149&1.671&0.00464&158& -      &3.7(3)&76(2)&109&87(2)\\
    \runn{160_a}&159&1.707&0.00421&109& -      &4.6(1)&42(1)& 70 &60(2)\\
    \runn{170_a}&168&1.740&0.00395& 83 & -      &4.9(1)&11(1)& 35 &35(2)\\
    \runn{180_a}&177&1.771&0.00367&170&23(1)&5.0(1)&  -     & 25 &23(2)\\
    \runn{190_a}&186&1.801&0.00341&171& 8(1) &-        &  -     &  8  & 6(1) \\
    \runn{200_a}&195&1.829&0.00314& 76 & 7(1) &-        &  -     &  6  & 6(2) \\
    \hline
  \end{tabular}
  \caption{
  \label{tab:cpmd}
A comparison of $\dpd$ measured from the difference of correlation functions with the three contributions computed from fitting the eigenvalue density to the expression in Eq.~\eqref{eq:expn} and with the result $\dpd^\text{ms}$ obtained from the mode sum given in Eq.~\eqref{eq:eigen_chi_sub_ms}, for the $32^3\times8$ ensembles.  All results are renormalized in the $\overline{\mathrm{MS}}(\mu=2\mathrm{GeV})$ scheme.}
\end{table}

\begin{table}[hbt]
  \centering
  \begin{tabular}{c|cccccc}\hline
    $N_+\backslash N_0$&0&1&2&3&4&5\\ \hline
    $N_0=1$    &40&29& -& -& -& -\\
    $N_0=2$    &11&20&12& -& -& -\\
    $N_0=3$    & 3&11& 6& 2& -& -\\
    $N_0=4$    & 0& 1& 2& 1& 0& -\\
    $N_0=5$    & 0& 2& 0& 0& 0& 0\\
    \hline
  \end{tabular}
  \caption{
  \label{tab:zero}
The number of configurations found in the $177$ MeV (\runref{180_a}) ensemble with given values for the total number ($N_0$) of near-zero modes and total number ($N_+$) of those modes with positive chirality.  We consider only modes with $\Lambda \le 12.5$ MeV and a chirality whose magnitude exceeds 0.7.  The distribution is clearly different from the bimodal distribution $N_+ = N_0$ or 0 expected if these near-zero modes were induced by non-zero global topology and the Atiyah-Singer theorem.}
\end{table}

\section{Conclusions}
\label{sec:concl}

We have extended earlier finite temperature QCD studies~\cite{Bazavov:2012qja} from $16^3\times8$ to larger $24^3\times8$ and $32^3\times8$ volumes, all performed using a 200 MeV pion mass and the chiral, DWF lattice action.   Significant dependence on volume is seen for both the chiral condensate, $\Sigma_l$, and the disconnected chiral susceptibility, $\chid$, for temperatures below $T_c$.  Most dramatic is the large decrease in $\chid$ below $T_c$ as the volume is increased from $16^3$ to $24^3$ and $32^3$ which is shown in the left panel of Fig.~\ref{fig:chi}.  Without data at one or more additional values of the light quark mass, we are unable to make a proper comparison of this finite volume dependence with the predictions of $O(4)$ universality.  However, on a qualitative level this behavior is predicted by finite-volume $O(4)$ scaling~\cite{Engels:2013xx} and was anticipated by the results given in Ref.~\cite{Braun:2010vd}.  Here a model calculation is presented using renormalization group methods applied to a theory including fundamental quarks, gluons and mesons.  Since the volume dependence of this theory should be consistent with $O(4)$ universal behavior, the results in Ref.~\cite{Braun:2010vd} can be viewed as a prediction of $O(4)$ universal finite volume behavior which is now evident in our lattice calculation.  We expect to make a quantitative comparison with finite volume $O(4)$ scaling when the HotQCD $32^3\times8$ and $64^3\times8$, $m_\pi=135$ MeV data can be included in the analysis.

A second result presented here is the observation of non-vanishing $\ua$ symmetry breaking above $T_c$ and its quantitative connection to the density of near-zero Dirac eigenvalues.  The volume dependence of these near-zero modes and the failure of their chiralities to be correlated per configuration matches precisely the expectation of the dilute instanton gas approximation.  This might also be called the dilute caloron gas approximation if we recognize the finite temperature distortions that are expected for instantons at finite temperature whose space-time extent approaches the length $1/T$~\cite{Harrington:1978ve, Harrington:1978ua, Kraan:1998kp, Kraan:1998pm, Lee:1998bb}.   (For a thorough review of the subject of instantons in QCD, including their effects at finite temperature, see Ref.~\cite{Schafer:1996wv}.)  While more study of the space-time structure of these zero modes is required to completely establish this picture of $\ua$ symmetry breaking, our results are all well-explained by this mechanism.

The possible $\ua$ symmetry breaking above $T_c$  was recently analyzed theoretically by Aoki, {\it et al.} using a lattice regularization, based on overlap fermions~\cite{Aoki:2012yj}.  We also refer the reader to this paper for a discussion of and references to earlier theoretical work on the question of $\ua$ symmetry breaking above $T_c$ and its relation to the Dirac eigenvalue spectrum.  Among the conclusions of Ref.~\cite{Aoki:2012yj} is that $\pmd$ vanishes in the limit of infinite volume and vanishing quark mass for $T > T_c$.   We have found a non-zero value for $\pmd$ on the smallest, $16^3$ volume which becomes larger when the volume was increased eight-fold to $32^3$.  While we have examined only a single quark mass, we believe that this mass is sufficiently small as to be a good approximation to zero.  We believe this to be the case because the explicit $\sua$ symmetry breaking effect of the quark mass on the difference $(\pmd) - (\sme)$ is significantly smaller than the scale of $\pmd$.   (We  are now studying a second, smaller mass to test this assertion.)  However, our results and the arguments presented in Ref.~\cite{Aoki:2012yj} can be made consistent if those arguments are reversed to conclude that the analyticity in $\mtot_l^2$ assumed above $T_c$ in Ref.~\cite{Aoki:2012yj} is not present.

There is also a potential conflict between our results and the conclusions of a recent 2-flavor  study of Cossu, {\it et al.}~\cite{Cossu:2013uua} on a  $16^3\times8$ volume using overlap fermions.  Reference~\cite{Cossu:2013uua} reaches the conclusion that there is a gap in the Dirac eigenvalue spectrum and degeneracy between the $\pi$ and $\delta$ correlators above $T_c$.  However,  the numerical evidence supporting their conclusion is strongest at relatively high temperatures where our results also show few small Dirac eigenvalues and small (but significant) results for $\pmd$.   Given our larger volumes and our smaller light quark mass, which is fixed in physical units, it is possible that the small effects which we are able to extract may not be visible in this first overlap study.  

Especially interesting is the failure of this overlap calculation to see the small peaks in the Dirac spectrum near $\lambda = 0$ found in our DWF work.  As is pointed out by Cossu, {\it et al.}, residual chiral symmetry breaking in a DWF calculation does distort the small eigenvalue region.  However, while this distortion may shift individual eigenvalues by a few MeV, it is not expected to create near-zero modes that are not present in the continuum theory.  Our detailed comparisons of the predictions of spectral formulae with the improved chiral condensate suggest that the averaged features of the Dirac spectrum, even for $\lambda \sim 1$ MeV, are accurate.   We believe that this absence of a near-zero mode peak in the overlap data has at least two possible explanations.  First since the size of these peaks is very temperature dependent, even a 10\% underestimate of the energy scale for the overlap relative to the DWF simulation could explain their absence in the former.  Second, the elimination of topology change in the overlap simulation results in a non-ergodic evolution algorithm which may distort the thermal distribution of near-zero modes, especially at weaker couplings and smaller dynamical quark masses, in spite of the evidence to the contrary.

The study of $16^3$, $24^3$ and $32^3$ volumes in this work gives us a good understanding of the effects of finite volume and a very interesting opportunity to compare with the predictions of $O(4)$ universality.  By working at relatively small light quark mass on a line of constant physics ($m_\pi = 200$ MeV), we believe that the effects of explicit chiral symmetry breaking present are small and that the evidence for anomalous symmetry breaking just above $T_c \approx 160$ MeV is strong. This symmetry breaking decreases rapidly as the temperature grows making the signal difficult to see at our highest temperature, 196 MeV.  The study of the Dirac eigenvalue spectrum suggests that this $\ua$ symmetry breaking results from near-zero modes whose characteristics match well with those predicted by the dilute instanton gas approximation.  However, it is important to verify this picture by extending the investigation to even smaller light quark mass and larger volumes.  Calculations currently being carried out by the HotQCD collaboration on $32^3\times8$ and $64^3\times8$ volumes with $m_\pi=135$ MeV should resolve these remaining uncertainties.

We would like to thank Peter Boyle whose high performance Blue Gene/Q code was essential to these calculations and Ron Soltz for his support of this project.  Code optimization has been supported through the Scientific Discovery through Advanced Computing (SciDAC) program funded by U.S. Department of Energy, Offices of Science, Advanced Scientific Computing Research, Nuclear Physics and High Energy Physics.  In addition, this work has been supported in part by contracts DE-AC02-98CH10886, DE-AC52-07NA27344, \ DE-FC06-ER41446, \ DE-FG02-92ER40699, \ DE-FG02-91ER-40628, DE- FG02-91ER-40661, DE-FG02-04ER-41298, DE-KA-14-01-02 with the U.S. Department of Energy, Lawrence Livermore National Laboratory under Contract DE-AC52-07NA27344, NSF grants PHY0903571, PHY08-57333, PHY07-57035, PHY07-57333 and PHY07-03296.  NC, ZL, RM and HY were supported in part by U.S. DOE grant DE-FG02-92ER40699.  MIB was supported in part by DE-FG02-00ER41132.  The numerical calculations have been performed on the Blue Gene/Q computers at the IBM T. J. Watson and the RIKEN BNL Reseach Centers, as well as the Blue Gene/L and Blue Gene/P computers at Lawrence Livermore National Laboratory (LLNL) and the New York Center for Computational Sciences (NYCCS) at Brookhaven National Laboratory.  We thank the LLNL Multiprogrammatic and Institutional Computing program for time on the LLNL Blue Gene/L and Blue Gene/P supercomputers.

\bibliography{main} % Produces the bibliography via BibTeX.

\end{document}